\documentclass[aps,prb,reprint,superscriptaddress]{revtex4-2}
\usepackage{graphicx}% Include figure files
\usepackage{dcolumn}% Align table columns on decimal point
\usepackage{bm}% bold math
\usepackage{xcolor}

\begin{document}

%\preprint{APS/123-QED}

\title{Enhanced two-dimensional ferromagnetism in van der Waals $\beta$-UTe$_3$ monolayers}

\author{S. M. Thomas}
\email{smthomas@lanl.gov}
\affiliation{Los Alamos National Laboratory, Los Alamos, NM 87545}
\author{A.E. Llacsahuanga}
\affiliation{Department of Physics and Astronomy, Purdue University, West
Lafayette, Indiana, 47907, USA}
\author{W. Simeth}
\affiliation{Los Alamos National Laboratory, Los Alamos, NM 87545}
\author{C. S. Kengle}
\affiliation{Los Alamos National Laboratory, Los Alamos, NM 87545}
\author{F. Orlandi}
\affiliation{ISIS Facility, Rutherford Appleton Laboratory, Chilton, Didcot, OX11 0QX, United Kingdom}
\author{D. Khalyavin}
\affiliation{ISIS Facility, Rutherford Appleton Laboratory, Chilton, Didcot, OX11 0QX, United Kingdom}
\author{P. Manuel}
\affiliation{ISIS Facility, Rutherford Appleton Laboratory, Chilton, Didcot, OX11 0QX, United Kingdom}
\author{F. Ronning}
\affiliation{Los Alamos National Laboratory, Los Alamos, NM 87545}
\author{E. D. Bauer}
\affiliation{Los Alamos National Laboratory, Los Alamos, NM 87545}
\author{J. D. Thompson}
\affiliation{Los Alamos National Laboratory, Los Alamos, NM 87545}
\author{Jian-Xin Zhu}
\affiliation{Los Alamos National Laboratory, Los Alamos, NM 87545}
\author{A. O. Scheie}
\affiliation{Los Alamos National Laboratory, Los Alamos, NM 87545}
\author{Yong P. Chen}
\affiliation{Department of Physics and Astronomy, Purdue University, West
Lafayette, Indiana, 47907, USA}
\affiliation{School of Electrical and Computer Engineering and Purdue Quantum
Science and Engineering Institute, Purdue University, West Lafayette,
Indiana, 47907, USA}
\author{P. F. S. Rosa}
\email{pfsrosa@lanl.gov}
\affiliation{Los Alamos National Laboratory, Los Alamos, NM 87545}

\date{\today}% It is always \today, today,
\maketitle

\textbf{The discovery of local-moment magnetism in van der Waals (vdW) \textit{semiconductors} down to the single-layer limit has led to a paradigm shift in the understanding of two-dimensional (2D) magnets and unleashed their potential for applications in microelectronic and optoelectronic devices~\cite{lee2016ising,wang2016raman,Gong2017,Huang2017,Burch2018,Wang2022,Ziebel2024}.
The incorporation of strong electronic and magnetic correlations in 2D vdW \textit{metals} remains a sought-after platform not only to enable control of emergent quantum phases, such as superconductivity, but also to achieve more theoretically tractable microscopic models of complex materials \cite{Monthoux2007, Mizukami2011, Cao2018, Artificial2021, LEDWITH2021, Simeth2023}.
To date, however, there is limited success in the discovery of such metallic vdW platforms~\cite{Han2018,Zhang2018,May2019}, and $f$-electron monolayers remain out of reach~\cite{Lei2020,Posey2024,Broyles2025}.
Here we demonstrate that the actinide $\beta$-UTe$_3$ can be exfoliated to the monolayer limit.
A sizable electronic specific heat coefficient provides the hallmark of strong correlations. Remarkably, $\beta$-UTe$_3$ remains ferromagnetic in the half-unit-cell limit with an enhanced ordering temperature of 35~K, a factor of two larger than its bulk counterpart.
Our work establishes $\beta$-UTe$_3$ as a novel materials platform for investigating and modeling correlated behavior in the monolayer limit and opens numerous avenues for quantum control with, \textit{e.g.}, strain engineering.}

%\section{Introduction}

Two decades ago, the exfoliation of graphite down to single-layer graphene sheets transformed the aspiration of atomic-scale materials design into reality~\cite{Novoselov2004}.
%Graphene's high flexibility, electron mobility, and biocompatibility are being actively investigated in microelectronics, energy storage~\cite{Raccichini2015}, and biomedical applications~\cite{Chung2013}. 
When two or more graphene layers are stacked, novel quantum states can be designed from strongly correlated electrons that were not present in a single layer~\cite{Andrei2020}.
Examples include fractional Chern insulator states in twisted bilayer graphene~\cite{Xie2021} and unconventional, chiral superconductivity in pentalayers~\cite{han2025}.
Since graphene, many atomically thin, layered vdW crystals have been discovered and now provide a vast materials platform for exceptional properties and applications in 2D~\cite{novoselov2016,tan2017,Castellanos2022}. 
A notable recent development is the observation of local-moment magnetism in \textit{semiconducting} monolayers~\cite{lee2016ising,wang2016raman,Gong2017,Huang2017,Burch2018,Wang2022}.
Though conventional wisdom does not expect ordering in the isotropic 2D limit due to the Mermin-Wagner theorem~\cite{Gibertini2019}, the presence of magnetic anisotropy and finite-size effects enables the stabilization of both ferromagnetic (FM) and antiferromagnetic (AFM) states~\cite{lee2016ising,wang2016raman,Gong2017,Huang2017,Burch2018,Wang2022,Ziebel2024}.
% The addition of robust local-moment degrees of freedom to the landscape of 2D vdW semimetals and semiconductors has opened entirely new avenues of inquiry from unprecedented tests of 2D theoretical spin models to novel, highly-tunable spintronic devices. 

In a \textit{metallic} 2D magnet, enhanced quantum fluctuations are expected and magnetic order may be destabilized in the 2D limit. As a result, sought-after states of matter may emerge in the monolayer limit ranging from a 2D strange metal that defies the concept of quasiparticles within Fermi-liquid theory \cite{varma2016quantum,Phillips2022} to unconventional superconductivity with topological excitations envisioned as the building block of quantum computing \cite{Alicea_2012}. 
%One key outstanding question is therefore the fate of magnetic excitations close to a nonmagnetic boundary in metallic vdW monolayers. 

Most known vdW magnets to date, however, are semiconducting and exhibit rather robust magnetic order with sizable ordered moments.
Rare examples of metallic vdW magnets include $d$-electron ferromagnets Fe$_{n}$GeTe$_{2}$ ($n=3,4,5$) and $f$-electron antiferromagnets CeSiI, UOTe, and \textit{R}Te$_{3}$ ($R =$ rare earth).
%~\cite{Fei2018, Deng2018, Wang2023, May2019, Broyles2025, Posey2024}. 
UOTe orders at $150$~K with a small electronic (Sommerfeld) contribution to the specific heat, $\gamma = 6$~mJ/mol.K$^{2}$, which indicates weak hybridization between local $4f$ moments and conduction electrons~\cite{Broyles2025}.
\textit{R}Te$_{3}$ members also show local-moment magnetism \cite{Tongay2021,Ru2006}, and GdTe$_{3}$ has been exfoliated only to the few-layer regime~\cite{Lei2020}.
Though Fe$_{n}$GeTe$_{2}$~\cite{zhu2016,May2019,Fei2018,Deng2018,Wang2023,dang2023} and CeSiI~\cite{Posey2024} do show signs of modest electronic correlations, as evident from their moderately enhanced Sommerfeld coefficients, these systems appear to be far from a magnetic instability.
In addition, UOTe and CeSiI have not been exfoliated down to the monolayer limit.

\begin{figure*}[!ht]
	\includegraphics[width=\textwidth]{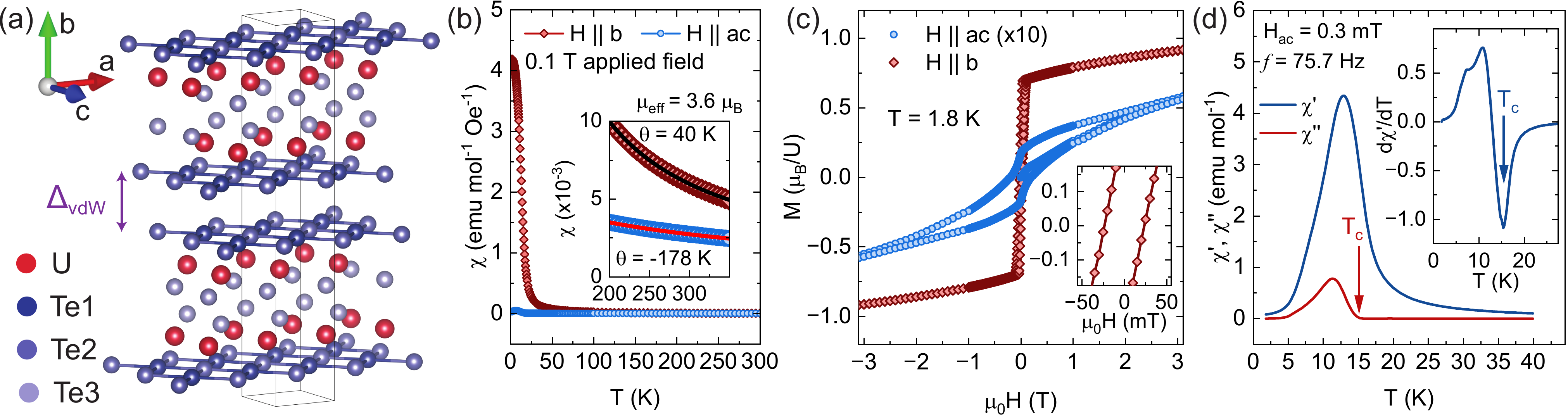}
	\caption{
        Crystal structure and magnetic measurements.
        (a)~vdW crystal structure of $\beta$-UTe$_{3}$.
        % This info is probably better to put in the text.
        %The van-der Waals gap is between the Te1/Te2 square-net layers.
        %Each half unit cell has two layers of U square nets with AB stacking.
        (b)~Magnetic susceptibility $vs.$ temperature for fields applied along the $b$~axis and in the $ac$~plane.
        The inset shows Curie-Weiss fits between 200~K and 350~K (solid lines). 
        (c)~Magnetization $vs.$ field at 1.8~K. The $ac$~plane data are multiplied by 10 for clarity. The inset shows a zoomed-in view of the $b$-axis data.
        (d)~Real ($\chi^{'}$) and imaginary ($\chi^{''}$) parts of the ac susceptibility $vs.$ temperature. Inset shows the derivative of $\chi^{''}$ with respect to temperature.
    }
	\label{fig:mag}
\end{figure*}

%{\color{red}Somewhere we should make some reference to other tri-tellurides. There is a nice review here (all other members are AFM or nonmagnetic)~\cite{Tongay2021}. In CeTe3, it is hard to see evidence of correlations in heat capacity due to the large magnetic anomaly~\cite{Ru2006}. Also maybe cite Schoop's Sci. Adv. article where they observe very high mobility, but are only able to exfoliate down to 3 1/2-layers (this could probably be put in the exfoliation discussion~\cite{Lei2020}}.
Here we investigate actinide vdW $\beta$-UTe$_{3}$ down to the half-unit-cell limit.
Bulk crystals show easy-axis magnetic anisotropy and a FM transition at $T_{C}= 15$~K in magnetic susceptibility data. Our specific heat and neutron diffraction measurements are consistent with a three-dimensional FM order wherein the interlayer magnetic exchange interaction is much smaller than the intralayer interaction, typical of quasi-2D magnetic systems \cite{peters1988,sun1991heat}. In fact, 2D correlations are clearly observed above $T_{C}$ in neutron diffraction experiments.
%{\color{red}Specific heat data, however, lack the $\lambda$-like peak at $T_{C}$ expected from a conventional three-dimensional (3D) second-order phase transition. Remarkably, our neutron diffraction measurements confirm a 3D FM ($k=0$) state with spins pointing out of the plane. The small entropy in specific heat at $T_{C}$ can be therefore explained by a tiny interlayer magnetic exchange interaction. The extracted critical exponent of $\beta = 0.23$ is not consistent with the 3D Heisenberg model and indicates that 2D contributions need to be considered. In fact, 2D correlations are clearly observed above $T_{C}$ in neutron diffraction experiments.}
A sizable Sommerfeld coefficient of $\gamma = 130$~mJ/mol.K$^{2}$ is observed in $\beta$-UTe$_{3}$, akin to other strongly correlated $5f$ Kondo metals and superconductors~\cite{Stewart1984,WHITE2015,Aoki2019,aoki2022unconventional}.
Remarkably, $\beta$-UTe$_3$ remains FM down to the half-unit-cell limit with an enhanced ordering temperature of $T^{ML}_{C}=35$~K, which stems from an `extraordinary' phase transition.
Our work establishes $\beta$-UTe$_3$ as a novel materials platform for investigating, controlling, and modeling the interplay between dimensional confinement, electronic correlations, and magnetism in 2D vdW materials.

%\section{Experimental Results}

$\beta$-UTe$_3$ crystallizes in space group $Cmcm$ (No. 63) in a pseudo-tetragonal structure with lattice parameters $a$~=~$c$~=~4.338~\AA$\:$ and $b$~=~24.743~\AA~\cite{Noel1989}.
The layered crystal structure, shown in Fig.~\ref{fig:mag}a, reveals a vdW gap ($\Delta_{\mathrm{vdW}}$) between the square nets of Te1 and Te2 atoms. The position of $\Delta_{\mathrm{vdW}}$ suggests the possibility of exfoliation down to a thickness of half of a unit cell, which is indeed consistent with the results presented below. Each half unit cell is comprised of two layers of U square nets with AB stacking. U atoms experience an environment that is locally noncentrosymmetric even though the unit cell has an inversion center that lies in the vdW gap.  \\

\noindent\textbf{3D ferromagnetic order in bulk $\beta$-UTe$_{3}$}

Magnetic susceptibility measurements, presented in Fig.~\ref{fig:mag}b, reveal a highly anisotropic response and a clear FM transition at low temperatures, in agreement with previous reports~\cite{Noel1989}.
For out-of-plane fields, a Curie-Weiss (CW) fit yields a positive Weiss temperature, $\theta_{W}=40$~K, consistent with interlayer FM interactions (inset of Fig.~\ref{fig:mag}b). Similar fits to the $ac$-plane data yield a larger negative value for $\theta_{W}$, which, in the absence of crystal field effects, points to the presence of competing intralayer AFM interactions. Our CW fit also yields an effective moment of $\mu_{eff} = 3.6~\mu_{B}$, which matches the expected Hund's values for either U$^{3+}$ (3.62 $\mu_{B}$) and U$^{4+}$ (3.58 $\mu_{B}$). Overall, our magnetic susceptibility data reveal that the easy magnetization axis is the $b$~axis, which in turn shows a tiny coercive field of 100~mT in magnetization loops (Fig.~\ref{fig:mag}c), consistent with soft FM. To precisely determine the critical FM temperature, $T_{C}$, we performed ac susceptibility measurements at low driving fields~\cite{balanda2013ac}. As shown in Fig.~\ref{fig:mag}d, $\chi^{''}$ sets in at $T_{C}=15$~K, consistent with the minimum in $d\chi^{'}/dT$ (inset).
%There is much larger hysteresis observed for fields applied in the $ac$~plane as well as a kink-like feature near zero field.
% I'm not sure if this is a correct statement, but I thought Wolfgang mentioned something about an AFM component
%PR: I would suggest not going into more details here. I'm afraid I'm going to much into the weeds of these measurements. For Nature journals, we should try to give a high-level description of what's going on and stay in the main message.

According to the Ginzburg-Landau mean-field theory, the magnetization squared ($M^{2}$) is linearly proportional to $H_{int}/M$, where $H_{int}$ is the applied field after taking into account demagnetization effects \cite{Parisi1988}.
An $M^{2}$ $vs.$ $H_{int}/M$ plot, known as the Arrott plot~\cite{Arrott1957}, should in turn yield linear curves close to $T_{C}$, consistent with mean-field critical exponents $\beta = 0.5$ and $\gamma = 1.0$.
The significant curvature found in Fig.~S7 readily reveals that the FM transition in $\beta$-UTe$_3$ is not mean field, and different exponents are at play.
Efforts to obtain critical exponents using a modified Arrott plot, however, did not yield self-consistent results~\cite{Pramanik2009}. 

\begin{figure*}
    \includegraphics[width=0.9\textwidth]{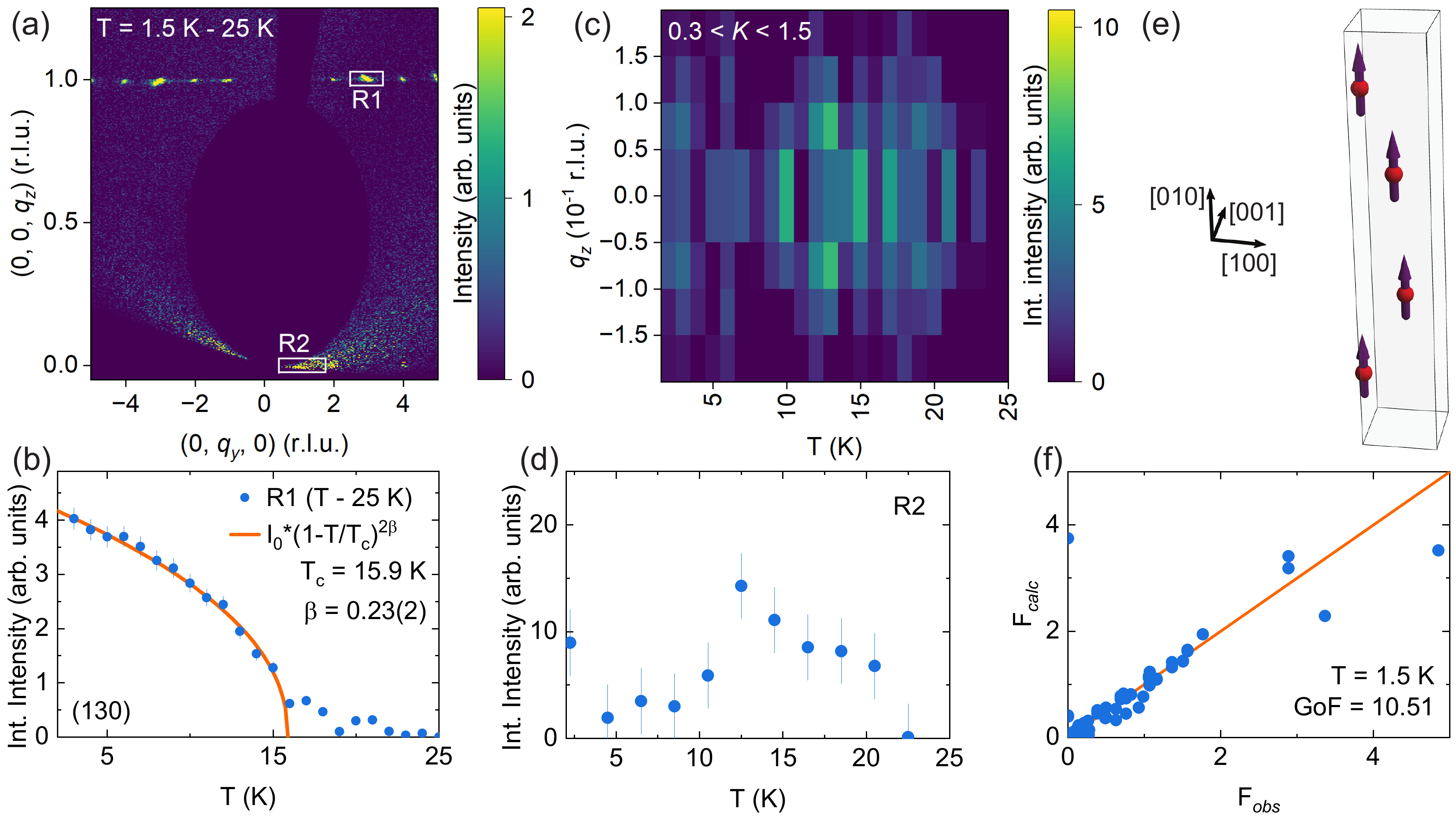} 
    \caption{Neutron diffraction study of bulk magnetic order in UTe$_3$.
    (a)~Momentum-space slice through magnetic diffraction intensity obtained by temperature subtraction (1.5\,K--25\,K). Color coding reflects scattering intensity in the plane $\boldsymbol{q}=(0,q_y,q_z)$, which is a superposition of two twins with swapped $H$ and $L$-axes: (0, K, L) and (H, K, 0) planes. Peaks on the line $q_z=1$ correspond to resolution limited magnetic Bragg peaks (see SI for further information). The two rectangles R1 and R2 indicate schematically integration windows for the integrated intensities presented in panels (b) and (d). 
    %Negative intensities, which arise due to thermal variation of lattice parameters, are not covered by the color-scale.
    (b)~Thermal variation of the magnetic Bragg peak $\bm{Q}=(130)$. Blue data points denote integrated intensity of the Bragg peak in the rectangular region of interest R1. The orange line corresponds to an order parameter fit with the indicated parameters.
    (c)~Variation of diffuse scattering on the $K$-axis in region R2 as a function of momentum space coordinate $q_z$ and temperature. 
    %The color coding reflects the intensity in the rectangular region R2 shifted along the coordinate $L$.
    Intensity was integrated within a momentum-space volume of 0.1 r.l.u along $q_z$ and $q_x$. For illustration purposes, data were symmetrized with a reflection around $q_z=0$.
    (d)~Temperature dependence of total integrated diffuse scattering intensity (blue data points) integrated over $0.3\leq K\leq 1.5$, $-0.05\leq q_x\leq 0.05$, and $-0.07\leq q_z\leq 0.13$. Measurements were performed at the same temperatures as in panel (c) and averaged over pairs of temperatures.
    (e) Ferromagnetic spin texture determined from magnetic structure refinements and a careful analysis of recorded magnetic structure factor at $T=1.5\,$K (see SI). The moments are essentially pointing along the $b$-axis.
    (f)~Comparison of calculated and observed structure factor for the combined magnetic and structural refinement carried out at $T=1.5\,$K.
    The refined magnetic moment obtained from this fit is $M=0.48(11)\,\mu_{\text{B}}$.
    }
    \label{fig:neutrons}
\end{figure*}

To address this challenge, we turn to neutron diffraction measurements of a bulk crystal.
The temperature-subtracted magnetic neutron diffraction intensity map, shown in Fig.~\ref{fig:neutrons}a, reveals Bragg peaks at integer valued momentum transfers (H, K, L), consistent with 3D FM order.
Figure~\ref{fig:neutrons}b presents the temperature dependence of the integrated intensity around R1, the magnetic Bragg peak at momentum transfer (1, 3, 0).
An order parameter fit (solid line) yields a critical exponent of $\beta = 0.23(2)$ and a critical temperature of $T_{C}=15.9(9)$~K.
The extracted exponent does not match the mean-field value ($\beta = 0.5$), in agreement with magnetization data.
The exponent also does not match the expected values from the 3D Heisenberg ($\beta = 0.365$), the 3D Ising ($\beta = 0.325$), or the 2D Ising ($\beta = 0.125$) models~\cite{Parisi1988,Gibertini2019}.
Instead, $\beta = 0.23$ is the expectation from a 2D XY phase transition in finite-sized samples~\cite{Bramwell1993}.
Such exponent has been observed in many layered vdW Heisenberg magnets with planar anisotropy because these materials can be treated as quasi-2D XY systems~\cite{Bedoya2021,scheie2022}.
$\beta$-UTe$_{3}$, however, displays easy-axis anisotropy with spins pointing out of the plane, which poses a puzzle to future investigations.

Evidence for 2D magnetic correlations is observed in the diffraction results within area R2 (see Fig.~\ref{fig:neutrons}a).
Figure~\ref{fig:neutrons}c shows a colormap of the integrated intensity as a function of temperature and momentum space coordinate L.
The corresponding temperature-subtracted diffraction intensity, shown in Fig.~\ref{fig:neutrons}d, reveals a significant diffuse scattering contribution along the reciprocal K axis at 15~K.
Notably, the diffuse signal survives well above $T_{C}$, consistent with the presence of 2D correlations in the $ac$ plane that peak around $T_{C}$. Magnetic structure refinements reveal a ferromagnetic structure with moments along the $b$ axis, as shown in Figs.~\ref{fig:neutrons}e,f, and a magnetic moment of $M=0.48(11)$~$\mu_{\mathrm{B}}$ (see SI for details). \\

\noindent\textbf{Evidence for 2D magnetism, electronic correlations, and metallicity} 

\begin{figure*}[!ht]
	\includegraphics[width=0.9\textwidth]{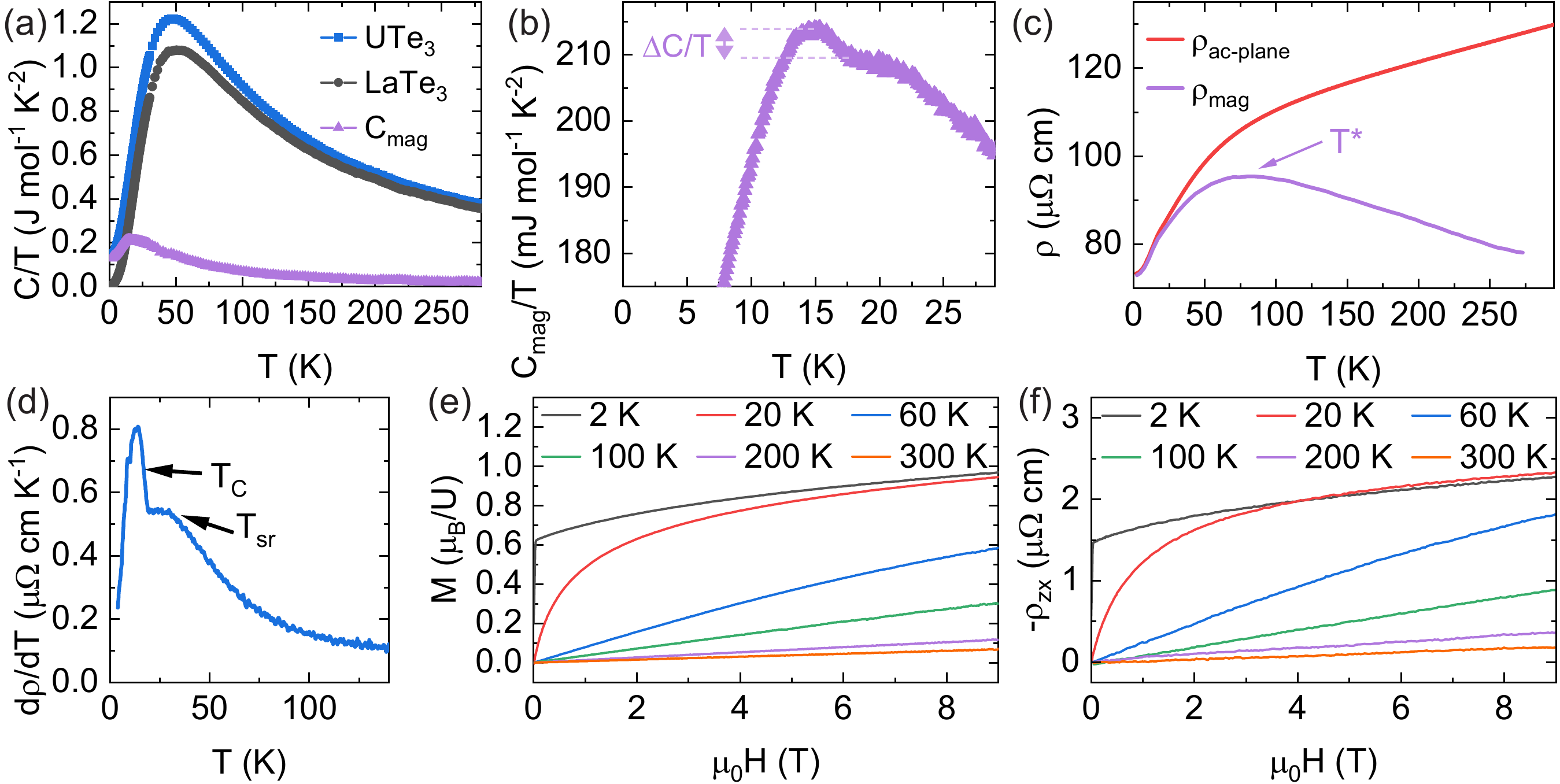}
	\caption{
        Physical property measurements.
        (a)~Specific heat over temperature versus temperature for $\beta$-UTe$_{3}$, LaTe$_3$, and the difference (C$_{\mathrm{mag}}$).
        (b)~A zoomed-in view of C$_{\mathrm{mag}}$/T versus temperature shows a slight feature (5 mJ/mol.K$^2$) at T$_{\mathrm{C}}$. 
        (c)~In-plane resistivity versus temperature. The magnetic contribution, $\rho_{\mathrm{mag}}$, shows a peak at $T^{*}=75$~K.
        (d)~Resistivity derivative $vs.$ temperature. The arrows indicate a jump at T$_{\mathrm{c}}$ and a broad shoulder near T$_{sr}$=30~K, attributable to the onset of short-range magnetic correlations.
        (e)~Magnetization versus field applied along the $b$ axis.
        (f)~Negative of Hall resistivity versus field applied along the $b$ axis.
    }
	\label{fig:phys}
\end{figure*}

Our thermodynamic data point to a highly correlated, 2D magnetic state in $\beta$-UTe$_{3}$. 
Figure~\ref{fig:phys}a shows the temperature-dependent specific heat, $C/T$, for both $\beta$-UTe$_{3}$ and its nonmagnetic analog LaTe$_{3}$, which is used as the phonon reference. Notably, $C/T$ data lack the expected $\lambda$-like peak characteristic of a three-dimensional (3D) second-order phase transition.
The magnetic contribution to the specific heat, $C_{\mathrm{mag}}/T$, increases on cooling as magnetic entropy builds up. At $T_{C}$, however, only a tiny anomaly is observed, whose entropy is about $15$~mJ/mol.K or 0.3\% of $R\ln{2}$, the entropy of a doublet ground state. The entropy reaches $R\ln{2}$ only at 30~K, consistent with the presence of short-range 2D correlations above $T_{C}$, as observed in neutron diffraction.
The small anomaly at $T_{C}$ is reminiscent of cuprate antiferromagnet La$_{2}$CuO$_{4}$, which shows no detectable entropy around its AFM transition at $T_{N} = 304$~K~\cite{sun1991heat}.
This apparent contradiction can be solved by recalling that the entropy of a 3D transition in a layered material will depend on the strength of the interlayer exchange interaction, $J_{\perp}$.
In La$_{2}$CuO$_{4}$, $J_{\perp}$ is about five orders of magnitude smaller than the intralayer exchange interaction, $J$, and the expected entropy is below the detection limit of typical calorimeters~\cite{peters1988,sun1991heat}.
We thus expect that $J_{\perp}$ is a few orders of magnitude smaller than $J$ in $\beta$-UTe$_{3}$.

Specific heat also allows the extraction of the Sommerfeld coefficient ($\gamma$) from the residual intercept in a $C/T$ vs $T^{2}$ plot (see Fig.~S4).
Within Fermi-liquid theory, $\gamma$ is proportional to the effective mass of electrons ($m^{*}$), and an enhanced $\gamma$ reflects strong underlying electronic correlations.
In $\beta$-UTe$_{3}$, we find $\gamma = 130$~mJ/mol.K$^{2}$, akin to other strongly correlated $5f$ Kondo metals and superconductors~\cite{Stewart1984,WHITE2015,Aoki2019,aoki2022unconventional}.
As a comparison, $\gamma$ is only 0.6~mJ/mol.K$^{2}$ in nonmagnetic LaTe$_{3}$, consistent with an uncorrelated metal.
Our results therefore point to a two-order-of-magnitude renormalization of $m^{*}$ in $\beta$-UTe$_{3}$.

To confirm the metallic state of $\beta$-UTe$_{3}$, we turn to electrical transport measurements and density functional theory (DFT) calculations.
The temperature-dependent in-plane electrical resistivity, $\rho_{ac}$(T), of a bulk $\beta$-UTe$_{3}$ crystal, shown in Fig.~\ref{fig:phys}c, reveals a monotonic decrease in $\rho_{ac}$ on cooling, consistent with metallic behavior (red curve). By subtracting the contribution from nonmagnetic LaTe$_{3}$~\cite{PhysRevB.73.033101}, we obtain the magnetic contribution to the electrical resistivity, $\rho_{\mathrm{mag}}$ (purple curve in Fig.~\ref{fig:phys}c).  Below $T^{*}= 75$~K, $\rho_{\mathrm{mag}}$ decreases more rapidly, and such a reduction in magnetic scattering is typical of the formation of a Kondo coherent state \cite{Yang2008}.
At low temperatures, a small kink is observed around $T_{C}$.
The transition is more clearly visualized in the derivative of $\rho_{ac}$ with respect to temperature, $d\rho_{ac}/dT$, shown in Fig.~\ref{fig:phys}d, which reveals a peak at $T_{C}$. A small hump is also apparent at $T_{sr}=30$~K, which matches the temperature at which the specific heat entropy reaches $R\ln{2}$ and marks the onset of short-range 2D magnetic correlations.

%Below $T_{C}$, $\rho_{ac}$ follows the $T^{2}$ dependence expected from a Fermi liquid (FL). A fit of the data to the FL expression, $\rho=\rho_{0} + A T^{2}$, yields a residual resistivity of $\rho_{0}=73$~$\mu\Omega$cm and a $T^{2}$ coefficient of $A = 0.03$~$\mu\Omega$cm/K$^{2}$.
%By combining the FL coefficient and the Sommerfeld coefficient, we extract the Kadowaki-Woods (KW) ratio $\frac{A}{\gamma^2} \approx 0.2\times 10^{-5}\mu\Omega\text{cm} (\frac{\text{mol K}}{\text{mJ}})^2$, which is approximately 5 times smaller than that observed in other heavy-fermion compounds \cite{Kadowaki}.

\begin{figure}[!hb]
	\includegraphics[width=1\columnwidth]{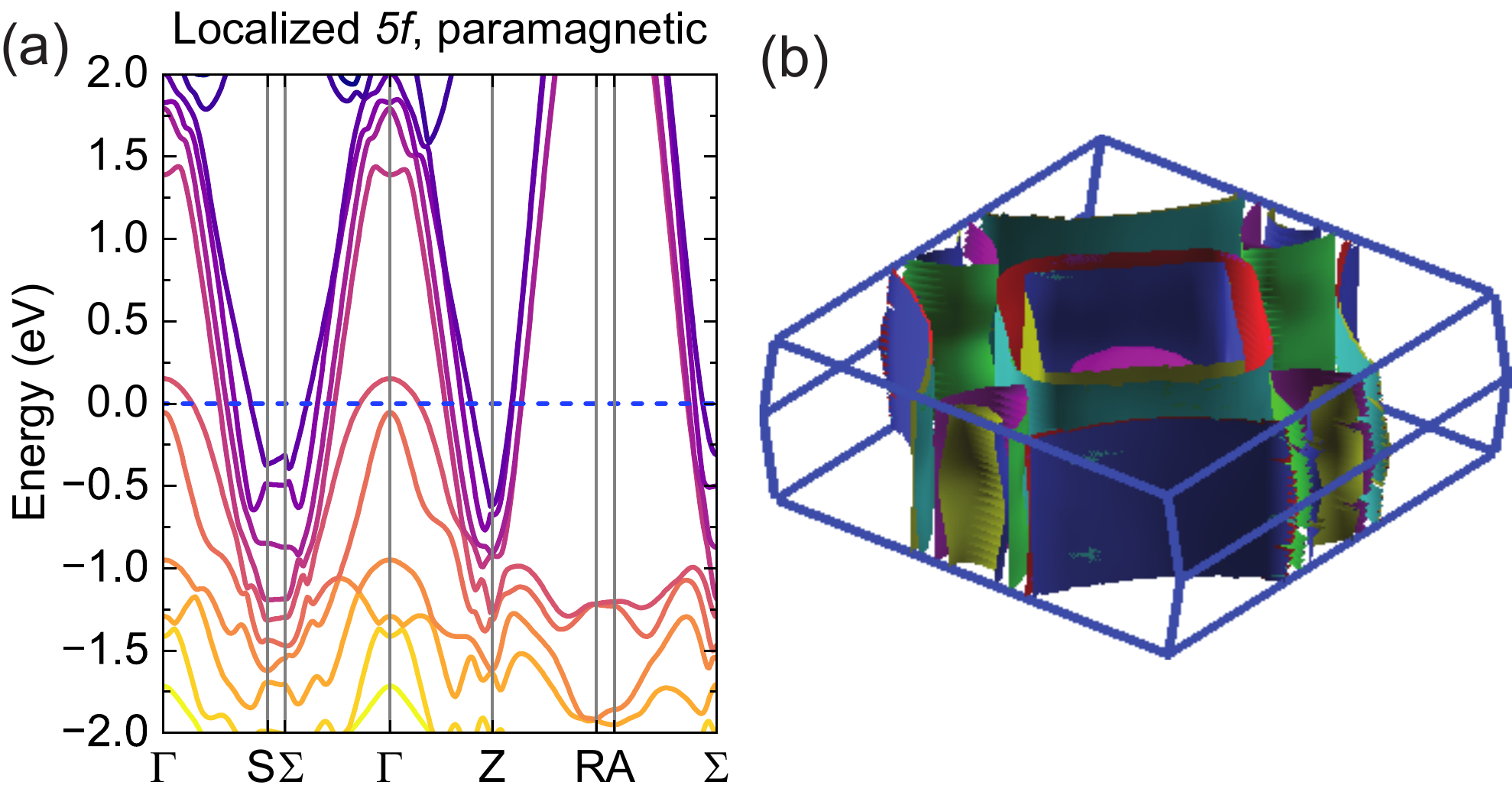}
	\caption{
        DFT calculations.
        (a)~DFT band structure calculation of $\beta$-UTe$_{3}$ in the nonmagnetic state with localized $f$ electrons and (b) its corresponding Fermi surface.
        Spin-orbit coupling is included.
    }
	\label{fig:DFT}
    \vskip -0.5cm
\end{figure}

\begin{figure*}[!ht]
	\includegraphics[width=\textwidth]{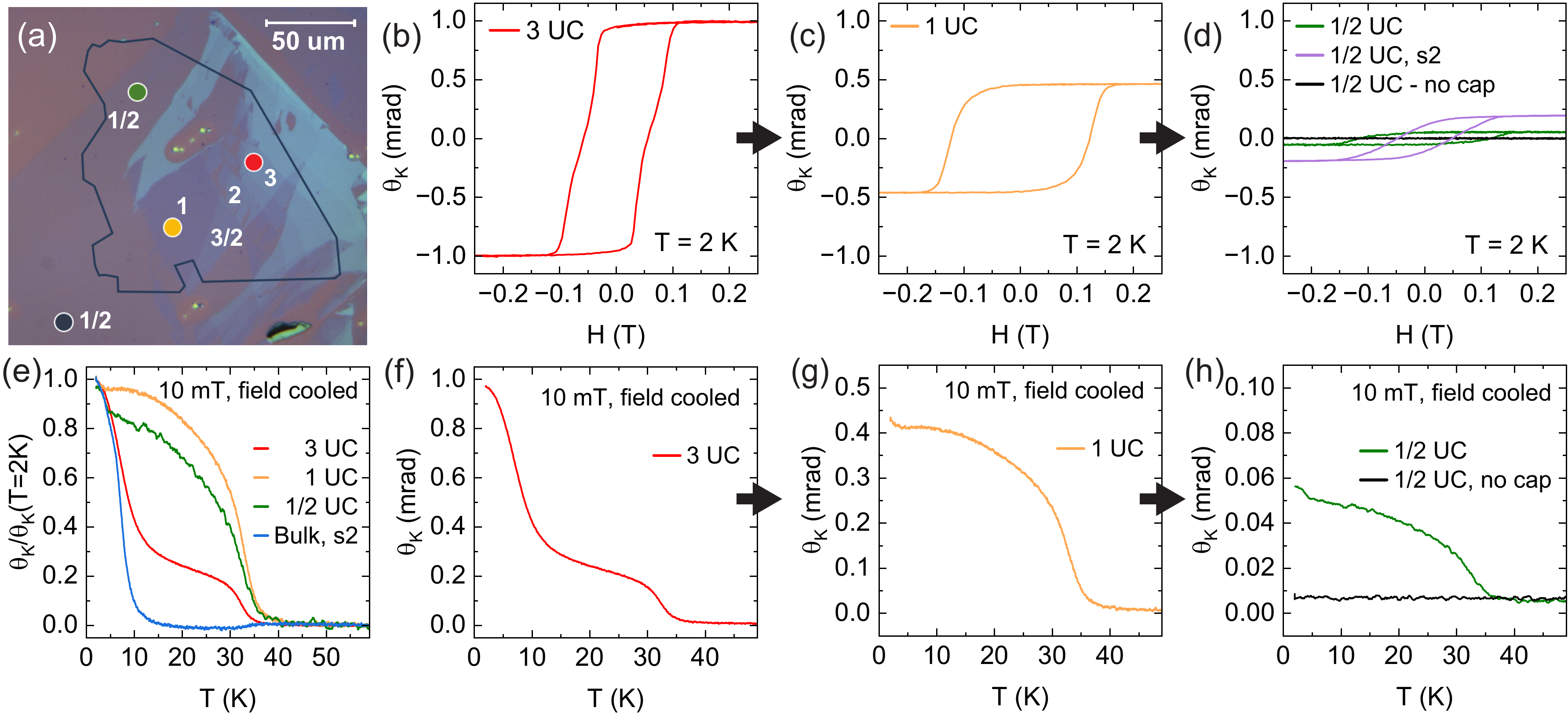}
	\caption{
        Magnetic measurements of exfoliated sample.
        (a)~Optical image of exfoliated UTe$_3$ before placing h-BN capping layer (indicated by gray outline). 
        Sample s2 is not shown.
        (b)--(d)~Kerr rotation versus field at 2~K for a number of different spots ranging from 3 unit cell to 1/2 unit cell thickness.
        (e)~Normalized kerr rotation versus temperature during cooling in a 10~mT applied field.
        (f)--(g)~Kerr rotation versus temperature during cooling in a 10~mT applied field for a number of different spots.
    }
	\label{fig:lowfield}
\end{figure*}
To determine the carrier density ($n$), we perform magnetization, $M$, and Hall resistivity, $\rho_{zx}$, measurements with fields applied along the $b$ axis, as shown in Figs.~\ref{fig:phys}e and \ref{fig:phys}f, respectively.
Generally, $\rho_{zx}$ can be written as $R_{0}H+R_{s}M$, where the first term is the ordinary Hall component due to the Lorentz force, and $R_{0}$ is inversely proportional to $n$ in a single-band approximation~\cite{nagaosa2010}.
Extracting $R_{0}$ from fits to the data proved challenging because the ordinary term is much smaller than the second term, $R_{s}M$, the Hall effect contribution due to the spontaneous magnetization, known as the anomalous Hall effect.
In fact, $\rho_{zx}$ displays a dominant anomalous component at all temperatures up to 300~K.
At low temperatures, however, the nonlinearity in the anomalous contribution allows the most reliable estimate of the carrier density.
At 2~K, $R_{0}=+2.6$~n$\Omega$cm$/$T, which corresponds to a carrier density of $n \sim  10^{23}$ holes/cm$^{3}$, consistent with a metallic state (see SI for details).

The electronic band structure of $\beta$-UTe$_{3}$ also supports a metallic correlated state. Our DFT calculations utilized the generalized gradient approximation (GGA) and the Perdew-Burke-Ernzerhof (PBE) exchange correlational function with the WIEN2k package \cite{SCHWARZ2003,Perdew1996}.
DFT calculations investigated two scenarios: Figure~\ref{fig:DFT}a shows the band structure of $\beta$-UTe$_{3}$ in the paramagnetic state assuming that three uranium $5f$ electrons are localized in the core, whereas Fig.~S8 shows the DFT+U band structure calculation in the FM state with a Coulomb term $U=5$~eV.
In the localized calculation, one small hole pocket and four larger pockets from Te $p$ bands are centered around $\Gamma=(0,0,0)$, whereas four electron pockets are present around $\Sigma=(0.258,0.258,0)$ and $Z=(0,0,1/2)$.
Importantly, dispersionless bands are observed along $S-\Sigma$ and $R-A$ directions, which imply a quasi-2D electronic structure along the $b$ direction.
Figure~\ref{fig:DFT}b shows the resulting quasi-2D Fermi surface of $\beta$-UTe$_{3}$.
The calculated density of states at the Fermi level, $N(E_{F})=0.56$~states/eV/f.u., yields $\gamma = 1.4$~mJ/mol.K$^{2}$.
This small Sommerfeld coefficient, extracted in the absence of $5f$ electrons at $E_{F}$, is consistent with the experimentally-measured coefficient for LaTe$_{3}$ and confirms the two-order-of-magnitude renormalization of m$^{*}$ in $\beta$-UTe$_{3}$. \\

\noindent\textbf{Enhanced magnetism in the 2D limit}\nopagebreak

We now turn to the evolution of the FM state down to the half-unit-cell limit. Figure~\ref{fig:lowfield}a shows an optical image of exfoliated $\beta$-UTe$_{3}$ flakes. The numbers indicate the thickness, in unit cells (UC), of each layer and reveal successful exfoliation down to half of a unit cell (see SI for details). The colored circles correspond to the regions investigated by the polar Kerr effect (PKE).

Figures~\ref{fig:lowfield}(b)-(d) show the polar Kerr angle, $\theta_{k}$, at 2~K as a function of fields applied along the $b$ axis on a three-UC flake (b), a single-UC flake (c), and a half-UC flake (d). $\theta_{k}$ measures the change in the light polarization upon reflection from a magnetized material and is sensitive to out-of-plane magnetic fields. Remarkably, magnetic hysteresis is observed down to the half-UC limit with a coercive field of $\sim 0.2$~T, a value twice as large as that in bulk samples. The uncapped half-UC spot [dark blue circle in Fig.~\ref{fig:lowfield}(a)] did not show a detectable Kerr signal due to flake degradation when exposed to air. Importantly, FM hysteresis loops in capped flakes are also observed at 20~K, as shown in Fig~S2.

To better understand the evolution of the FM state, we perform temperature-dependent PKE measurements. The flakes are cooled in a $b$-axis field of 10~mT, and measurements are performed under zero-field warming conditions. Figures~\ref{fig:lowfield}(f)-(h) show $\theta_{k}$ as a function of temperature for a three-UC flake (f), a single-UC flake (g), and a half-UC flake (h). Notably, the three-UC flake reveals another FM transition at higher temperatures in addition to the anomaly at the expected bulk $T_{\mathrm{C}}$. At the single-UC limit, the bulk transition becomes undetectable, and a single prototypical FM transition is observed at $T^{ML}_{C}\sim 35$~K [Fig.~\ref{fig:lowfield}(g)], a factor of two larger than the bulk counterpart. At the half-UC limit, we observe a similar transition at $T^{ML}_{C}$ in the h-BN capped flake, whereas the uncapped flake shows no signal [Fig.~\ref{fig:lowfield}(h)]. Notably, bulk single crystals also exhibit a small feature at $T^{ML}_{C}$, as shown in the normalized $\theta_{k}(T)/\theta_{k}(2\mathrm{K})$ plot in Fig.~\ref{fig:lowfield}(e), as well as a tiny anomaly in magnetic susceptibility measurements that is consistent with moments from just a couple of $\beta$-UTe$_{3}$ layers (see Fig.~S3b and SI for details). This result implies that the top and bottom layers of $\beta$-UTe$_{3}$ intrinsically display enhanced magnetism with a higher transition temperature compared to the bulk. Importantly, $T^{ML}_{C}$ matches the onset of short-range 2D magnetic correlations observed in electrical resistivity, specific heat, and neutron diffraction measurements.

Quantum confinement is generally expected to boost electronic correlations and quantum fluctuations, which act to destabilize a magnetically ordered ground state. In fact, the ordering temperature of most vdW magnets at the few-layer limit (e.g., FePS$_{3}$~\cite{lee2016ising,wang2016raman}, NiPS$_{3}$~\cite{kim2019suppression}, MnPSe$_{3}$~\cite{liu2020exploring}, Cr$_{2}$Ge$_{2}$Te$_{6}$~\cite{Gong2017}, CrI$_{3}$~\cite{Huang2017}, Fe$_{3}$GeTe$_{2}$~\cite{Deng2018}, CeSiI~\cite{Posey2024}) is either reduced or remains unchanged compared to their bulk counterpart. Our surprising findings point to precisely the opposite: magnetic order thrives in the 2D limit in $\beta$-UTe$_{3}$ flakes and reaches a transition temperature of $T^{ML}_{C}= 35$~K, a factor of two larger than its bulk counterpart. 

Given our experimental evidence for 2D magnetic correlations that set in around 30~K in $\beta$-UTe$_{3}$, the most likely scenario for our observations is the stabilization of 2D correlations into a higher-temperature ordered state at the surface.
Previous theoretical work has established three distinct phase transitions in magnetic systems that display distinct surface ($J_{s}$) and bulk ($J_{b}$) exchange interactions: an `ordinary' transition when $J_{s} \ll J_{b}$, which is the typical transition wherein bulk and surface order at the same temperature; a less common `surface' transition when $J_{s} \gg J_{b}$ and only the surface orders; and a highly uncommon `extraordinary' transition, which also occurs when $J_{s} \gg J_{b}$ but with bulk order happening at a lower temperature compared to the surface~\cite{binder1974surface,binder1983critical}.
An extraordinary transition requires that the exchange interaction responsible for magnetic ordering is enhanced at the surface compared to the bulk. In 3D magnetic materials, ordinary transitions are commonly observed because their in-plane exchange interactions ($J$) are comparable to the exchange interactions between neighboring planes ($J_{\perp}$).
In this case, $J_{s}$ is therefore expected to be smaller than $J_{b}$ because of the missing $J_{\perp}$ contribution from the top layer.
In contrast, in quasi-2D magnetic materials wherein $J \gg J_{\perp}$, small changes on the surface layers may overcome the missing contribution from $J_{\perp}$.
As pointed out recently by Guo \textit{et al.}, quasi-2D systems such as vdW materials are prime candidates for realizing split surface-extraordinary phase transitions, and antiferromagnetic CrSBr was found to exhibit a 10\% increase in ordering temperature in the monolayer limit~\cite{guo2024extraordinary}.

Our results not only support the condition $J_{s} \gg J_{b}$ in $\beta$-UTe$_{3}$ but also demonstrate a remarkable increase in the ordering temperature by 130\% in the monolayer limit.
The enhanced magnetic exchange interaction at the surface may be either due to small changes in the exchange interaction due to inversion symmetry breaking at the surface or to a small relaxation of the lattice due to surface strain.
Even though ferromagnetic order is enhanced in $\beta$-UTe$_{3}$, our thermal expansion measurements  along the $b$ axis (see SI) reveal that $T_{C}$ can be quickly suppressed with uniaxial out-of-plane compression with a rapid rate of -91~K/GPa, which provides a powerful avenue for destabilizing magnetism.
Our results therefore open numerous avenues for quantum control of electronic correlations, including strain engineering at the monolayer limit.
More broadly, our work establishes $\beta$-UTe$_3$ as a novel materials platform for investigating, modeling, and controlling the interplay between dimensional confinement, electronic correlations, and magnetism in 2D vdW materials.

%In summary, we report the synthesis of large single crystals of actinide vdW $\beta$-UTe$_{3}$ and their exfoliation down to the single-layer limit. $\beta$-UTe$_{3}$ displays significant easy-axis magnetic anisotropy and a ferromagnetic (FM) transition at $T_{C}\sim 13$~K. Specific heat and neutron diffraction data show evidence for 3D canted FM magnetic order with a tiny interlayer magnetic exchange interaction. The extracted critical exponent of $\beta = 0.23$, however, is not consistent with the 3D Heisenberg, XY, or Ising models and indicates that 2D contributions need to be considered. In fact, 2D correlations are clearly observed above $T_{c}$ in neutron diffraction experiments. Density functional theory calculations also support a 2D Fermi surface. Importantly, $\beta$-UTe$_{3}$ displays a Sommerfeld coefficient of $\gamma \sim 130$~mJ/mol.K$^{2}$, akin to other strongly correlated $5f$ Kondo metals and superconductors. Surprisingly, $\beta$-UTe$_3$ remains ferromagnetic down to the half-layer limit with an enhanced ordering temperature of $T_{C}=30$~K. Our work establishes $\beta$-UTe$_3$ as the ideal materials platform for investigating, modeling, and controlling the interplay between dimensional confinement, electronic correlations, and magnetism in 2D vdW materials.

\begin{acknowledgments}
We acknowledge helpful discussions with Andrew J. Millis.
Work at Los Alamos National Laboratory was performed under the auspices of the U.S. Department of Energy, Office of Basic Energy Sciences, Division of Materials Science and Engineering.
 WS and CSK acknowledge support from the Los Alamos Laboratory Directed Research and Development program and the Seaborg Institute.
Scanning electron microscope and energy dispersive x-ray measurements were performed at the Electron Microscopy Lab and supported by the Center for Integrated Nanotechnologies, an Office of Science User Facility operated for the U.S. Department of Energy Office of Science. 
The electronic structure calculations were supported in part by the Center for Integrated Nanotechnologies in partnership with the LANL Institutional Computing Program for computational resources.
Work at Purdue University was supported by the U.S. Department of Energy, Office of Science, National Quantum Information Science Research Centers.
We acknowledge the Science and Technology Facilities Council (STFC) for access to neutron beamtime at the WISH diffractometer at ISIS (RB2410150).
A. E. L. and Y. P. Chen also acknowledge Arnab Banerjee for allowing the utilization of the experimental resources funded by the US-DOE Quantum Science Center in his laboratory.
\end{acknowledgments}

\section*{ Methods}

\subsubsection*{Materials synthesis}

Single crystals of $\beta$-UTe$_3$ were prepared using the self-flux technique. U (99.99\%) and Te (99.9999\%) pieces in a 1:15 ratio were loaded into an alumina crucible and sealed under vacuum in a quartz ampule.
The reagents were heated to 875~$^{\circ}$C, held at 875~$^{\circ}$C for 100~h, and slow cooled at 2~$^{\circ}$C/h to 550~$^{\circ}$C. The ampule was then inverted, and the flux was removed $via$ centrifugation. The resultant plate-like crystals were up to 1.2~cm on a side and 1~mm thick (see Fig.~S9). Scanning electron microscope measurements reveal residual Te flux on the surface of the crystals, and energy dispersive X-ray data yield a stoichiometry of UTe$_{3.8(9)}$. We attribute the large error bars and the apparent excess of Te to a combination of residual flux on the surface and the overlap between Te $L$ (3.768~keV) and U $M$ (3.17~keV) edges.

\subsubsection*{Bulk characterization}
Magnetic properties of $\beta$-UTe$_3$ single crystals were collected on a Quantum Design Magnetic Properties Measurement System (MPMS3) with a 7~T magnet and a Quantum Design Physical Property Measurement System (PPMS) with a 16~T and vibrating sample magnetometer options.
Magnetic susceptibility and isothermal magnetization measurements were collected with fields parallel and perpendicular to the crystallographic $b$ axis.

Specific heat measurements were obtained in a PPMS with a $^3$He insert capable of reaching 0.35 K.
Measurements to 100~mK were performed in an Oxford Proteox dilution refrigerator.
Nonmagnetic analogue LaTe$_3$ was measured between $T = 2$ to $300$~K, and the magnetic entropy of UTe$_3$ was obtained by integrating $C_p/T$ after subtracting off the lattice contribution of LaTe$_3$.

Thermal expansion and magnetostriction measurements were performed using a capacitance dilatometer described in Ref.~\cite{Schmiedeshoff2006}.
All thermal expansion and magnetostriction measurements were performed using a slow continuous temperature or field ramp, respectively.

\subsubsection*{Exfoliated sample fabrication and measurement}
The sample fabrication process involved exfoliating UTe$_3$ onto gold-coated silicon substrates (15 nm Au/2 nm Ti/285 nm SiO$_2$/Si) immediately after ion polishing the substrate surface using a JEOL IB-19500CP Cross Section Polisher.
This method consistently yielded large monolayer and few-layer UTe$_3$ flakes (on the order of 100 $\mu$m in size), comparable to those obtained by exfoliation onto freshly deposited gold surfaces.
In more detail, the substrate was mounted on a rotational stage with its rotation axis near the center of the substrate, and the substrate's surface aligned parallel to the broad ion beam.
Ion polishing was carried out at an accelerating voltage of 4 kV for 1.5 minutes. 
Immediately afterward, low-residue Nitto tape carrying freshly cleaved UTe$_3$ was applied to the polished substrate.
Tape removal was performed inside an argon-filled glovebox with oxygen and water levels below 0.01 ppm. Finally, selected UTe$_3$ flakes were capped with h-BN to further reduce degradation.

To confirm the thickness of the exfoliated flakes we used a HR-2D AFM system from AFM Workshop inside the argon filled glovebox with oxygen and water content less than 0.01 ppm to avoid air degradation. The scans were performed in semi-contact mode with a line frequency of 0.5 Hz.

The PKE measurements were performed using a fiber-based, zero-area, Sagnac interferometer with a center frequency of 1550 nm~\cite{Xia2006,Fried2014}.
The measurements were performed in a Quantum Design PPMS cryostat with the optical multi-function probe option that provides XYZ sample positioning and a high-resolution camera for imaging the samples.
After sample alignment, a shutter can be removed to enable the PKE measurement.
The spot size for the PKE measurement was confirmed using a USAF resolution target.

\subsubsection*{Neutron diffraction}

Neutron diffraction on bulk UTe$_3$ was carried out on the time-of-flight diffractometer WISH (STFC, Rutherford Appleton Laboratory)~\cite{2011_Chapon_NeutronNews}. WISH is equipped with a solid methane moderator that provides neutron pulses of high brilliance of a broad wavelength bandwidth covering the range between 1\,\AA \, and 10\,\AA. In combination with the large angular coverage of detector banks this enables to access a large volume in momentum space. 

A UTe$_3$ sample of mass $0.26\,$g was studied at a fixed orientation at different temperatures. The sample comprises of a twin structure associated with a $90\,$deg rotation around the van-der-Waals axis $(010)$ (see SI for further information). In addition, the sample displays grain distribution with a mosaicity of the order 3 deg.

The sample was oriented such that the $(010)$ axis and $(001)$ axis (or $(100)$ axis for the rotated twin) coincided with the horizontal scattering plane. Data were analyzed with the package Mantid~\cite{2014_Arnold_NuclearInstrumentsandMethodsinPhysicsResearchSectionAAcceleratorsSpectrometersDetectorsandAssociatedEquipment}. Diffraction data were described in the reciprocal space of an orthorhombic lattice with lattice parameters $a=4.338\,$\AA, $b=24.743\,$\AA, and $c=4.338\,$\AA. Momentum transfers along $(100)$, $(010)$, and $(001)$ are denoted with a capital $\bm{Q}$ and characterized in reciprocal space units (r.l.u.), which along the three axes are given by $\frac{2\pi}{a}$, $\frac{2\pi}{b}$, and $\frac{2\pi}{c}$, respectively. 
%For simplification of the presentation, the data in the manuscript are presented in a twin- and grain-independent momentum space coordinate system $(q_x,q_y,q_z)$ with reciprocal lattice units (r.l.u.) given by $\frac{2\pi}{a}$, $\frac{2\pi}{b}$, and $\frac{2\pi}{a}$. The axis $q_x$ corresponds to the vertical direction, $q_y$ is horizontal and parallel to the $K$-axis, and $q_z$ is horizontal and parallel to either $L$ or $H$, depending on the twin that we consider.
The diffraction data in the manuscript are presented in momentum space coordinate system $(q_x,q_y,q_z)$ with reciprocal lattice units (r.l.u.) given by  $\frac{2\pi}{a}$, $\frac{2\pi}{b}$, and $\frac{2\pi}{a}$. The axis $q_x$ corresponds to the vertical direction (referring to the horizontal scattering plane), $q_y$ is horizontal and parallel to the $K$-axis common for both twins, and $q_z$ is horizontal and parallel to either $L$ or $H$, depending on the twin that we consider.

Error bars of neutron diffraction data points correspond to uncertainties due to statistical errors. Their size was calculated assuming Poisson counting statistics.

\bibliography{lib}

\begin{thebibliography}{66}%
\makeatletter
\providecommand \@ifxundefined [1]{%
 \@ifx{#1\undefined}
}%
\providecommand \@ifnum [1]{%
 \ifnum #1\expandafter \@firstoftwo
 \else \expandafter \@secondoftwo
 \fi
}%
\providecommand \@ifx [1]{%
 \ifx #1\expandafter \@firstoftwo
 \else \expandafter \@secondoftwo
 \fi
}%
\providecommand \natexlab [1]{#1}%
\providecommand \enquote  [1]{``#1''}%
\providecommand \bibnamefont  [1]{#1}%
\providecommand \bibfnamefont [1]{#1}%
\providecommand \citenamefont [1]{#1}%
\providecommand \href@noop [0]{\@secondoftwo}%
\providecommand \href [0]{\begingroup \@sanitize@url \@href}%
\providecommand \@href[1]{\@@startlink{#1}\@@href}%
\providecommand \@@href[1]{\endgroup#1\@@endlink}%
\providecommand \@sanitize@url [0]{\catcode `\\12\catcode `\$12\catcode `\&12\catcode `\#12\catcode `\^12\catcode `\_12\catcode `\%12\relax}%
\providecommand \@@startlink[1]{}%
\providecommand \@@endlink[0]{}%
\providecommand \url  [0]{\begingroup\@sanitize@url \@url }%
\providecommand \@url [1]{\endgroup\@href {#1}{\urlprefix }}%
\providecommand \urlprefix  [0]{URL }%
\providecommand \Eprint [0]{\href }%
\providecommand \doibase [0]{https://doi.org/}%
\providecommand \selectlanguage [0]{\@gobble}%
\providecommand \bibinfo  [0]{\@secondoftwo}%
\providecommand \bibfield  [0]{\@secondoftwo}%
\providecommand \translation [1]{[#1]}%
\providecommand \BibitemOpen [0]{}%
\providecommand \bibitemStop [0]{}%
\providecommand \bibitemNoStop [0]{.\EOS\space}%
\providecommand \EOS [0]{\spacefactor3000\relax}%
\providecommand \BibitemShut  [1]{\csname bibitem#1\endcsname}%
\let\auto@bib@innerbib\@empty
%</preamble>
\bibitem [{\citenamefont {Lee}\ \emph {et~al.}(2016)\citenamefont {Lee}, \citenamefont {Lee}, \citenamefont {Ryoo}, \citenamefont {Kang}, \citenamefont {Kim}, \citenamefont {Kim}, \citenamefont {Park}, \citenamefont {Park},\ and\ \citenamefont {Cheong}}]{lee2016ising}%
  \BibitemOpen
  \bibfield  {author} {\bibinfo {author} {\bibfnamefont {J.-U.}\ \bibnamefont {Lee}}, \bibinfo {author} {\bibfnamefont {S.}~\bibnamefont {Lee}}, \bibinfo {author} {\bibfnamefont {J.~H.}\ \bibnamefont {Ryoo}}, \bibinfo {author} {\bibfnamefont {S.}~\bibnamefont {Kang}}, \bibinfo {author} {\bibfnamefont {T.~Y.}\ \bibnamefont {Kim}}, \bibinfo {author} {\bibfnamefont {P.}~\bibnamefont {Kim}}, \bibinfo {author} {\bibfnamefont {C.-H.}\ \bibnamefont {Park}}, \bibinfo {author} {\bibfnamefont {J.-G.}\ \bibnamefont {Park}},\ and\ \bibinfo {author} {\bibfnamefont {H.}~\bibnamefont {Cheong}},\ }\bibfield  {title} {\bibinfo {title} {Ising-type magnetic ordering in atomically thin {(FePS$_{3}$)}},\ }\href@noop {} {\bibfield  {journal} {\bibinfo  {journal} {Nano letters}\ }\textbf {\bibinfo {volume} {16}},\ \bibinfo {pages} {7433} (\bibinfo {year} {2016})}\BibitemShut {NoStop}%
\bibitem [{\citenamefont {Wang}\ \emph {et~al.}(2016)\citenamefont {Wang}, \citenamefont {Du}, \citenamefont {Liu}, \citenamefont {Hu}, \citenamefont {Zhang}, \citenamefont {Zhang}, \citenamefont {Owen}, \citenamefont {Lu}, \citenamefont {Gan}, \citenamefont {Sengupta} \emph {et~al.}}]{wang2016raman}%
  \BibitemOpen
  \bibfield  {author} {\bibinfo {author} {\bibfnamefont {X.}~\bibnamefont {Wang}}, \bibinfo {author} {\bibfnamefont {K.}~\bibnamefont {Du}}, \bibinfo {author} {\bibfnamefont {Y.~Y.~F.}\ \bibnamefont {Liu}}, \bibinfo {author} {\bibfnamefont {P.}~\bibnamefont {Hu}}, \bibinfo {author} {\bibfnamefont {J.}~\bibnamefont {Zhang}}, \bibinfo {author} {\bibfnamefont {Q.}~\bibnamefont {Zhang}}, \bibinfo {author} {\bibfnamefont {M.~H.~S.}\ \bibnamefont {Owen}}, \bibinfo {author} {\bibfnamefont {X.}~\bibnamefont {Lu}}, \bibinfo {author} {\bibfnamefont {C.~K.}\ \bibnamefont {Gan}}, \bibinfo {author} {\bibfnamefont {P.}~\bibnamefont {Sengupta}}, \emph {et~al.},\ }\bibfield  {title} {\bibinfo {title} {Raman spectroscopy of atomically thin two-dimensional magnetic iron phosphorus trisulfide {(FePS$_{3}$)} crystals},\ }\href@noop {} {\bibfield  {journal} {\bibinfo  {journal} {2D Materials}\ }\textbf {\bibinfo {volume} {3}},\ \bibinfo {pages} {031009} (\bibinfo {year} {2016})}\BibitemShut {NoStop}%
\bibitem [{\citenamefont {Gong}\ \emph {et~al.}(2017)\citenamefont {Gong}, \citenamefont {Li}, \citenamefont {Li}, \citenamefont {Ji}, \citenamefont {Stern}, \citenamefont {Xia}, \citenamefont {Cao}, \citenamefont {Bao}, \citenamefont {Wang}, \citenamefont {Wang}, \citenamefont {Qiu}, \citenamefont {Cava}, \citenamefont {Louie}, \citenamefont {Xia},\ and\ \citenamefont {Zhang}}]{Gong2017}%
  \BibitemOpen
  \bibfield  {author} {\bibinfo {author} {\bibfnamefont {C.}~\bibnamefont {Gong}}, \bibinfo {author} {\bibfnamefont {L.}~\bibnamefont {Li}}, \bibinfo {author} {\bibfnamefont {Z.}~\bibnamefont {Li}}, \bibinfo {author} {\bibfnamefont {H.}~\bibnamefont {Ji}}, \bibinfo {author} {\bibfnamefont {A.}~\bibnamefont {Stern}}, \bibinfo {author} {\bibfnamefont {Y.}~\bibnamefont {Xia}}, \bibinfo {author} {\bibfnamefont {T.}~\bibnamefont {Cao}}, \bibinfo {author} {\bibfnamefont {W.}~\bibnamefont {Bao}}, \bibinfo {author} {\bibfnamefont {C.}~\bibnamefont {Wang}}, \bibinfo {author} {\bibfnamefont {Y.}~\bibnamefont {Wang}}, \bibinfo {author} {\bibfnamefont {Z.~Q.}\ \bibnamefont {Qiu}}, \bibinfo {author} {\bibfnamefont {R.~J.}\ \bibnamefont {Cava}}, \bibinfo {author} {\bibfnamefont {S.~G.}\ \bibnamefont {Louie}}, \bibinfo {author} {\bibfnamefont {J.}~\bibnamefont {Xia}},\ and\ \bibinfo {author} {\bibfnamefont {X.}~\bibnamefont {Zhang}},\ }\bibfield  {title} {\bibinfo {title} {Discovery of intrinsic ferromagnetism in
  two-dimensional van der waals crystals},\ }\href@noop {} {\bibfield  {journal} {\bibinfo  {journal} {Nature}\ }\textbf {\bibinfo {volume} {546}},\ \bibinfo {pages} {265} (\bibinfo {year} {2017})}\BibitemShut {NoStop}%
\bibitem [{\citenamefont {Huang}\ \emph {et~al.}(2017)\citenamefont {Huang}, \citenamefont {Clark}, \citenamefont {Navarro-Moratalla}, \citenamefont {Klein}, \citenamefont {Cheng}, \citenamefont {Seyler}, \citenamefont {Zhong}, \citenamefont {Schmidgall}, \citenamefont {McGuire}, \citenamefont {Cobden}, \citenamefont {Yao}, \citenamefont {Xiao}, \citenamefont {Jarillo-Herrero},\ and\ \citenamefont {Xu}}]{Huang2017}%
  \BibitemOpen
  \bibfield  {author} {\bibinfo {author} {\bibfnamefont {B.}~\bibnamefont {Huang}}, \bibinfo {author} {\bibfnamefont {G.}~\bibnamefont {Clark}}, \bibinfo {author} {\bibfnamefont {E.}~\bibnamefont {Navarro-Moratalla}}, \bibinfo {author} {\bibfnamefont {D.~R.}\ \bibnamefont {Klein}}, \bibinfo {author} {\bibfnamefont {R.}~\bibnamefont {Cheng}}, \bibinfo {author} {\bibfnamefont {K.~L.}\ \bibnamefont {Seyler}}, \bibinfo {author} {\bibfnamefont {D.}~\bibnamefont {Zhong}}, \bibinfo {author} {\bibfnamefont {E.}~\bibnamefont {Schmidgall}}, \bibinfo {author} {\bibfnamefont {M.~A.}\ \bibnamefont {McGuire}}, \bibinfo {author} {\bibfnamefont {D.~H.}\ \bibnamefont {Cobden}}, \bibinfo {author} {\bibfnamefont {W.}~\bibnamefont {Yao}}, \bibinfo {author} {\bibfnamefont {D.}~\bibnamefont {Xiao}}, \bibinfo {author} {\bibfnamefont {P.}~\bibnamefont {Jarillo-Herrero}},\ and\ \bibinfo {author} {\bibfnamefont {X.}~\bibnamefont {Xu}},\ }\bibfield  {title} {\bibinfo {title} {{Layer-dependent ferromagnetism in a van der Waals crystal
  down to the monolayer limit}},\ }\href@noop {} {\bibfield  {journal} {\bibinfo  {journal} {Nature}\ }\textbf {\bibinfo {volume} {546}},\ \bibinfo {pages} {270} (\bibinfo {year} {2017})}\BibitemShut {NoStop}%
\bibitem [{\citenamefont {Burch}\ \emph {et~al.}(2018)\citenamefont {Burch}, \citenamefont {Mandrus},\ and\ \citenamefont {Park}}]{Burch2018}%
  \BibitemOpen
  \bibfield  {author} {\bibinfo {author} {\bibfnamefont {K.~S.}\ \bibnamefont {Burch}}, \bibinfo {author} {\bibfnamefont {D.}~\bibnamefont {Mandrus}},\ and\ \bibinfo {author} {\bibfnamefont {J.-G.}\ \bibnamefont {Park}},\ }\bibfield  {title} {\bibinfo {title} {{Magnetism in two-dimensional van der Waals materials}},\ }\href@noop {} {\bibfield  {journal} {\bibinfo  {journal} {Nature}\ }\textbf {\bibinfo {volume} {563}},\ \bibinfo {pages} {47} (\bibinfo {year} {2018})}\BibitemShut {NoStop}%
\bibitem [{\citenamefont {Wang}\ \emph {et~al.}(2022)\citenamefont {Wang}, \citenamefont {Bedoya-Pinto}, \citenamefont {Blei}, \citenamefont {Dismukes}, \citenamefont {Hamo}, \citenamefont {Jenkins}, \citenamefont {Koperski}, \citenamefont {Liu}, \citenamefont {Sun}, \citenamefont {Telford}, \citenamefont {Kim}, \citenamefont {Augustin}, \citenamefont {Vool}, \citenamefont {Yin}, \citenamefont {Li}, \citenamefont {Falin}, \citenamefont {Dean}, \citenamefont {Casanova}, \citenamefont {Evans}, \citenamefont {Chshiev}, \citenamefont {Mishchenko}, \citenamefont {Petrovic}, \citenamefont {He}, \citenamefont {Zhao}, \citenamefont {Tsen}, \citenamefont {Gerardot}, \citenamefont {Brotons-Gisbert}, \citenamefont {Guguchia}, \citenamefont {Roy}, \citenamefont {Tongay}, \citenamefont {Wang}, \citenamefont {Hasan}, \citenamefont {Wrachtrup}, \citenamefont {Yacoby}, \citenamefont {Fert}, \citenamefont {Parkin}, \citenamefont {Novoselov}, \citenamefont {Dai}, \citenamefont {Balicas},\ and\ \citenamefont
  {Santos}}]{Wang2022}%
  \BibitemOpen
  \bibfield  {author} {\bibinfo {author} {\bibfnamefont {Q.~H.}\ \bibnamefont {Wang}}, \bibinfo {author} {\bibfnamefont {A.}~\bibnamefont {Bedoya-Pinto}}, \bibinfo {author} {\bibfnamefont {M.}~\bibnamefont {Blei}}, \bibinfo {author} {\bibfnamefont {A.~H.}\ \bibnamefont {Dismukes}}, \bibinfo {author} {\bibfnamefont {A.}~\bibnamefont {Hamo}}, \bibinfo {author} {\bibfnamefont {S.}~\bibnamefont {Jenkins}}, \bibinfo {author} {\bibfnamefont {M.}~\bibnamefont {Koperski}}, \bibinfo {author} {\bibfnamefont {Y.}~\bibnamefont {Liu}}, \bibinfo {author} {\bibfnamefont {Q.-C.}\ \bibnamefont {Sun}}, \bibinfo {author} {\bibfnamefont {E.~J.}\ \bibnamefont {Telford}}, \bibinfo {author} {\bibfnamefont {H.~H.}\ \bibnamefont {Kim}}, \bibinfo {author} {\bibfnamefont {M.}~\bibnamefont {Augustin}}, \bibinfo {author} {\bibfnamefont {U.}~\bibnamefont {Vool}}, \bibinfo {author} {\bibfnamefont {J.-X.}\ \bibnamefont {Yin}}, \bibinfo {author} {\bibfnamefont {L.~H.}\ \bibnamefont {Li}}, \bibinfo {author} {\bibfnamefont {A.}~\bibnamefont
  {Falin}}, \bibinfo {author} {\bibfnamefont {C.~R.}\ \bibnamefont {Dean}}, \bibinfo {author} {\bibfnamefont {F.}~\bibnamefont {Casanova}}, \bibinfo {author} {\bibfnamefont {R.~F.~L.}\ \bibnamefont {Evans}}, \bibinfo {author} {\bibfnamefont {M.}~\bibnamefont {Chshiev}}, \bibinfo {author} {\bibfnamefont {A.}~\bibnamefont {Mishchenko}}, \bibinfo {author} {\bibfnamefont {C.}~\bibnamefont {Petrovic}}, \bibinfo {author} {\bibfnamefont {R.}~\bibnamefont {He}}, \bibinfo {author} {\bibfnamefont {L.}~\bibnamefont {Zhao}}, \bibinfo {author} {\bibfnamefont {A.~W.}\ \bibnamefont {Tsen}}, \bibinfo {author} {\bibfnamefont {B.~D.}\ \bibnamefont {Gerardot}}, \bibinfo {author} {\bibfnamefont {M.}~\bibnamefont {Brotons-Gisbert}}, \bibinfo {author} {\bibfnamefont {Z.}~\bibnamefont {Guguchia}}, \bibinfo {author} {\bibfnamefont {X.}~\bibnamefont {Roy}}, \bibinfo {author} {\bibfnamefont {S.}~\bibnamefont {Tongay}}, \bibinfo {author} {\bibfnamefont {Z.}~\bibnamefont {Wang}}, \bibinfo {author} {\bibfnamefont {M.~Z.}\ \bibnamefont
  {Hasan}}, \bibinfo {author} {\bibfnamefont {J.}~\bibnamefont {Wrachtrup}}, \bibinfo {author} {\bibfnamefont {A.}~\bibnamefont {Yacoby}}, \bibinfo {author} {\bibfnamefont {A.}~\bibnamefont {Fert}}, \bibinfo {author} {\bibfnamefont {S.}~\bibnamefont {Parkin}}, \bibinfo {author} {\bibfnamefont {K.~S.}\ \bibnamefont {Novoselov}}, \bibinfo {author} {\bibfnamefont {P.}~\bibnamefont {Dai}}, \bibinfo {author} {\bibfnamefont {L.}~\bibnamefont {Balicas}},\ and\ \bibinfo {author} {\bibfnamefont {E.~J.~G.}\ \bibnamefont {Santos}},\ }\bibfield  {title} {\bibinfo {title} {{The Magnetic Genome of Two-Dimensional van der Waals Materials}},\ }\href {https://doi.org/10.1021/acsnano.1c09150} {\bibfield  {journal} {\bibinfo  {journal} {ACS Nano}\ }\textbf {\bibinfo {volume} {16}},\ \bibinfo {pages} {6960} (\bibinfo {year} {2022})}\BibitemShut {NoStop}%
\bibitem [{\citenamefont {Ziebel}\ \emph {et~al.}(2024)\citenamefont {Ziebel}, \citenamefont {Feuer}, \citenamefont {Cox}, \citenamefont {Zhu}, \citenamefont {Dean},\ and\ \citenamefont {Roy}}]{Ziebel2024}%
  \BibitemOpen
  \bibfield  {author} {\bibinfo {author} {\bibfnamefont {M.~E.}\ \bibnamefont {Ziebel}}, \bibinfo {author} {\bibfnamefont {M.~L.}\ \bibnamefont {Feuer}}, \bibinfo {author} {\bibfnamefont {J.}~\bibnamefont {Cox}}, \bibinfo {author} {\bibfnamefont {X.}~\bibnamefont {Zhu}}, \bibinfo {author} {\bibfnamefont {C.~R.}\ \bibnamefont {Dean}},\ and\ \bibinfo {author} {\bibfnamefont {X.}~\bibnamefont {Roy}},\ }\bibfield  {title} {\bibinfo {title} {{CrSBr}: An {Air-Stable}, {Two-Dimensional} magnetic semiconductor},\ }\href@noop {} {\bibfield  {journal} {\bibinfo  {journal} {Nano Lett.}\ }\textbf {\bibinfo {volume} {24}},\ \bibinfo {pages} {4319} (\bibinfo {year} {2024})}\BibitemShut {NoStop}%
\bibitem [{\citenamefont {Monthoux}\ \emph {et~al.}(2007)\citenamefont {Monthoux}, \citenamefont {Pines},\ and\ \citenamefont {Lonzarich}}]{Monthoux2007}%
  \BibitemOpen
  \bibfield  {author} {\bibinfo {author} {\bibfnamefont {P.}~\bibnamefont {Monthoux}}, \bibinfo {author} {\bibfnamefont {D.}~\bibnamefont {Pines}},\ and\ \bibinfo {author} {\bibfnamefont {G.~G.}\ \bibnamefont {Lonzarich}},\ }\bibfield  {title} {\bibinfo {title} {Superconductivity without phonons},\ }\href {https://doi.org/10.1038/nature06480} {\bibfield  {journal} {\bibinfo  {journal} {Nature}\ }\textbf {\bibinfo {volume} {450}},\ \bibinfo {pages} {1177} (\bibinfo {year} {2007})}\BibitemShut {NoStop}%
\bibitem [{\citenamefont {Mizukami}\ \emph {et~al.}(2011)\citenamefont {Mizukami}, \citenamefont {Shishido}, \citenamefont {Shibauchi}, \citenamefont {Shimozawa}, \citenamefont {Yasumoto}, \citenamefont {Watanabe}, \citenamefont {Yamashita}, \citenamefont {Ikeda}, \citenamefont {Terashima}, \citenamefont {Kontani},\ and\ \citenamefont {Matsuda}}]{Mizukami2011}%
  \BibitemOpen
  \bibfield  {author} {\bibinfo {author} {\bibfnamefont {Y.}~\bibnamefont {Mizukami}}, \bibinfo {author} {\bibfnamefont {H.}~\bibnamefont {Shishido}}, \bibinfo {author} {\bibfnamefont {T.}~\bibnamefont {Shibauchi}}, \bibinfo {author} {\bibfnamefont {M.}~\bibnamefont {Shimozawa}}, \bibinfo {author} {\bibfnamefont {S.}~\bibnamefont {Yasumoto}}, \bibinfo {author} {\bibfnamefont {D.}~\bibnamefont {Watanabe}}, \bibinfo {author} {\bibfnamefont {M.}~\bibnamefont {Yamashita}}, \bibinfo {author} {\bibfnamefont {H.}~\bibnamefont {Ikeda}}, \bibinfo {author} {\bibfnamefont {T.}~\bibnamefont {Terashima}}, \bibinfo {author} {\bibfnamefont {H.}~\bibnamefont {Kontani}},\ and\ \bibinfo {author} {\bibfnamefont {Y.}~\bibnamefont {Matsuda}},\ }\bibfield  {title} {\bibinfo {title} {Extremely strong-coupling superconductivity in artificial two-dimensional {Kondo} lattices},\ }\href {https://doi.org/10.1038/nphys2112} {\bibfield  {journal} {\bibinfo  {journal} {Nature Physics}\ }\textbf {\bibinfo {volume} {7}},\ \bibinfo {pages}
  {849} (\bibinfo {year} {2011})}\BibitemShut {NoStop}%
\bibitem [{\citenamefont {Cao}\ \emph {et~al.}(2018)\citenamefont {Cao}, \citenamefont {Fatemi}, \citenamefont {Fang}, \citenamefont {Watanabe}, \citenamefont {Taniguchi}, \citenamefont {Kaxiras},\ and\ \citenamefont {Jarillo-Herrero}}]{Cao2018}%
  \BibitemOpen
  \bibfield  {author} {\bibinfo {author} {\bibfnamefont {Y.}~\bibnamefont {Cao}}, \bibinfo {author} {\bibfnamefont {V.}~\bibnamefont {Fatemi}}, \bibinfo {author} {\bibfnamefont {S.}~\bibnamefont {Fang}}, \bibinfo {author} {\bibfnamefont {K.}~\bibnamefont {Watanabe}}, \bibinfo {author} {\bibfnamefont {T.}~\bibnamefont {Taniguchi}}, \bibinfo {author} {\bibfnamefont {E.}~\bibnamefont {Kaxiras}},\ and\ \bibinfo {author} {\bibfnamefont {P.}~\bibnamefont {Jarillo-Herrero}},\ }\bibfield  {title} {\bibinfo {title} {{Unconventional superconductivity in magic-angle graphene superlattices}},\ }\href@noop {} {\bibfield  {journal} {\bibinfo  {journal} {Nature}\ }\textbf {\bibinfo {volume} {556}},\ \bibinfo {pages} {43} (\bibinfo {year} {2018})}\BibitemShut {NoStop}%
\bibitem [{\citenamefont {Va{\v n}o}\ \emph {et~al.}(2021)\citenamefont {Va{\v n}o}, \citenamefont {Amini}, \citenamefont {Ganguli}, \citenamefont {Chen}, \citenamefont {Lado}, \citenamefont {Kezilebieke},\ and\ \citenamefont {Liljeroth}}]{Artificial2021}%
  \BibitemOpen
  \bibfield  {author} {\bibinfo {author} {\bibfnamefont {V.}~\bibnamefont {Va{\v n}o}}, \bibinfo {author} {\bibfnamefont {M.}~\bibnamefont {Amini}}, \bibinfo {author} {\bibfnamefont {S.~C.}\ \bibnamefont {Ganguli}}, \bibinfo {author} {\bibfnamefont {G.}~\bibnamefont {Chen}}, \bibinfo {author} {\bibfnamefont {J.~L.}\ \bibnamefont {Lado}}, \bibinfo {author} {\bibfnamefont {S.}~\bibnamefont {Kezilebieke}},\ and\ \bibinfo {author} {\bibfnamefont {P.}~\bibnamefont {Liljeroth}},\ }\bibfield  {title} {\bibinfo {title} {{Artificial heavy fermions in a van der Waals heterostructure}},\ }\href {https://doi.org/10.1038/s41586-021-04021-0} {\bibfield  {journal} {\bibinfo  {journal} {Nature}\ }\textbf {\bibinfo {volume} {599}},\ \bibinfo {pages} {582} (\bibinfo {year} {2021})}\BibitemShut {NoStop}%
\bibitem [{\citenamefont {Ledwith}\ \emph {et~al.}(2021)\citenamefont {Ledwith}, \citenamefont {Khalaf},\ and\ \citenamefont {Vishwanath}}]{LEDWITH2021}%
  \BibitemOpen
  \bibfield  {author} {\bibinfo {author} {\bibfnamefont {P.~J.}\ \bibnamefont {Ledwith}}, \bibinfo {author} {\bibfnamefont {E.}~\bibnamefont {Khalaf}},\ and\ \bibinfo {author} {\bibfnamefont {A.}~\bibnamefont {Vishwanath}},\ }\bibfield  {title} {\bibinfo {title} {Strong coupling theory of magic-angle graphene: A pedagogical introduction},\ }\href {https://doi.org/https://doi.org/10.1016/j.aop.2021.168646} {\bibfield  {journal} {\bibinfo  {journal} {Annals of Physics}\ }\textbf {\bibinfo {volume} {435}},\ \bibinfo {pages} {168646} (\bibinfo {year} {2021})},\ \bibinfo {note} {special issue on Philip W. Anderson}\BibitemShut {NoStop}%
\bibitem [{\citenamefont {Simeth}\ \emph {et~al.}(2023)\citenamefont {Simeth}, \citenamefont {Wang}, \citenamefont {Ghioldi}, \citenamefont {Fobes}, \citenamefont {Podlesnyak}, \citenamefont {Sung}, \citenamefont {Bauer}, \citenamefont {Lass}, \citenamefont {Flury}, \citenamefont {Vonka}, \citenamefont {Mazzone}, \citenamefont {Niedermayer}, \citenamefont {Nomura}, \citenamefont {Arita}, \citenamefont {Batista}, \citenamefont {Ronning},\ and\ \citenamefont {Janoschek}}]{Simeth2023}%
  \BibitemOpen
  \bibfield  {author} {\bibinfo {author} {\bibfnamefont {W.}~\bibnamefont {Simeth}}, \bibinfo {author} {\bibfnamefont {Z.}~\bibnamefont {Wang}}, \bibinfo {author} {\bibfnamefont {E.~A.}\ \bibnamefont {Ghioldi}}, \bibinfo {author} {\bibfnamefont {D.~M.}\ \bibnamefont {Fobes}}, \bibinfo {author} {\bibfnamefont {A.}~\bibnamefont {Podlesnyak}}, \bibinfo {author} {\bibfnamefont {N.~H.}\ \bibnamefont {Sung}}, \bibinfo {author} {\bibfnamefont {E.~D.}\ \bibnamefont {Bauer}}, \bibinfo {author} {\bibfnamefont {J.}~\bibnamefont {Lass}}, \bibinfo {author} {\bibfnamefont {S.}~\bibnamefont {Flury}}, \bibinfo {author} {\bibfnamefont {J.}~\bibnamefont {Vonka}}, \bibinfo {author} {\bibfnamefont {D.~G.}\ \bibnamefont {Mazzone}}, \bibinfo {author} {\bibfnamefont {C.}~\bibnamefont {Niedermayer}}, \bibinfo {author} {\bibfnamefont {Y.}~\bibnamefont {Nomura}}, \bibinfo {author} {\bibfnamefont {R.}~\bibnamefont {Arita}}, \bibinfo {author} {\bibfnamefont {C.~D.}\ \bibnamefont {Batista}}, \bibinfo {author} {\bibfnamefont
  {F.}~\bibnamefont {Ronning}},\ and\ \bibinfo {author} {\bibfnamefont {M.}~\bibnamefont {Janoschek}},\ }\bibfield  {title} {\bibinfo {title} {A microscopic kondo lattice model for the heavy fermion antiferromagnet {CeIn$_{3}$}},\ }\href {https://doi.org/10.1038/s41467-023-43947-z} {\bibfield  {journal} {\bibinfo  {journal} {Nature Communications}\ }\textbf {\bibinfo {volume} {14}},\ \bibinfo {pages} {8239} (\bibinfo {year} {2023})}\BibitemShut {NoStop}%
\bibitem [{\citenamefont {Han}\ \emph {et~al.}(2018)\citenamefont {Han}, \citenamefont {Duong}, \citenamefont {Keum}, \citenamefont {Yun},\ and\ \citenamefont {Lee}}]{Han2018}%
  \BibitemOpen
  \bibfield  {author} {\bibinfo {author} {\bibfnamefont {G.~H.}\ \bibnamefont {Han}}, \bibinfo {author} {\bibfnamefont {D.~L.}\ \bibnamefont {Duong}}, \bibinfo {author} {\bibfnamefont {D.~H.}\ \bibnamefont {Keum}}, \bibinfo {author} {\bibfnamefont {S.~J.}\ \bibnamefont {Yun}},\ and\ \bibinfo {author} {\bibfnamefont {Y.~H.}\ \bibnamefont {Lee}},\ }\bibfield  {title} {\bibinfo {title} {van der waals metallic transition metal dichalcogenides},\ }\href {https://doi.org/10.1021/acs.chemrev.7b00618} {\bibfield  {journal} {\bibinfo  {journal} {Chemical Reviews}\ }\textbf {\bibinfo {volume} {118}},\ \bibinfo {pages} {6297} (\bibinfo {year} {2018})}\BibitemShut {NoStop}%
\bibitem [{\citenamefont {Zhang}\ \emph {et~al.}(2018)\citenamefont {Zhang}, \citenamefont {Lu}, \citenamefont {Zhu}, \citenamefont {Tan}, \citenamefont {Feng}, \citenamefont {Liu}, \citenamefont {Zhang}, \citenamefont {Chen}, \citenamefont {Liu}, \citenamefont {Luo}, \citenamefont {Xie}, \citenamefont {Luo}, \citenamefont {Zhang},\ and\ \citenamefont {Lai}}]{Zhang2018}%
  \BibitemOpen
  \bibfield  {author} {\bibinfo {author} {\bibfnamefont {Y.}~\bibnamefont {Zhang}}, \bibinfo {author} {\bibfnamefont {H.}~\bibnamefont {Lu}}, \bibinfo {author} {\bibfnamefont {X.}~\bibnamefont {Zhu}}, \bibinfo {author} {\bibfnamefont {S.}~\bibnamefont {Tan}}, \bibinfo {author} {\bibfnamefont {W.}~\bibnamefont {Feng}}, \bibinfo {author} {\bibfnamefont {Q.}~\bibnamefont {Liu}}, \bibinfo {author} {\bibfnamefont {W.}~\bibnamefont {Zhang}}, \bibinfo {author} {\bibfnamefont {Q.}~\bibnamefont {Chen}}, \bibinfo {author} {\bibfnamefont {Y.}~\bibnamefont {Liu}}, \bibinfo {author} {\bibfnamefont {X.}~\bibnamefont {Luo}}, \bibinfo {author} {\bibfnamefont {D.}~\bibnamefont {Xie}}, \bibinfo {author} {\bibfnamefont {L.}~\bibnamefont {Luo}}, \bibinfo {author} {\bibfnamefont {Z.}~\bibnamefont {Zhang}},\ and\ \bibinfo {author} {\bibfnamefont {X.}~\bibnamefont {Lai}},\ }\bibfield  {title} {\bibinfo {title} {Emergence of kondo lattice behavior in a van der waals itinerant ferromagnet, {Fe$_{3}$GeTe$_{2}$}},\ }\href
  {https://doi.org/10.1126/sciadv.aao6791} {\bibfield  {journal} {\bibinfo  {journal} {Science Advances}\ }\textbf {\bibinfo {volume} {4}},\ \bibinfo {pages} {eaao6791} (\bibinfo {year} {2018})},\ \Eprint {https://arxiv.org/abs/https://www.science.org/doi/pdf/10.1126/sciadv.aao6791} {https://www.science.org/doi/pdf/10.1126/sciadv.aao6791} \BibitemShut {NoStop}%
\bibitem [{\citenamefont {May}\ \emph {et~al.}(2019)\citenamefont {May}, \citenamefont {Ovchinnikov}, \citenamefont {Zheng}, \citenamefont {Hermann}, \citenamefont {Calder}, \citenamefont {Huang}, \citenamefont {Fei}, \citenamefont {Liu}, \citenamefont {Xu},\ and\ \citenamefont {McGuire}}]{May2019}%
  \BibitemOpen
  \bibfield  {author} {\bibinfo {author} {\bibfnamefont {A.~F.}\ \bibnamefont {May}}, \bibinfo {author} {\bibfnamefont {D.}~\bibnamefont {Ovchinnikov}}, \bibinfo {author} {\bibfnamefont {Q.}~\bibnamefont {Zheng}}, \bibinfo {author} {\bibfnamefont {R.}~\bibnamefont {Hermann}}, \bibinfo {author} {\bibfnamefont {S.}~\bibnamefont {Calder}}, \bibinfo {author} {\bibfnamefont {B.}~\bibnamefont {Huang}}, \bibinfo {author} {\bibfnamefont {Z.}~\bibnamefont {Fei}}, \bibinfo {author} {\bibfnamefont {Y.}~\bibnamefont {Liu}}, \bibinfo {author} {\bibfnamefont {X.}~\bibnamefont {Xu}},\ and\ \bibinfo {author} {\bibfnamefont {M.~A.}\ \bibnamefont {McGuire}},\ }\bibfield  {title} {\bibinfo {title} {Ferromagnetism near room temperature in the cleavable van der waals crystal {Fe$_5$GeTe$_2$}},\ }\href@noop {} {\bibfield  {journal} {\bibinfo  {journal} {ACS Nano}\ }\textbf {\bibinfo {volume} {13}},\ \bibinfo {pages} {4436} (\bibinfo {year} {2019})}\BibitemShut {NoStop}%
\bibitem [{\citenamefont {Lei}\ \emph {et~al.}(2020)\citenamefont {Lei}, \citenamefont {Lin}, \citenamefont {Jia}, \citenamefont {Gray}, \citenamefont {Topp}, \citenamefont {Farahi}, \citenamefont {Klemenz}, \citenamefont {Gao}, \citenamefont {Rodolakis}, \citenamefont {McChesney}, \citenamefont {Ast}, \citenamefont {Yazdani}, \citenamefont {Burch}, \citenamefont {Wu}, \citenamefont {Ong},\ and\ \citenamefont {Schoop}}]{Lei2020}%
  \BibitemOpen
  \bibfield  {author} {\bibinfo {author} {\bibfnamefont {S.}~\bibnamefont {Lei}}, \bibinfo {author} {\bibfnamefont {J.}~\bibnamefont {Lin}}, \bibinfo {author} {\bibfnamefont {Y.}~\bibnamefont {Jia}}, \bibinfo {author} {\bibfnamefont {M.}~\bibnamefont {Gray}}, \bibinfo {author} {\bibfnamefont {A.}~\bibnamefont {Topp}}, \bibinfo {author} {\bibfnamefont {G.}~\bibnamefont {Farahi}}, \bibinfo {author} {\bibfnamefont {S.}~\bibnamefont {Klemenz}}, \bibinfo {author} {\bibfnamefont {T.}~\bibnamefont {Gao}}, \bibinfo {author} {\bibfnamefont {F.}~\bibnamefont {Rodolakis}}, \bibinfo {author} {\bibfnamefont {J.~L.}\ \bibnamefont {McChesney}}, \bibinfo {author} {\bibfnamefont {C.~R.}\ \bibnamefont {Ast}}, \bibinfo {author} {\bibfnamefont {A.}~\bibnamefont {Yazdani}}, \bibinfo {author} {\bibfnamefont {K.~S.}\ \bibnamefont {Burch}}, \bibinfo {author} {\bibfnamefont {S.}~\bibnamefont {Wu}}, \bibinfo {author} {\bibfnamefont {N.~P.}\ \bibnamefont {Ong}},\ and\ \bibinfo {author} {\bibfnamefont {L.~M.}\ \bibnamefont {Schoop}},\
  }\bibfield  {title} {\bibinfo {title} {{High mobility in a van der Waals layered antiferromagnetic metal}},\ }\href {https://doi.org/10.1126/sciadv.aay6407} {\bibfield  {journal} {\bibinfo  {journal} {Science Advances}\ }\textbf {\bibinfo {volume} {6}},\ \bibinfo {pages} {eaay6407} (\bibinfo {year} {2020})}\BibitemShut {NoStop}%
\bibitem [{\citenamefont {Posey}\ \emph {et~al.}(2024)\citenamefont {Posey}, \citenamefont {Turkel}, \citenamefont {Rezaee}, \citenamefont {Devarakonda}, \citenamefont {Kundu}, \citenamefont {Ong}, \citenamefont {Thinel}, \citenamefont {Chica}, \citenamefont {Vitalone}, \citenamefont {Jing}, \citenamefont {Xu}, \citenamefont {Needell}, \citenamefont {Meirzadeh}, \citenamefont {Feuer}, \citenamefont {Jindal}, \citenamefont {Cui}, \citenamefont {Valla}, \citenamefont {Thunstr{\"o}m}, \citenamefont {Yilmaz}, \citenamefont {Vescovo}, \citenamefont {Graf}, \citenamefont {Zhu}, \citenamefont {Scheie}, \citenamefont {May}, \citenamefont {Eriksson}, \citenamefont {Basov}, \citenamefont {Dean}, \citenamefont {Rubio}, \citenamefont {Kim}, \citenamefont {Ziebel}, \citenamefont {Millis}, \citenamefont {Pasupathy},\ and\ \citenamefont {Roy}}]{Posey2024}%
  \BibitemOpen
  \bibfield  {author} {\bibinfo {author} {\bibfnamefont {V.~A.}\ \bibnamefont {Posey}}, \bibinfo {author} {\bibfnamefont {S.}~\bibnamefont {Turkel}}, \bibinfo {author} {\bibfnamefont {M.}~\bibnamefont {Rezaee}}, \bibinfo {author} {\bibfnamefont {A.}~\bibnamefont {Devarakonda}}, \bibinfo {author} {\bibfnamefont {A.~K.}\ \bibnamefont {Kundu}}, \bibinfo {author} {\bibfnamefont {C.~S.}\ \bibnamefont {Ong}}, \bibinfo {author} {\bibfnamefont {M.}~\bibnamefont {Thinel}}, \bibinfo {author} {\bibfnamefont {D.~G.}\ \bibnamefont {Chica}}, \bibinfo {author} {\bibfnamefont {R.~A.}\ \bibnamefont {Vitalone}}, \bibinfo {author} {\bibfnamefont {R.}~\bibnamefont {Jing}}, \bibinfo {author} {\bibfnamefont {S.}~\bibnamefont {Xu}}, \bibinfo {author} {\bibfnamefont {D.~R.}\ \bibnamefont {Needell}}, \bibinfo {author} {\bibfnamefont {E.}~\bibnamefont {Meirzadeh}}, \bibinfo {author} {\bibfnamefont {M.~L.}\ \bibnamefont {Feuer}}, \bibinfo {author} {\bibfnamefont {A.}~\bibnamefont {Jindal}}, \bibinfo {author} {\bibfnamefont
  {X.}~\bibnamefont {Cui}}, \bibinfo {author} {\bibfnamefont {T.}~\bibnamefont {Valla}}, \bibinfo {author} {\bibfnamefont {P.}~\bibnamefont {Thunstr{\"o}m}}, \bibinfo {author} {\bibfnamefont {T.}~\bibnamefont {Yilmaz}}, \bibinfo {author} {\bibfnamefont {E.}~\bibnamefont {Vescovo}}, \bibinfo {author} {\bibfnamefont {D.}~\bibnamefont {Graf}}, \bibinfo {author} {\bibfnamefont {X.}~\bibnamefont {Zhu}}, \bibinfo {author} {\bibfnamefont {A.}~\bibnamefont {Scheie}}, \bibinfo {author} {\bibfnamefont {A.~F.}\ \bibnamefont {May}}, \bibinfo {author} {\bibfnamefont {O.}~\bibnamefont {Eriksson}}, \bibinfo {author} {\bibfnamefont {D.~N.}\ \bibnamefont {Basov}}, \bibinfo {author} {\bibfnamefont {C.~R.}\ \bibnamefont {Dean}}, \bibinfo {author} {\bibfnamefont {A.}~\bibnamefont {Rubio}}, \bibinfo {author} {\bibfnamefont {P.}~\bibnamefont {Kim}}, \bibinfo {author} {\bibfnamefont {M.~E.}\ \bibnamefont {Ziebel}}, \bibinfo {author} {\bibfnamefont {A.~J.}\ \bibnamefont {Millis}}, \bibinfo {author} {\bibfnamefont {A.~N.}\
  \bibnamefont {Pasupathy}},\ and\ \bibinfo {author} {\bibfnamefont {X.}~\bibnamefont {Roy}},\ }\bibfield  {title} {\bibinfo {title} {{Two-dimensional heavy fermions in the van der Waals metal {CeSiI}}},\ }\href@noop {} {\bibfield  {journal} {\bibinfo  {journal} {Nature}\ }\textbf {\bibinfo {volume} {625}},\ \bibinfo {pages} {483} (\bibinfo {year} {2024})}\BibitemShut {NoStop}%
\bibitem [{\citenamefont {Broyles}\ \emph {et~al.}(2025)\citenamefont {Broyles}, \citenamefont {Mardanya}, \citenamefont {Liu}, \citenamefont {Ahn}, \citenamefont {Dinh}, \citenamefont {Alqasseri}, \citenamefont {Garner}, \citenamefont {Rehfuss}, \citenamefont {Guo}, \citenamefont {Zhu}, \citenamefont {Martinez}, \citenamefont {Li}, \citenamefont {Hao}, \citenamefont {Cao}, \citenamefont {Boswell}, \citenamefont {Xie}, \citenamefont {Philbrick}, \citenamefont {Kong}, \citenamefont {Yang}, \citenamefont {Vishwanath}, \citenamefont {Kim}, \citenamefont {Xu}, \citenamefont {Hoffman}, \citenamefont {Denlinger}, \citenamefont {Chowdhury},\ and\ \citenamefont {Ran}}]{Broyles2025}%
  \BibitemOpen
  \bibfield  {author} {\bibinfo {author} {\bibfnamefont {C.}~\bibnamefont {Broyles}}, \bibinfo {author} {\bibfnamefont {S.}~\bibnamefont {Mardanya}}, \bibinfo {author} {\bibfnamefont {M.}~\bibnamefont {Liu}}, \bibinfo {author} {\bibfnamefont {J.}~\bibnamefont {Ahn}}, \bibinfo {author} {\bibfnamefont {T.}~\bibnamefont {Dinh}}, \bibinfo {author} {\bibfnamefont {G.}~\bibnamefont {Alqasseri}}, \bibinfo {author} {\bibfnamefont {J.}~\bibnamefont {Garner}}, \bibinfo {author} {\bibfnamefont {Z.}~\bibnamefont {Rehfuss}}, \bibinfo {author} {\bibfnamefont {K.}~\bibnamefont {Guo}}, \bibinfo {author} {\bibfnamefont {J.}~\bibnamefont {Zhu}}, \bibinfo {author} {\bibfnamefont {D.}~\bibnamefont {Martinez}}, \bibinfo {author} {\bibfnamefont {D.}~\bibnamefont {Li}}, \bibinfo {author} {\bibfnamefont {Y.}~\bibnamefont {Hao}}, \bibinfo {author} {\bibfnamefont {H.}~\bibnamefont {Cao}}, \bibinfo {author} {\bibfnamefont {M.}~\bibnamefont {Boswell}}, \bibinfo {author} {\bibfnamefont {W.}~\bibnamefont {Xie}}, \bibinfo {author}
  {\bibfnamefont {J.~G.}\ \bibnamefont {Philbrick}}, \bibinfo {author} {\bibfnamefont {T.}~\bibnamefont {Kong}}, \bibinfo {author} {\bibfnamefont {L.}~\bibnamefont {Yang}}, \bibinfo {author} {\bibfnamefont {A.}~\bibnamefont {Vishwanath}}, \bibinfo {author} {\bibfnamefont {P.}~\bibnamefont {Kim}}, \bibinfo {author} {\bibfnamefont {S.-Y.}\ \bibnamefont {Xu}}, \bibinfo {author} {\bibfnamefont {J.~E.}\ \bibnamefont {Hoffman}}, \bibinfo {author} {\bibfnamefont {J.~D.}\ \bibnamefont {Denlinger}}, \bibinfo {author} {\bibfnamefont {S.}~\bibnamefont {Chowdhury}},\ and\ \bibinfo {author} {\bibfnamefont {S.}~\bibnamefont {Ran}},\ }\bibfield  {title} {\bibinfo {title} {{UOTe: Kondo-Interacting Topological Antiferromagnet in a Van der Waals Lattice}},\ }\href {https://doi.org/https://doi.org/10.1002/adma.202414966} {\bibfield  {journal} {\bibinfo  {journal} {Advanced Materials}\ }\textbf {\bibinfo {volume} {37}},\ \bibinfo {pages} {2414966} (\bibinfo {year} {2025})}\BibitemShut {NoStop}%
\bibitem [{\citenamefont {Novoselov}\ \emph {et~al.}(2004)\citenamefont {Novoselov}, \citenamefont {Geim}, \citenamefont {Morozov}, \citenamefont {Jiang}, \citenamefont {Zhang}, \citenamefont {Dubonos}, \citenamefont {Grigorieva},\ and\ \citenamefont {Firsov}}]{Novoselov2004}%
  \BibitemOpen
  \bibfield  {author} {\bibinfo {author} {\bibfnamefont {K.~S.}\ \bibnamefont {Novoselov}}, \bibinfo {author} {\bibfnamefont {A.~K.}\ \bibnamefont {Geim}}, \bibinfo {author} {\bibfnamefont {S.~V.}\ \bibnamefont {Morozov}}, \bibinfo {author} {\bibfnamefont {D.}~\bibnamefont {Jiang}}, \bibinfo {author} {\bibfnamefont {Y.}~\bibnamefont {Zhang}}, \bibinfo {author} {\bibfnamefont {S.~V.}\ \bibnamefont {Dubonos}}, \bibinfo {author} {\bibfnamefont {I.~V.}\ \bibnamefont {Grigorieva}},\ and\ \bibinfo {author} {\bibfnamefont {A.~A.}\ \bibnamefont {Firsov}},\ }\bibfield  {title} {\bibinfo {title} {{Electric Field Effect in Atomically Thin Carbon Films}},\ }\href {https://doi.org/10.1126/science.1102896} {\bibfield  {journal} {\bibinfo  {journal} {Science}\ }\textbf {\bibinfo {volume} {306}},\ \bibinfo {pages} {666} (\bibinfo {year} {2004})}\BibitemShut {NoStop}%
\bibitem [{\citenamefont {Andrei}\ and\ \citenamefont {MacDonald}(2020)}]{Andrei2020}%
  \BibitemOpen
  \bibfield  {author} {\bibinfo {author} {\bibfnamefont {E.~Y.}\ \bibnamefont {Andrei}}\ and\ \bibinfo {author} {\bibfnamefont {A.~H.}\ \bibnamefont {MacDonald}},\ }\bibfield  {title} {\bibinfo {title} {{Graphene bilayers with a twist}},\ }\href@noop {} {\bibfield  {journal} {\bibinfo  {journal} {Nature Materials}\ }\textbf {\bibinfo {volume} {19}},\ \bibinfo {pages} {1265} (\bibinfo {year} {2020})}\BibitemShut {NoStop}%
\bibitem [{\citenamefont {Xie}\ \emph {et~al.}(2021)\citenamefont {Xie}, \citenamefont {Pierce}, \citenamefont {Park}, \citenamefont {Parker}, \citenamefont {Khalaf}, \citenamefont {Ledwith}, \citenamefont {Cao}, \citenamefont {Lee}, \citenamefont {Chen}, \citenamefont {Forrester}, \citenamefont {Watanabe}, \citenamefont {Taniguchi}, \citenamefont {Vishwanath}, \citenamefont {Jarillo-Herrero},\ and\ \citenamefont {Yacoby}}]{Xie2021}%
  \BibitemOpen
  \bibfield  {author} {\bibinfo {author} {\bibfnamefont {Y.}~\bibnamefont {Xie}}, \bibinfo {author} {\bibfnamefont {A.~T.}\ \bibnamefont {Pierce}}, \bibinfo {author} {\bibfnamefont {J.~M.}\ \bibnamefont {Park}}, \bibinfo {author} {\bibfnamefont {D.~E.}\ \bibnamefont {Parker}}, \bibinfo {author} {\bibfnamefont {E.}~\bibnamefont {Khalaf}}, \bibinfo {author} {\bibfnamefont {P.}~\bibnamefont {Ledwith}}, \bibinfo {author} {\bibfnamefont {Y.}~\bibnamefont {Cao}}, \bibinfo {author} {\bibfnamefont {S.~H.}\ \bibnamefont {Lee}}, \bibinfo {author} {\bibfnamefont {S.}~\bibnamefont {Chen}}, \bibinfo {author} {\bibfnamefont {P.~R.}\ \bibnamefont {Forrester}}, \bibinfo {author} {\bibfnamefont {K.}~\bibnamefont {Watanabe}}, \bibinfo {author} {\bibfnamefont {T.}~\bibnamefont {Taniguchi}}, \bibinfo {author} {\bibfnamefont {A.}~\bibnamefont {Vishwanath}}, \bibinfo {author} {\bibfnamefont {P.}~\bibnamefont {Jarillo-Herrero}},\ and\ \bibinfo {author} {\bibfnamefont {A.}~\bibnamefont {Yacoby}},\ }\bibfield  {title} {\bibinfo
  {title} {{Fractional Chern insulators in magic-angle twisted bilayer graphene}},\ }\href@noop {} {\bibfield  {journal} {\bibinfo  {journal} {Nature}\ }\textbf {\bibinfo {volume} {600}},\ \bibinfo {pages} {439} (\bibinfo {year} {2021})}\BibitemShut {NoStop}%
\bibitem [{\citenamefont {Han}\ \emph {et~al.}(2025)\citenamefont {Han}, \citenamefont {Lu}, \citenamefont {Hadjri}, \citenamefont {Shi}, \citenamefont {Wu}, \citenamefont {Xu}, \citenamefont {Yao}, \citenamefont {Cotten}, \citenamefont {Sedeh}, \citenamefont {Weldeyesus}, \citenamefont {Yang}, \citenamefont {Seo}, \citenamefont {Ye}, \citenamefont {Zhou}, \citenamefont {Liu}, \citenamefont {Shi}, \citenamefont {Hua}, \citenamefont {Watanabe}, \citenamefont {Taniguchi}, \citenamefont {Xiong}, \citenamefont {Zumbühl}, \citenamefont {Fu},\ and\ \citenamefont {Ju}}]{han2025}%
  \BibitemOpen
  \bibfield  {author} {\bibinfo {author} {\bibfnamefont {T.}~\bibnamefont {Han}}, \bibinfo {author} {\bibfnamefont {Z.}~\bibnamefont {Lu}}, \bibinfo {author} {\bibfnamefont {Z.}~\bibnamefont {Hadjri}}, \bibinfo {author} {\bibfnamefont {L.}~\bibnamefont {Shi}}, \bibinfo {author} {\bibfnamefont {Z.}~\bibnamefont {Wu}}, \bibinfo {author} {\bibfnamefont {W.}~\bibnamefont {Xu}}, \bibinfo {author} {\bibfnamefont {Y.}~\bibnamefont {Yao}}, \bibinfo {author} {\bibfnamefont {A.~A.}\ \bibnamefont {Cotten}}, \bibinfo {author} {\bibfnamefont {O.~S.}\ \bibnamefont {Sedeh}}, \bibinfo {author} {\bibfnamefont {H.}~\bibnamefont {Weldeyesus}}, \bibinfo {author} {\bibfnamefont {J.}~\bibnamefont {Yang}}, \bibinfo {author} {\bibfnamefont {J.}~\bibnamefont {Seo}}, \bibinfo {author} {\bibfnamefont {S.}~\bibnamefont {Ye}}, \bibinfo {author} {\bibfnamefont {M.}~\bibnamefont {Zhou}}, \bibinfo {author} {\bibfnamefont {H.}~\bibnamefont {Liu}}, \bibinfo {author} {\bibfnamefont {G.}~\bibnamefont {Shi}}, \bibinfo {author} {\bibfnamefont
  {Z.}~\bibnamefont {Hua}}, \bibinfo {author} {\bibfnamefont {K.}~\bibnamefont {Watanabe}}, \bibinfo {author} {\bibfnamefont {T.}~\bibnamefont {Taniguchi}}, \bibinfo {author} {\bibfnamefont {P.}~\bibnamefont {Xiong}}, \bibinfo {author} {\bibfnamefont {D.~M.}\ \bibnamefont {Zumbühl}}, \bibinfo {author} {\bibfnamefont {L.}~\bibnamefont {Fu}},\ and\ \bibinfo {author} {\bibfnamefont {L.}~\bibnamefont {Ju}},\ }\href {https://arxiv.org/abs/2408.15233} {\bibinfo {title} {{Signatures of Chiral Superconductivity in Rhombohedral Graphene}}} (\bibinfo {year} {2025}),\ \Eprint {https://arxiv.org/abs/2408.15233} {arXiv:2408.15233 [cond-mat.mes-hall]} \BibitemShut {NoStop}%
\bibitem [{\citenamefont {Novoselov}\ \emph {et~al.}(2016)\citenamefont {Novoselov}, \citenamefont {Mishchenko}, \citenamefont {Carvalho},\ and\ \citenamefont {Neto}}]{novoselov2016}%
  \BibitemOpen
  \bibfield  {author} {\bibinfo {author} {\bibfnamefont {K.~S.}\ \bibnamefont {Novoselov}}, \bibinfo {author} {\bibfnamefont {A.}~\bibnamefont {Mishchenko}}, \bibinfo {author} {\bibfnamefont {A.}~\bibnamefont {Carvalho}},\ and\ \bibinfo {author} {\bibfnamefont {A.~H.~C.}\ \bibnamefont {Neto}},\ }\bibfield  {title} {\bibinfo {title} {{2D materials and van der Waals heterostructures}},\ }\href {https://doi.org/10.1126/science.aac9439} {\bibfield  {journal} {\bibinfo  {journal} {Science}\ }\textbf {\bibinfo {volume} {353}},\ \bibinfo {pages} {aac9439} (\bibinfo {year} {2016})}\BibitemShut {NoStop}%
\bibitem [{\citenamefont {Tan}\ \emph {et~al.}(2017)\citenamefont {Tan}, \citenamefont {Cao}, \citenamefont {Wu}, \citenamefont {He}, \citenamefont {Yang}, \citenamefont {Zhang}, \citenamefont {Chen}, \citenamefont {Zhao}, \citenamefont {Han}, \citenamefont {Nam} \emph {et~al.}}]{tan2017}%
  \BibitemOpen
  \bibfield  {author} {\bibinfo {author} {\bibfnamefont {C.}~\bibnamefont {Tan}}, \bibinfo {author} {\bibfnamefont {X.}~\bibnamefont {Cao}}, \bibinfo {author} {\bibfnamefont {X.-J.}\ \bibnamefont {Wu}}, \bibinfo {author} {\bibfnamefont {Q.}~\bibnamefont {He}}, \bibinfo {author} {\bibfnamefont {J.}~\bibnamefont {Yang}}, \bibinfo {author} {\bibfnamefont {X.}~\bibnamefont {Zhang}}, \bibinfo {author} {\bibfnamefont {J.}~\bibnamefont {Chen}}, \bibinfo {author} {\bibfnamefont {W.}~\bibnamefont {Zhao}}, \bibinfo {author} {\bibfnamefont {S.}~\bibnamefont {Han}}, \bibinfo {author} {\bibfnamefont {G.-H.}\ \bibnamefont {Nam}}, \emph {et~al.},\ }\bibfield  {title} {\bibinfo {title} {{Recent advances in ultrathin two-dimensional nanomaterials}},\ }\href@noop {} {\bibfield  {journal} {\bibinfo  {journal} {Chemical reviews}\ }\textbf {\bibinfo {volume} {117}},\ \bibinfo {pages} {6225} (\bibinfo {year} {2017})}\BibitemShut {NoStop}%
\bibitem [{\citenamefont {Castellanos-Gomez}\ \emph {et~al.}(2022)\citenamefont {Castellanos-Gomez}, \citenamefont {Duan}, \citenamefont {Fei}, \citenamefont {Gutierrez}, \citenamefont {Huang}, \citenamefont {Huang}, \citenamefont {Quereda}, \citenamefont {Qian}, \citenamefont {Sutter},\ and\ \citenamefont {Sutter}}]{Castellanos2022}%
  \BibitemOpen
  \bibfield  {author} {\bibinfo {author} {\bibfnamefont {A.}~\bibnamefont {Castellanos-Gomez}}, \bibinfo {author} {\bibfnamefont {X.}~\bibnamefont {Duan}}, \bibinfo {author} {\bibfnamefont {Z.}~\bibnamefont {Fei}}, \bibinfo {author} {\bibfnamefont {H.~R.}\ \bibnamefont {Gutierrez}}, \bibinfo {author} {\bibfnamefont {Y.}~\bibnamefont {Huang}}, \bibinfo {author} {\bibfnamefont {X.}~\bibnamefont {Huang}}, \bibinfo {author} {\bibfnamefont {J.}~\bibnamefont {Quereda}}, \bibinfo {author} {\bibfnamefont {Q.}~\bibnamefont {Qian}}, \bibinfo {author} {\bibfnamefont {E.}~\bibnamefont {Sutter}},\ and\ \bibinfo {author} {\bibfnamefont {P.}~\bibnamefont {Sutter}},\ }\bibfield  {title} {\bibinfo {title} {Van der waals heterostructures},\ }\href {https://doi.org/10.1038/s43586-022-00139-1} {\bibfield  {journal} {\bibinfo  {journal} {Nature Reviews Methods Primers}\ }\textbf {\bibinfo {volume} {2}},\ \bibinfo {pages} {58} (\bibinfo {year} {2022})}\BibitemShut {NoStop}%
\bibitem [{\citenamefont {Gibertini}\ \emph {et~al.}(2019)\citenamefont {Gibertini}, \citenamefont {Koperski}, \citenamefont {Morpurgo},\ and\ \citenamefont {Novoselov}}]{Gibertini2019}%
  \BibitemOpen
  \bibfield  {author} {\bibinfo {author} {\bibfnamefont {M.}~\bibnamefont {Gibertini}}, \bibinfo {author} {\bibfnamefont {M.}~\bibnamefont {Koperski}}, \bibinfo {author} {\bibfnamefont {A.~F.}\ \bibnamefont {Morpurgo}},\ and\ \bibinfo {author} {\bibfnamefont {K.~S.}\ \bibnamefont {Novoselov}},\ }\bibfield  {title} {\bibinfo {title} {Magnetic {2D} materials and heterostructures},\ }\href@noop {} {\bibfield  {journal} {\bibinfo  {journal} {Nature Nanotechnology}\ }\textbf {\bibinfo {volume} {14}},\ \bibinfo {pages} {408} (\bibinfo {year} {2019})}\BibitemShut {NoStop}%
\bibitem [{\citenamefont {Varma}(2016)}]{varma2016quantum}%
  \BibitemOpen
  \bibfield  {author} {\bibinfo {author} {\bibfnamefont {C.~M.}\ \bibnamefont {Varma}},\ }\bibfield  {title} {\bibinfo {title} {Quantum-critical fluctuations in 2d metals: strange metals and superconductivity in antiferromagnets and in cuprates},\ }\href@noop {} {\bibfield  {journal} {\bibinfo  {journal} {Reports on Progress in Physics}\ }\textbf {\bibinfo {volume} {79}},\ \bibinfo {pages} {082501} (\bibinfo {year} {2016})}\BibitemShut {NoStop}%
\bibitem [{\citenamefont {Phillips}\ \emph {et~al.}(2022)\citenamefont {Phillips}, \citenamefont {Hussey},\ and\ \citenamefont {Abbamonte}}]{Phillips2022}%
  \BibitemOpen
  \bibfield  {author} {\bibinfo {author} {\bibfnamefont {P.~W.}\ \bibnamefont {Phillips}}, \bibinfo {author} {\bibfnamefont {N.~E.}\ \bibnamefont {Hussey}},\ and\ \bibinfo {author} {\bibfnamefont {P.}~\bibnamefont {Abbamonte}},\ }\bibfield  {title} {\bibinfo {title} {Stranger than metals},\ }\href {https://doi.org/10.1126/science.abh4273} {\bibfield  {journal} {\bibinfo  {journal} {Science}\ }\textbf {\bibinfo {volume} {377}},\ \bibinfo {pages} {eabh4273} (\bibinfo {year} {2022})}\BibitemShut {NoStop}%
\bibitem [{\citenamefont {Alicea}(2012)}]{Alicea_2012}%
  \BibitemOpen
  \bibfield  {author} {\bibinfo {author} {\bibfnamefont {J.}~\bibnamefont {Alicea}},\ }\bibfield  {title} {\bibinfo {title} {New directions in the pursuit of majorana fermions in solid state systems},\ }\href {https://doi.org/10.1088/0034-4885/75/7/076501} {\bibfield  {journal} {\bibinfo  {journal} {Reports on Progress in Physics}\ }\textbf {\bibinfo {volume} {75}},\ \bibinfo {pages} {076501} (\bibinfo {year} {2012})}\BibitemShut {NoStop}%
\bibitem [{\citenamefont {Yumigeta}\ \emph {et~al.}(2021)\citenamefont {Yumigeta}, \citenamefont {Qin}, \citenamefont {Li}, \citenamefont {Blei}, \citenamefont {Attarde}, \citenamefont {Kopas},\ and\ \citenamefont {Tongay}}]{Tongay2021}%
  \BibitemOpen
  \bibfield  {author} {\bibinfo {author} {\bibfnamefont {K.}~\bibnamefont {Yumigeta}}, \bibinfo {author} {\bibfnamefont {Y.}~\bibnamefont {Qin}}, \bibinfo {author} {\bibfnamefont {H.}~\bibnamefont {Li}}, \bibinfo {author} {\bibfnamefont {M.}~\bibnamefont {Blei}}, \bibinfo {author} {\bibfnamefont {Y.}~\bibnamefont {Attarde}}, \bibinfo {author} {\bibfnamefont {C.}~\bibnamefont {Kopas}},\ and\ \bibinfo {author} {\bibfnamefont {S.}~\bibnamefont {Tongay}},\ }\bibfield  {title} {\bibinfo {title} {{Advances in Rare-Earth Tritelluride Quantum Materials: Structure, Properties, and Synthesis}},\ }\href {https://doi.org/https://doi.org/10.1002/advs.202004762} {\bibfield  {journal} {\bibinfo  {journal} {Advanced Science}\ }\textbf {\bibinfo {volume} {8}},\ \bibinfo {pages} {2004762} (\bibinfo {year} {2021})}\BibitemShut {NoStop}%
\bibitem [{\citenamefont {Ru}\ and\ \citenamefont {Fisher}(2006{\natexlab{a}})}]{Ru2006}%
  \BibitemOpen
  \bibfield  {author} {\bibinfo {author} {\bibfnamefont {N.}~\bibnamefont {Ru}}\ and\ \bibinfo {author} {\bibfnamefont {I.~R.}\ \bibnamefont {Fisher}},\ }\bibfield  {title} {\bibinfo {title} {{Thermodynamic and transport properties of $\mathrm{Y}{\mathrm{Te}}_{3}$, $\mathrm{La}{\mathrm{Te}}_{3}$, and $\mathrm{Ce}{\mathrm{Te}}_{3}$}},\ }\href {https://doi.org/10.1103/PhysRevB.73.033101} {\bibfield  {journal} {\bibinfo  {journal} {Phys. Rev. B}\ }\textbf {\bibinfo {volume} {73}},\ \bibinfo {pages} {033101} (\bibinfo {year} {2006}{\natexlab{a}})}\BibitemShut {NoStop}%
\bibitem [{\citenamefont {Zhu}\ \emph {et~al.}(2016)\citenamefont {Zhu}, \citenamefont {Janoschek}, \citenamefont {Chaves}, \citenamefont {Cezar}, \citenamefont {Durakiewicz}, \citenamefont {Ronning}, \citenamefont {Sassa}, \citenamefont {Mansson}, \citenamefont {Scott}, \citenamefont {Wakeham} \emph {et~al.}}]{zhu2016}%
  \BibitemOpen
  \bibfield  {author} {\bibinfo {author} {\bibfnamefont {J.-X.}\ \bibnamefont {Zhu}}, \bibinfo {author} {\bibfnamefont {M.}~\bibnamefont {Janoschek}}, \bibinfo {author} {\bibfnamefont {D.}~\bibnamefont {Chaves}}, \bibinfo {author} {\bibfnamefont {J.}~\bibnamefont {Cezar}}, \bibinfo {author} {\bibfnamefont {T.}~\bibnamefont {Durakiewicz}}, \bibinfo {author} {\bibfnamefont {F.}~\bibnamefont {Ronning}}, \bibinfo {author} {\bibfnamefont {Y.}~\bibnamefont {Sassa}}, \bibinfo {author} {\bibfnamefont {M.}~\bibnamefont {Mansson}}, \bibinfo {author} {\bibfnamefont {B.}~\bibnamefont {Scott}}, \bibinfo {author} {\bibfnamefont {N.}~\bibnamefont {Wakeham}}, \emph {et~al.},\ }\bibfield  {title} {\bibinfo {title} {Electronic correlation and magnetism in the ferromagnetic metal fe 3 gete 2},\ }\href@noop {} {\bibfield  {journal} {\bibinfo  {journal} {Physical Review B}\ }\textbf {\bibinfo {volume} {93}},\ \bibinfo {pages} {144404} (\bibinfo {year} {2016})}\BibitemShut {NoStop}%
\bibitem [{\citenamefont {Fei}\ \emph {et~al.}(2018)\citenamefont {Fei}, \citenamefont {Huang}, \citenamefont {Malinowski}, \citenamefont {Wang}, \citenamefont {Song}, \citenamefont {Sanchez}, \citenamefont {Yao}, \citenamefont {Xiao}, \citenamefont {Zhu}, \citenamefont {May}, \citenamefont {Wu}, \citenamefont {Cobden}, \citenamefont {Chu},\ and\ \citenamefont {Xu}}]{Fei2018}%
  \BibitemOpen
  \bibfield  {author} {\bibinfo {author} {\bibfnamefont {Z.}~\bibnamefont {Fei}}, \bibinfo {author} {\bibfnamefont {B.}~\bibnamefont {Huang}}, \bibinfo {author} {\bibfnamefont {P.}~\bibnamefont {Malinowski}}, \bibinfo {author} {\bibfnamefont {W.}~\bibnamefont {Wang}}, \bibinfo {author} {\bibfnamefont {T.}~\bibnamefont {Song}}, \bibinfo {author} {\bibfnamefont {J.}~\bibnamefont {Sanchez}}, \bibinfo {author} {\bibfnamefont {W.}~\bibnamefont {Yao}}, \bibinfo {author} {\bibfnamefont {D.}~\bibnamefont {Xiao}}, \bibinfo {author} {\bibfnamefont {X.}~\bibnamefont {Zhu}}, \bibinfo {author} {\bibfnamefont {A.~F.}\ \bibnamefont {May}}, \bibinfo {author} {\bibfnamefont {W.}~\bibnamefont {Wu}}, \bibinfo {author} {\bibfnamefont {D.~H.}\ \bibnamefont {Cobden}}, \bibinfo {author} {\bibfnamefont {J.-H.}\ \bibnamefont {Chu}},\ and\ \bibinfo {author} {\bibfnamefont {X.}~\bibnamefont {Xu}},\ }\bibfield  {title} {\bibinfo {title} {Two-dimensional itinerant ferromagnetism in atomically thin {Fe$_{3}$GeTe$_{2}$}},\ }\href@noop {}
  {\bibfield  {journal} {\bibinfo  {journal} {Nature Materials}\ }\textbf {\bibinfo {volume} {17}},\ \bibinfo {pages} {778} (\bibinfo {year} {2018})}\BibitemShut {NoStop}%
\bibitem [{\citenamefont {Deng}\ \emph {et~al.}(2018)\citenamefont {Deng}, \citenamefont {Yu}, \citenamefont {Song}, \citenamefont {Zhang}, \citenamefont {Wang}, \citenamefont {Sun}, \citenamefont {Yi}, \citenamefont {Wu}, \citenamefont {Wu}, \citenamefont {Zhu}, \citenamefont {Wang}, \citenamefont {Chen},\ and\ \citenamefont {Zhang}}]{Deng2018}%
  \BibitemOpen
  \bibfield  {author} {\bibinfo {author} {\bibfnamefont {Y.}~\bibnamefont {Deng}}, \bibinfo {author} {\bibfnamefont {Y.}~\bibnamefont {Yu}}, \bibinfo {author} {\bibfnamefont {Y.}~\bibnamefont {Song}}, \bibinfo {author} {\bibfnamefont {J.}~\bibnamefont {Zhang}}, \bibinfo {author} {\bibfnamefont {N.~Z.}\ \bibnamefont {Wang}}, \bibinfo {author} {\bibfnamefont {Z.}~\bibnamefont {Sun}}, \bibinfo {author} {\bibfnamefont {Y.}~\bibnamefont {Yi}}, \bibinfo {author} {\bibfnamefont {Y.~Z.}\ \bibnamefont {Wu}}, \bibinfo {author} {\bibfnamefont {S.}~\bibnamefont {Wu}}, \bibinfo {author} {\bibfnamefont {J.}~\bibnamefont {Zhu}}, \bibinfo {author} {\bibfnamefont {J.}~\bibnamefont {Wang}}, \bibinfo {author} {\bibfnamefont {X.~H.}\ \bibnamefont {Chen}},\ and\ \bibinfo {author} {\bibfnamefont {Y.}~\bibnamefont {Zhang}},\ }\bibfield  {title} {\bibinfo {title} {Gate-tunable room-temperature ferromagnetism in two-dimensional {Fe$_{3}$GeTe$_{2}$}},\ }\href@noop {} {\bibfield  {journal} {\bibinfo  {journal} {Nature}\ }\textbf
  {\bibinfo {volume} {563}},\ \bibinfo {pages} {94} (\bibinfo {year} {2018})}\BibitemShut {NoStop}%
\bibitem [{\citenamefont {Wang}\ \emph {et~al.}(2023)\citenamefont {Wang}, \citenamefont {Lu}, \citenamefont {Guo}, \citenamefont {Li}, \citenamefont {Wu}, \citenamefont {Li}, \citenamefont {Xie}, \citenamefont {Sun}, \citenamefont {Li}, \citenamefont {Damas}, \citenamefont {Friedel}, \citenamefont {Migot}, \citenamefont {Ghanbaja}, \citenamefont {Moreau}, \citenamefont {Fagot-Revurat}, \citenamefont {Petit-Watelot}, \citenamefont {Hauet}, \citenamefont {Robertson}, \citenamefont {Mangin}, \citenamefont {Zhao},\ and\ \citenamefont {Nie}}]{Wang2023}%
  \BibitemOpen
  \bibfield  {author} {\bibinfo {author} {\bibfnamefont {H.}~\bibnamefont {Wang}}, \bibinfo {author} {\bibfnamefont {H.}~\bibnamefont {Lu}}, \bibinfo {author} {\bibfnamefont {Z.}~\bibnamefont {Guo}}, \bibinfo {author} {\bibfnamefont {A.}~\bibnamefont {Li}}, \bibinfo {author} {\bibfnamefont {P.}~\bibnamefont {Wu}}, \bibinfo {author} {\bibfnamefont {J.}~\bibnamefont {Li}}, \bibinfo {author} {\bibfnamefont {W.}~\bibnamefont {Xie}}, \bibinfo {author} {\bibfnamefont {Z.}~\bibnamefont {Sun}}, \bibinfo {author} {\bibfnamefont {P.}~\bibnamefont {Li}}, \bibinfo {author} {\bibfnamefont {H.}~\bibnamefont {Damas}}, \bibinfo {author} {\bibfnamefont {A.~M.}\ \bibnamefont {Friedel}}, \bibinfo {author} {\bibfnamefont {S.}~\bibnamefont {Migot}}, \bibinfo {author} {\bibfnamefont {J.}~\bibnamefont {Ghanbaja}}, \bibinfo {author} {\bibfnamefont {L.}~\bibnamefont {Moreau}}, \bibinfo {author} {\bibfnamefont {Y.}~\bibnamefont {Fagot-Revurat}}, \bibinfo {author} {\bibfnamefont {S.}~\bibnamefont {Petit-Watelot}}, \bibinfo {author}
  {\bibfnamefont {T.}~\bibnamefont {Hauet}}, \bibinfo {author} {\bibfnamefont {J.}~\bibnamefont {Robertson}}, \bibinfo {author} {\bibfnamefont {S.}~\bibnamefont {Mangin}}, \bibinfo {author} {\bibfnamefont {W.}~\bibnamefont {Zhao}},\ and\ \bibinfo {author} {\bibfnamefont {T.}~\bibnamefont {Nie}},\ }\bibfield  {title} {\bibinfo {title} {Interfacial engineering of ferromagnetism in wafer-scale van der waals {Fe$_{4}$GeTe$_{2}$} far above room temperature},\ }\href@noop {} {\bibfield  {journal} {\bibinfo  {journal} {Nature Communications}\ }\textbf {\bibinfo {volume} {14}},\ \bibinfo {pages} {2483} (\bibinfo {year} {2023})}\BibitemShut {NoStop}%
\bibitem [{\citenamefont {Dang}\ \emph {et~al.}(2023)\citenamefont {Dang}, \citenamefont {Kozlenko}, \citenamefont {Lis}, \citenamefont {Kichanov}, \citenamefont {Lukin}, \citenamefont {Golosova}, \citenamefont {Savenko}, \citenamefont {Duong}, \citenamefont {Phan}, \citenamefont {Tran} \emph {et~al.}}]{dang2023}%
  \BibitemOpen
  \bibfield  {author} {\bibinfo {author} {\bibfnamefont {N.-T.}\ \bibnamefont {Dang}}, \bibinfo {author} {\bibfnamefont {D.~P.}\ \bibnamefont {Kozlenko}}, \bibinfo {author} {\bibfnamefont {O.~N.}\ \bibnamefont {Lis}}, \bibinfo {author} {\bibfnamefont {S.~E.}\ \bibnamefont {Kichanov}}, \bibinfo {author} {\bibfnamefont {Y.~V.}\ \bibnamefont {Lukin}}, \bibinfo {author} {\bibfnamefont {N.~O.}\ \bibnamefont {Golosova}}, \bibinfo {author} {\bibfnamefont {B.~N.}\ \bibnamefont {Savenko}}, \bibinfo {author} {\bibfnamefont {D.-L.}\ \bibnamefont {Duong}}, \bibinfo {author} {\bibfnamefont {T.-L.}\ \bibnamefont {Phan}}, \bibinfo {author} {\bibfnamefont {T.-A.}\ \bibnamefont {Tran}}, \emph {et~al.},\ }\bibfield  {title} {\bibinfo {title} {High pressure-driven magnetic disorder and structural transformation in {Fe$_{3}$GeTe$_{2}$}: Emergence of a magnetic quantum critical point},\ }\href@noop {} {\bibfield  {journal} {\bibinfo  {journal} {Advanced Science}\ }\textbf {\bibinfo {volume} {10}},\ \bibinfo {pages} {2206842}
  (\bibinfo {year} {2023})}\BibitemShut {NoStop}%
\bibitem [{\citenamefont {Peters}\ \emph {et~al.}(1988)\citenamefont {Peters}, \citenamefont {Birgeneau}, \citenamefont {Kastner}, \citenamefont {Yoshizawa}, \citenamefont {Endoh}, \citenamefont {Tranquada}, \citenamefont {Shirane}, \citenamefont {Hidaka}, \citenamefont {Oda}, \citenamefont {Suzuki} \emph {et~al.}}]{peters1988}%
  \BibitemOpen
  \bibfield  {author} {\bibinfo {author} {\bibfnamefont {C.}~\bibnamefont {Peters}}, \bibinfo {author} {\bibfnamefont {R.}~\bibnamefont {Birgeneau}}, \bibinfo {author} {\bibfnamefont {M.}~\bibnamefont {Kastner}}, \bibinfo {author} {\bibfnamefont {H.}~\bibnamefont {Yoshizawa}}, \bibinfo {author} {\bibfnamefont {Y.}~\bibnamefont {Endoh}}, \bibinfo {author} {\bibfnamefont {J.}~\bibnamefont {Tranquada}}, \bibinfo {author} {\bibfnamefont {G.}~\bibnamefont {Shirane}}, \bibinfo {author} {\bibfnamefont {Y.}~\bibnamefont {Hidaka}}, \bibinfo {author} {\bibfnamefont {M.}~\bibnamefont {Oda}}, \bibinfo {author} {\bibfnamefont {M.}~\bibnamefont {Suzuki}}, \emph {et~al.},\ }\bibfield  {title} {\bibinfo {title} {Two-dimensional zone-center spin-wave excitations in {La$_{2}$CuO$_{4}$}},\ }\href@noop {} {\bibfield  {journal} {\bibinfo  {journal} {Physical Review B}\ }\textbf {\bibinfo {volume} {37}},\ \bibinfo {pages} {9761} (\bibinfo {year} {1988})}\BibitemShut {NoStop}%
\bibitem [{\citenamefont {Sun}\ \emph {et~al.}(1991)\citenamefont {Sun}, \citenamefont {Cho}, \citenamefont {Chou}, \citenamefont {Lee}, \citenamefont {Miller}, \citenamefont {Johnston}, \citenamefont {Hidaka},\ and\ \citenamefont {Murakami}}]{sun1991heat}%
  \BibitemOpen
  \bibfield  {author} {\bibinfo {author} {\bibfnamefont {K.}~\bibnamefont {Sun}}, \bibinfo {author} {\bibfnamefont {J.}~\bibnamefont {Cho}}, \bibinfo {author} {\bibfnamefont {F.}~\bibnamefont {Chou}}, \bibinfo {author} {\bibfnamefont {W.}~\bibnamefont {Lee}}, \bibinfo {author} {\bibfnamefont {L.}~\bibnamefont {Miller}}, \bibinfo {author} {\bibfnamefont {D.}~\bibnamefont {Johnston}}, \bibinfo {author} {\bibfnamefont {Y.}~\bibnamefont {Hidaka}},\ and\ \bibinfo {author} {\bibfnamefont {T.}~\bibnamefont {Murakami}},\ }\bibfield  {title} {\bibinfo {title} {Heat capacity of single-crystal {La$_{2}$CuO$_{4}$} and polycrystalline {La$_{2-x}$Sr$_{x}$CuO$_{4}$} {($0 \leq x \leq 0.20$)} from 110 to 600 k},\ }\href@noop {} {\bibfield  {journal} {\bibinfo  {journal} {Physical Review B}\ }\textbf {\bibinfo {volume} {43}},\ \bibinfo {pages} {239} (\bibinfo {year} {1991})}\BibitemShut {NoStop}%
\bibitem [{\citenamefont {Stewart}(1984)}]{Stewart1984}%
  \BibitemOpen
  \bibfield  {author} {\bibinfo {author} {\bibfnamefont {G.~R.}\ \bibnamefont {Stewart}},\ }\bibfield  {title} {\bibinfo {title} {{Heavy-fermion systems}},\ }\href {https://doi.org/10.1103/RevModPhys.56.755} {\bibfield  {journal} {\bibinfo  {journal} {Rev. Mod. Phys.}\ }\textbf {\bibinfo {volume} {56}},\ \bibinfo {pages} {755} (\bibinfo {year} {1984})}\BibitemShut {NoStop}%
\bibitem [{\citenamefont {White}\ \emph {et~al.}(2015)\citenamefont {White}, \citenamefont {Thompson},\ and\ \citenamefont {Maple}}]{WHITE2015}%
  \BibitemOpen
  \bibfield  {author} {\bibinfo {author} {\bibfnamefont {B.}~\bibnamefont {White}}, \bibinfo {author} {\bibfnamefont {J.}~\bibnamefont {Thompson}},\ and\ \bibinfo {author} {\bibfnamefont {M.}~\bibnamefont {Maple}},\ }\bibfield  {title} {\bibinfo {title} {Unconventional superconductivity in heavy-fermion compounds},\ }\href {https://doi.org/https://doi.org/10.1016/j.physc.2015.02.044} {\bibfield  {journal} {\bibinfo  {journal} {Physica C: Superconductivity and its Applications}\ }\textbf {\bibinfo {volume} {514}},\ \bibinfo {pages} {246} (\bibinfo {year} {2015})}\BibitemShut {NoStop}%
\bibitem [{\citenamefont {Aoki}\ \emph {et~al.}(2019)\citenamefont {Aoki}, \citenamefont {Ishida},\ and\ \citenamefont {Flouquet}}]{Aoki2019}%
  \BibitemOpen
  \bibfield  {author} {\bibinfo {author} {\bibfnamefont {D.}~\bibnamefont {Aoki}}, \bibinfo {author} {\bibfnamefont {K.}~\bibnamefont {Ishida}},\ and\ \bibinfo {author} {\bibfnamefont {J.}~\bibnamefont {Flouquet}},\ }\bibfield  {title} {\bibinfo {title} {Review of u-based ferromagnetic superconductors: Comparison between {UGe$_{2}$}, {URhGe}, and {UCoGe}},\ }\href {https://doi.org/10.7566/JPSJ.88.022001} {\bibfield  {journal} {\bibinfo  {journal} {Journal of the Physical Society of Japan}\ }\textbf {\bibinfo {volume} {88}},\ \bibinfo {pages} {022001} (\bibinfo {year} {2019})}\BibitemShut {NoStop}%
\bibitem [{\citenamefont {Aoki}\ \emph {et~al.}(2022)\citenamefont {Aoki}, \citenamefont {Brison}, \citenamefont {Flouquet}, \citenamefont {Ishida}, \citenamefont {Knebel}, \citenamefont {Tokunaga},\ and\ \citenamefont {Yanase}}]{aoki2022unconventional}%
  \BibitemOpen
  \bibfield  {author} {\bibinfo {author} {\bibfnamefont {D.}~\bibnamefont {Aoki}}, \bibinfo {author} {\bibfnamefont {J.-P.}\ \bibnamefont {Brison}}, \bibinfo {author} {\bibfnamefont {J.}~\bibnamefont {Flouquet}}, \bibinfo {author} {\bibfnamefont {K.}~\bibnamefont {Ishida}}, \bibinfo {author} {\bibfnamefont {G.}~\bibnamefont {Knebel}}, \bibinfo {author} {\bibfnamefont {Y.}~\bibnamefont {Tokunaga}},\ and\ \bibinfo {author} {\bibfnamefont {Y.}~\bibnamefont {Yanase}},\ }\bibfield  {title} {\bibinfo {title} {Unconventional superconductivity in {UTe$_{2}$}},\ }\href@noop {} {\bibfield  {journal} {\bibinfo  {journal} {Journal of Physics: Condensed Matter}\ }\textbf {\bibinfo {volume} {34}},\ \bibinfo {pages} {243002} (\bibinfo {year} {2022})}\BibitemShut {NoStop}%
\bibitem [{\citenamefont {Noel}\ and\ \citenamefont {Levet}(1989)}]{Noel1989}%
  \BibitemOpen
  \bibfield  {author} {\bibinfo {author} {\bibfnamefont {H.}~\bibnamefont {Noel}}\ and\ \bibinfo {author} {\bibfnamefont {J.}~\bibnamefont {Levet}},\ }\bibfield  {title} {\bibinfo {title} {{Caractérisation d'un tritellurure d'uranium: {$\beta$-UTe$_{3}$} de structure type {NdTe$_{3}$}}},\ }\href {https://doi.org/https://doi.org/10.1016/0022-4596(89)90246-6} {\bibfield  {journal} {\bibinfo  {journal} {Journal of Solid State Chemistry}\ }\textbf {\bibinfo {volume} {79}},\ \bibinfo {pages} {28} (\bibinfo {year} {1989})}\BibitemShut {NoStop}%
\bibitem [{\citenamefont {Ba{\l}anda}(2013)}]{balanda2013ac}%
  \BibitemOpen
  \bibfield  {author} {\bibinfo {author} {\bibfnamefont {M.}~\bibnamefont {Ba{\l}anda}},\ }\bibfield  {title} {\bibinfo {title} {Ac susceptibility studies of phase transitions and magnetic relaxation: Conventional, molecular and low-dimensional magnets},\ }\href@noop {} {\bibfield  {journal} {\bibinfo  {journal} {Acta physica polonica A}\ }\textbf {\bibinfo {volume} {124}},\ \bibinfo {pages} {964} (\bibinfo {year} {2013})}\BibitemShut {NoStop}%
\bibitem [{\citenamefont {Parisi}(1988)}]{Parisi1988}%
  \BibitemOpen
  \bibfield  {author} {\bibinfo {author} {\bibfnamefont {G.}~\bibnamefont {Parisi}},\ }\href {https://cds.cern.ch/record/111935} {\emph {\bibinfo {title} {{Statistical field theory}}}},\ Frontiers in physics\ (\bibinfo  {publisher} {Addison-Wesley},\ \bibinfo {address} {Redwood City, CA},\ \bibinfo {year} {1988})\BibitemShut {NoStop}%
\bibitem [{\citenamefont {Arrott}(1957)}]{Arrott1957}%
  \BibitemOpen
  \bibfield  {author} {\bibinfo {author} {\bibfnamefont {A.}~\bibnamefont {Arrott}},\ }\bibfield  {title} {\bibinfo {title} {Criterion for ferromagnetism from observations of magnetic isotherms},\ }\href {https://doi.org/10.1103/PhysRev.108.1394} {\bibfield  {journal} {\bibinfo  {journal} {Phys. Rev.}\ }\textbf {\bibinfo {volume} {108}},\ \bibinfo {pages} {1394} (\bibinfo {year} {1957})}\BibitemShut {NoStop}%
\bibitem [{\citenamefont {Pramanik}\ and\ \citenamefont {Banerjee}(2009)}]{Pramanik2009}%
  \BibitemOpen
  \bibfield  {author} {\bibinfo {author} {\bibfnamefont {A.~K.}\ \bibnamefont {Pramanik}}\ and\ \bibinfo {author} {\bibfnamefont {A.}~\bibnamefont {Banerjee}},\ }\bibfield  {title} {\bibinfo {title} {Critical behavior at paramagnetic to ferromagnetic phase transition in {${\text{Pr}}_{0.5}{\text{Sr}}_{0.5}{\text{MnO}}_{3}$}: A bulk magnetization study},\ }\href {https://doi.org/10.1103/PhysRevB.79.214426} {\bibfield  {journal} {\bibinfo  {journal} {Phys. Rev. B}\ }\textbf {\bibinfo {volume} {79}},\ \bibinfo {pages} {214426} (\bibinfo {year} {2009})}\BibitemShut {NoStop}%
\bibitem [{\citenamefont {Bramwell}\ and\ \citenamefont {Holdsworth}(1993)}]{Bramwell1993}%
  \BibitemOpen
  \bibfield  {author} {\bibinfo {author} {\bibfnamefont {S.~T.}\ \bibnamefont {Bramwell}}\ and\ \bibinfo {author} {\bibfnamefont {P.~C.~W.}\ \bibnamefont {Holdsworth}},\ }\bibfield  {title} {\bibinfo {title} {{Magnetization and universal sub-critical behaviour in two-dimensional XY magnets}},\ }\href@noop {} {\bibfield  {journal} {\bibinfo  {journal} {Journal of Physics: Condensed Matter}\ }\textbf {\bibinfo {volume} {5}},\ \bibinfo {pages} {L53} (\bibinfo {year} {1993})}\BibitemShut {NoStop}%
\bibitem [{\citenamefont {Bedoya-Pinto}\ \emph {et~al.}(2021)\citenamefont {Bedoya-Pinto}, \citenamefont {Ji}, \citenamefont {Pandeya}, \citenamefont {Gargiani}, \citenamefont {Valvidares}, \citenamefont {Sessi}, \citenamefont {Taylor}, \citenamefont {Radu}, \citenamefont {Chang},\ and\ \citenamefont {Parkin}}]{Bedoya2021}%
  \BibitemOpen
  \bibfield  {author} {\bibinfo {author} {\bibfnamefont {A.}~\bibnamefont {Bedoya-Pinto}}, \bibinfo {author} {\bibfnamefont {J.-R.}\ \bibnamefont {Ji}}, \bibinfo {author} {\bibfnamefont {A.~K.}\ \bibnamefont {Pandeya}}, \bibinfo {author} {\bibfnamefont {P.}~\bibnamefont {Gargiani}}, \bibinfo {author} {\bibfnamefont {M.}~\bibnamefont {Valvidares}}, \bibinfo {author} {\bibfnamefont {P.}~\bibnamefont {Sessi}}, \bibinfo {author} {\bibfnamefont {J.~M.}\ \bibnamefont {Taylor}}, \bibinfo {author} {\bibfnamefont {F.}~\bibnamefont {Radu}}, \bibinfo {author} {\bibfnamefont {K.}~\bibnamefont {Chang}},\ and\ \bibinfo {author} {\bibfnamefont {S.~S.~P.}\ \bibnamefont {Parkin}},\ }\bibfield  {title} {\bibinfo {title} {{Intrinsic 2D-XY ferromagnetism in a van der Waals monolayer}},\ }\href {https://doi.org/10.1126/science.abd5146} {\bibfield  {journal} {\bibinfo  {journal} {Science}\ }\textbf {\bibinfo {volume} {374}},\ \bibinfo {pages} {616} (\bibinfo {year} {2021})}\BibitemShut {NoStop}%
\bibitem [{\citenamefont {Scheie}\ \emph {et~al.}(2022)\citenamefont {Scheie}, \citenamefont {Ziebel}, \citenamefont {Chica}, \citenamefont {Bae}, \citenamefont {Wang}, \citenamefont {Kolesnikov}, \citenamefont {Zhu},\ and\ \citenamefont {Roy}}]{scheie2022}%
  \BibitemOpen
  \bibfield  {author} {\bibinfo {author} {\bibfnamefont {A.}~\bibnamefont {Scheie}}, \bibinfo {author} {\bibfnamefont {M.}~\bibnamefont {Ziebel}}, \bibinfo {author} {\bibfnamefont {D.~G.}\ \bibnamefont {Chica}}, \bibinfo {author} {\bibfnamefont {Y.~J.}\ \bibnamefont {Bae}}, \bibinfo {author} {\bibfnamefont {X.}~\bibnamefont {Wang}}, \bibinfo {author} {\bibfnamefont {A.~I.}\ \bibnamefont {Kolesnikov}}, \bibinfo {author} {\bibfnamefont {X.}~\bibnamefont {Zhu}},\ and\ \bibinfo {author} {\bibfnamefont {X.}~\bibnamefont {Roy}},\ }\bibfield  {title} {\bibinfo {title} {Spin waves and magnetic exchange hamiltonian in {CrSBr}},\ }\href@noop {} {\bibfield  {journal} {\bibinfo  {journal} {Advanced Science}\ }\textbf {\bibinfo {volume} {9}},\ \bibinfo {pages} {2202467} (\bibinfo {year} {2022})}\BibitemShut {NoStop}%
\bibitem [{\citenamefont {Ru}\ and\ \citenamefont {Fisher}(2006{\natexlab{b}})}]{PhysRevB.73.033101}%
  \BibitemOpen
  \bibfield  {author} {\bibinfo {author} {\bibfnamefont {N.}~\bibnamefont {Ru}}\ and\ \bibinfo {author} {\bibfnamefont {I.~R.}\ \bibnamefont {Fisher}},\ }\bibfield  {title} {\bibinfo {title} {Thermodynamic and transport properties of $\mathrm{Y}{\mathrm{te}}_{3}$, $\mathrm{La}{\mathrm{te}}_{3}$, and $\mathrm{Ce}{\mathrm{te}}_{3}$},\ }\href {https://doi.org/10.1103/PhysRevB.73.033101} {\bibfield  {journal} {\bibinfo  {journal} {Phys. Rev. B}\ }\textbf {\bibinfo {volume} {73}},\ \bibinfo {pages} {033101} (\bibinfo {year} {2006}{\natexlab{b}})}\BibitemShut {NoStop}%
\bibitem [{\citenamefont {Yang}\ \emph {et~al.}(2008)\citenamefont {Yang}, \citenamefont {Fisk}, \citenamefont {Lee}, \citenamefont {Thompson},\ and\ \citenamefont {Pines}}]{Yang2008}%
  \BibitemOpen
  \bibfield  {author} {\bibinfo {author} {\bibfnamefont {Y.-f.}\ \bibnamefont {Yang}}, \bibinfo {author} {\bibfnamefont {Z.}~\bibnamefont {Fisk}}, \bibinfo {author} {\bibfnamefont {H.-O.}\ \bibnamefont {Lee}}, \bibinfo {author} {\bibfnamefont {J.~D.}\ \bibnamefont {Thompson}},\ and\ \bibinfo {author} {\bibfnamefont {D.}~\bibnamefont {Pines}},\ }\bibfield  {title} {\bibinfo {title} {Scaling the kondo lattice},\ }\href {https://doi.org/10.1038/nature07157} {\bibfield  {journal} {\bibinfo  {journal} {Nature}\ }\textbf {\bibinfo {volume} {454}},\ \bibinfo {pages} {611} (\bibinfo {year} {2008})}\BibitemShut {NoStop}%
\bibitem [{\citenamefont {Nagaosa}\ \emph {et~al.}(2010)\citenamefont {Nagaosa}, \citenamefont {Sinova}, \citenamefont {Onoda}, \citenamefont {MacDonald},\ and\ \citenamefont {Ong}}]{nagaosa2010}%
  \BibitemOpen
  \bibfield  {author} {\bibinfo {author} {\bibfnamefont {N.}~\bibnamefont {Nagaosa}}, \bibinfo {author} {\bibfnamefont {J.}~\bibnamefont {Sinova}}, \bibinfo {author} {\bibfnamefont {S.}~\bibnamefont {Onoda}}, \bibinfo {author} {\bibfnamefont {A.~H.}\ \bibnamefont {MacDonald}},\ and\ \bibinfo {author} {\bibfnamefont {N.~P.}\ \bibnamefont {Ong}},\ }\bibfield  {title} {\bibinfo {title} {Anomalous {Hall} effect},\ }\href@noop {} {\bibfield  {journal} {\bibinfo  {journal} {Reviews of Modern Physics}\ }\textbf {\bibinfo {volume} {82}},\ \bibinfo {pages} {1539} (\bibinfo {year} {2010})}\BibitemShut {NoStop}%
\bibitem [{\citenamefont {Schwarz}\ and\ \citenamefont {Blaha}(2003)}]{SCHWARZ2003}%
  \BibitemOpen
  \bibfield  {author} {\bibinfo {author} {\bibfnamefont {K.}~\bibnamefont {Schwarz}}\ and\ \bibinfo {author} {\bibfnamefont {P.}~\bibnamefont {Blaha}},\ }\bibfield  {title} {\bibinfo {title} {Solid state calculations using wien2k},\ }\href {https://doi.org/https://doi.org/10.1016/S0927-0256(03)00112-5} {\bibfield  {journal} {\bibinfo  {journal} {Computational Materials Science}\ }\textbf {\bibinfo {volume} {28}},\ \bibinfo {pages} {259} (\bibinfo {year} {2003})},\ \bibinfo {note} {proceedings of the Symposium on Software Development for Process and Materials Design}\BibitemShut {NoStop}%
\bibitem [{\citenamefont {Perdew}\ \emph {et~al.}(1996)\citenamefont {Perdew}, \citenamefont {Burke},\ and\ \citenamefont {Ernzerhof}}]{Perdew1996}%
  \BibitemOpen
  \bibfield  {author} {\bibinfo {author} {\bibfnamefont {J.~P.}\ \bibnamefont {Perdew}}, \bibinfo {author} {\bibfnamefont {K.}~\bibnamefont {Burke}},\ and\ \bibinfo {author} {\bibfnamefont {M.}~\bibnamefont {Ernzerhof}},\ }\bibfield  {title} {\bibinfo {title} {Generalized gradient approximation made simple},\ }\href {https://doi.org/10.1103/PhysRevLett.77.3865} {\bibfield  {journal} {\bibinfo  {journal} {Phys. Rev. Lett.}\ }\textbf {\bibinfo {volume} {77}},\ \bibinfo {pages} {3865} (\bibinfo {year} {1996})}\BibitemShut {NoStop}%
\bibitem [{\citenamefont {Kim}\ \emph {et~al.}(2019)\citenamefont {Kim}, \citenamefont {Lim}, \citenamefont {Lee}, \citenamefont {Lee}, \citenamefont {Kim}, \citenamefont {Park}, \citenamefont {Jeon}, \citenamefont {Park}, \citenamefont {Park},\ and\ \citenamefont {Cheong}}]{kim2019suppression}%
  \BibitemOpen
  \bibfield  {author} {\bibinfo {author} {\bibfnamefont {K.}~\bibnamefont {Kim}}, \bibinfo {author} {\bibfnamefont {S.~Y.}\ \bibnamefont {Lim}}, \bibinfo {author} {\bibfnamefont {J.-U.}\ \bibnamefont {Lee}}, \bibinfo {author} {\bibfnamefont {S.}~\bibnamefont {Lee}}, \bibinfo {author} {\bibfnamefont {T.~Y.}\ \bibnamefont {Kim}}, \bibinfo {author} {\bibfnamefont {K.}~\bibnamefont {Park}}, \bibinfo {author} {\bibfnamefont {G.~S.}\ \bibnamefont {Jeon}}, \bibinfo {author} {\bibfnamefont {C.-H.}\ \bibnamefont {Park}}, \bibinfo {author} {\bibfnamefont {J.-G.}\ \bibnamefont {Park}},\ and\ \bibinfo {author} {\bibfnamefont {H.}~\bibnamefont {Cheong}},\ }\bibfield  {title} {\bibinfo {title} {Suppression of magnetic ordering in xxz-type antiferromagnetic monolayer nips3},\ }\href@noop {} {\bibfield  {journal} {\bibinfo  {journal} {Nature communications}\ }\textbf {\bibinfo {volume} {10}},\ \bibinfo {pages} {345} (\bibinfo {year} {2019})}\BibitemShut {NoStop}%
\bibitem [{\citenamefont {Liu}\ \emph {et~al.}(2020)\citenamefont {Liu}, \citenamefont {Xu}, \citenamefont {Huang}, \citenamefont {Li}, \citenamefont {Feng}, \citenamefont {Huang}, \citenamefont {Zhu}, \citenamefont {Wang}, \citenamefont {Zhang}, \citenamefont {Hou} \emph {et~al.}}]{liu2020exploring}%
  \BibitemOpen
  \bibfield  {author} {\bibinfo {author} {\bibfnamefont {P.}~\bibnamefont {Liu}}, \bibinfo {author} {\bibfnamefont {Z.}~\bibnamefont {Xu}}, \bibinfo {author} {\bibfnamefont {H.}~\bibnamefont {Huang}}, \bibinfo {author} {\bibfnamefont {J.}~\bibnamefont {Li}}, \bibinfo {author} {\bibfnamefont {C.}~\bibnamefont {Feng}}, \bibinfo {author} {\bibfnamefont {M.}~\bibnamefont {Huang}}, \bibinfo {author} {\bibfnamefont {M.}~\bibnamefont {Zhu}}, \bibinfo {author} {\bibfnamefont {Z.}~\bibnamefont {Wang}}, \bibinfo {author} {\bibfnamefont {Z.}~\bibnamefont {Zhang}}, \bibinfo {author} {\bibfnamefont {D.}~\bibnamefont {Hou}}, \emph {et~al.},\ }\bibfield  {title} {\bibinfo {title} {Exploring the magnetic ordering in atomically thin antiferromagnetic mnpse3 by raman spectroscopy},\ }\href@noop {} {\bibfield  {journal} {\bibinfo  {journal} {Journal of Alloys and Compounds}\ }\textbf {\bibinfo {volume} {828}},\ \bibinfo {pages} {154432} (\bibinfo {year} {2020})}\BibitemShut {NoStop}%
\bibitem [{\citenamefont {Binder}\ and\ \citenamefont {Hohenberg}(1974)}]{binder1974surface}%
  \BibitemOpen
  \bibfield  {author} {\bibinfo {author} {\bibfnamefont {K.}~\bibnamefont {Binder}}\ and\ \bibinfo {author} {\bibfnamefont {P.}~\bibnamefont {Hohenberg}},\ }\bibfield  {title} {\bibinfo {title} {Surface effects on magnetic phase transitions},\ }\href@noop {} {\bibfield  {journal} {\bibinfo  {journal} {Physical Review B}\ }\textbf {\bibinfo {volume} {9}},\ \bibinfo {pages} {2194} (\bibinfo {year} {1974})}\BibitemShut {NoStop}%
\bibitem [{\citenamefont {Binder}(1983)}]{binder1983critical}%
  \BibitemOpen
  \bibfield  {author} {\bibinfo {author} {\bibfnamefont {K.}~\bibnamefont {Binder}},\ }\href@noop {} {\bibinfo {title} {Critical behaviour at surfaces, in “phase transitions and critical phenomena”, vol. 8, c. domb and jl lebowitz, eds}} (\bibinfo {year} {1983})\BibitemShut {NoStop}%
\bibitem [{\citenamefont {Guo}\ \emph {et~al.}(2024)\citenamefont {Guo}, \citenamefont {Liu}, \citenamefont {Schwartz}, \citenamefont {Sung}, \citenamefont {Zhang}, \citenamefont {Shimizu}, \citenamefont {Kondusamy}, \citenamefont {Li}, \citenamefont {Sun}, \citenamefont {Deng} \emph {et~al.}}]{guo2024extraordinary}%
  \BibitemOpen
  \bibfield  {author} {\bibinfo {author} {\bibfnamefont {X.}~\bibnamefont {Guo}}, \bibinfo {author} {\bibfnamefont {W.}~\bibnamefont {Liu}}, \bibinfo {author} {\bibfnamefont {J.}~\bibnamefont {Schwartz}}, \bibinfo {author} {\bibfnamefont {S.~H.}\ \bibnamefont {Sung}}, \bibinfo {author} {\bibfnamefont {D.}~\bibnamefont {Zhang}}, \bibinfo {author} {\bibfnamefont {M.}~\bibnamefont {Shimizu}}, \bibinfo {author} {\bibfnamefont {A.~L.}\ \bibnamefont {Kondusamy}}, \bibinfo {author} {\bibfnamefont {L.}~\bibnamefont {Li}}, \bibinfo {author} {\bibfnamefont {K.}~\bibnamefont {Sun}}, \bibinfo {author} {\bibfnamefont {H.}~\bibnamefont {Deng}}, \emph {et~al.},\ }\bibfield  {title} {\bibinfo {title} {Extraordinary phase transition revealed in a van der waals antiferromagnet},\ }\href@noop {} {\bibfield  {journal} {\bibinfo  {journal} {Nature communications}\ }\textbf {\bibinfo {volume} {15}},\ \bibinfo {pages} {6472} (\bibinfo {year} {2024})}\BibitemShut {NoStop}%
\bibitem [{\citenamefont {Schmiedeshoff}\ \emph {et~al.}(2006)\citenamefont {Schmiedeshoff}, \citenamefont {Lounsbury}, \citenamefont {Luna}, \citenamefont {Tracy}, \citenamefont {Schramm}, \citenamefont {Tozer}, \citenamefont {Correa}, \citenamefont {Hannahs}, \citenamefont {Murphy}, \citenamefont {Palm}, \citenamefont {Lacerda}, \citenamefont {Bud'Ko}, \citenamefont {Canfield}, \citenamefont {Smith}, \citenamefont {Lashley},\ and\ \citenamefont {Cooley}}]{Schmiedeshoff2006}%
  \BibitemOpen
  \bibfield  {author} {\bibinfo {author} {\bibfnamefont {G.~M.}\ \bibnamefont {Schmiedeshoff}}, \bibinfo {author} {\bibfnamefont {A.~W.}\ \bibnamefont {Lounsbury}}, \bibinfo {author} {\bibfnamefont {D.~J.}\ \bibnamefont {Luna}}, \bibinfo {author} {\bibfnamefont {S.~J.}\ \bibnamefont {Tracy}}, \bibinfo {author} {\bibfnamefont {A.~J.}\ \bibnamefont {Schramm}}, \bibinfo {author} {\bibfnamefont {S.~W.}\ \bibnamefont {Tozer}}, \bibinfo {author} {\bibfnamefont {V.~F.}\ \bibnamefont {Correa}}, \bibinfo {author} {\bibfnamefont {S.~T.}\ \bibnamefont {Hannahs}}, \bibinfo {author} {\bibfnamefont {T.~P.}\ \bibnamefont {Murphy}}, \bibinfo {author} {\bibfnamefont {E.~C.}\ \bibnamefont {Palm}}, \bibinfo {author} {\bibfnamefont {A.~H.}\ \bibnamefont {Lacerda}}, \bibinfo {author} {\bibfnamefont {S.~L.}\ \bibnamefont {Bud'Ko}}, \bibinfo {author} {\bibfnamefont {P.~C.}\ \bibnamefont {Canfield}}, \bibinfo {author} {\bibfnamefont {J.~L.}\ \bibnamefont {Smith}}, \bibinfo {author} {\bibfnamefont {J.~C.}\ \bibnamefont {Lashley}},\
  and\ \bibinfo {author} {\bibfnamefont {J.~C.}\ \bibnamefont {Cooley}},\ }\bibfield  {title} {\bibinfo {title} {{Versatile and compact capacitive dilatometer}},\ }\href {https://doi.org/10.1063/1.2403088} {\bibfield  {journal} {\bibinfo  {journal} {Review of Scientific Instruments}\ }\textbf {\bibinfo {volume} {77}},\ \bibinfo {pages} {123907} (\bibinfo {year} {2006})}\BibitemShut {NoStop}%
\bibitem [{\citenamefont {Xia}\ \emph {et~al.}(2006)\citenamefont {Xia}, \citenamefont {Beyersdorf}, \citenamefont {Fejer},\ and\ \citenamefont {Kapitulnik}}]{Xia2006}%
  \BibitemOpen
  \bibfield  {author} {\bibinfo {author} {\bibfnamefont {J.}~\bibnamefont {Xia}}, \bibinfo {author} {\bibfnamefont {P.~T.}\ \bibnamefont {Beyersdorf}}, \bibinfo {author} {\bibfnamefont {M.~M.}\ \bibnamefont {Fejer}},\ and\ \bibinfo {author} {\bibfnamefont {A.}~\bibnamefont {Kapitulnik}},\ }\bibfield  {title} {\bibinfo {title} {Modified sagnac interferometer for high-sensitivity magneto-optic measurements at cryogenic temperatures},\ }\href {https://doi.org/10.1063/1.2336620} {\bibfield  {journal} {\bibinfo  {journal} {Applied Physics Letters}\ }\textbf {\bibinfo {volume} {89}},\ \bibinfo {pages} {062508} (\bibinfo {year} {2006})}\BibitemShut {NoStop}%
\bibitem [{\citenamefont {Fried}\ \emph {et~al.}(2014)\citenamefont {Fried}, \citenamefont {Fejer},\ and\ \citenamefont {Kapitulnik}}]{Fried2014}%
  \BibitemOpen
  \bibfield  {author} {\bibinfo {author} {\bibfnamefont {A.}~\bibnamefont {Fried}}, \bibinfo {author} {\bibfnamefont {M.}~\bibnamefont {Fejer}},\ and\ \bibinfo {author} {\bibfnamefont {A.}~\bibnamefont {Kapitulnik}},\ }\bibfield  {title} {\bibinfo {title} {A scanning, all-fiber sagnac interferometer for high resolution magneto-optic measurements at 820 nm},\ }\href {https://doi.org/10.1063/1.4897184} {\bibfield  {journal} {\bibinfo  {journal} {Review of Scientific Instruments}\ }\textbf {\bibinfo {volume} {85}},\ \bibinfo {pages} {103707} (\bibinfo {year} {2014})}\BibitemShut {NoStop}%
\bibitem [{\citenamefont {Chapon}\ \emph {et~al.}(2011)\citenamefont {Chapon}, \citenamefont {Manuel}, \citenamefont {Radaelli}, \citenamefont {Benson}, \citenamefont {Perrott}, \citenamefont {Ansell}, \citenamefont {Rhodes}, \citenamefont {Raspino}, \citenamefont {Duxbury}, \citenamefont {Spill},\ and\ \citenamefont {Norris}}]{2011_Chapon_NeutronNews}%
  \BibitemOpen
  \bibfield  {author} {\bibinfo {author} {\bibfnamefont {L.~C.}\ \bibnamefont {Chapon}}, \bibinfo {author} {\bibfnamefont {P.}~\bibnamefont {Manuel}}, \bibinfo {author} {\bibfnamefont {P.~G.}\ \bibnamefont {Radaelli}}, \bibinfo {author} {\bibfnamefont {C.}~\bibnamefont {Benson}}, \bibinfo {author} {\bibfnamefont {L.}~\bibnamefont {Perrott}}, \bibinfo {author} {\bibfnamefont {S.}~\bibnamefont {Ansell}}, \bibinfo {author} {\bibfnamefont {N.~J.}\ \bibnamefont {Rhodes}}, \bibinfo {author} {\bibfnamefont {D.}~\bibnamefont {Raspino}}, \bibinfo {author} {\bibfnamefont {D.}~\bibnamefont {Duxbury}}, \bibinfo {author} {\bibfnamefont {E.}~\bibnamefont {Spill}},\ and\ \bibinfo {author} {\bibfnamefont {J.}~\bibnamefont {Norris}},\ }\bibfield  {title} {\bibinfo {title} {Wish: {{The New Powder}} and {{Single Crystal Magnetic Diffractometer}} on the {{Second Target Station}}},\ }\href {https://doi.org/10.1080/10448632.2011.569650} {\bibfield  {journal} {\bibinfo  {journal} {Neutron News}\ }\textbf {\bibinfo {volume} {22}},\
  \bibinfo {pages} {22} (\bibinfo {year} {2011})}\BibitemShut {NoStop}%
\bibitem [{\citenamefont {Arnold}\ \emph {et~al.}(2014)\citenamefont {Arnold}, \citenamefont {Bilheux}, \citenamefont {Borreguero}, \citenamefont {Buts}, \citenamefont {Campbell}, \citenamefont {Chapon}, \citenamefont {Doucet}, \citenamefont {Draper}, \citenamefont {Ferraz~Leal}, \citenamefont {Gigg}, \citenamefont {Lynch}, \citenamefont {Markvardsen}, \citenamefont {Mikkelson}, \citenamefont {Mikkelson}, \citenamefont {Miller}, \citenamefont {Palmen}, \citenamefont {Parker}, \citenamefont {Passos}, \citenamefont {Perring}, \citenamefont {Peterson}, \citenamefont {Ren}, \citenamefont {Reuter}, \citenamefont {Savici}, \citenamefont {Taylor}, \citenamefont {Taylor}, \citenamefont {Tolchenov}, \citenamefont {Zhou},\ and\ \citenamefont {Zikovsky}}]{2014_Arnold_NuclearInstrumentsandMethodsinPhysicsResearchSectionAAcceleratorsSpectrometersDetectorsandAssociatedEquipment}%
  \BibitemOpen
  \bibfield  {author} {\bibinfo {author} {\bibfnamefont {O.}~\bibnamefont {Arnold}}, \bibinfo {author} {\bibfnamefont {J.}~\bibnamefont {Bilheux}}, \bibinfo {author} {\bibfnamefont {J.}~\bibnamefont {Borreguero}}, \bibinfo {author} {\bibfnamefont {A.}~\bibnamefont {Buts}}, \bibinfo {author} {\bibfnamefont {S.}~\bibnamefont {Campbell}}, \bibinfo {author} {\bibfnamefont {L.}~\bibnamefont {Chapon}}, \bibinfo {author} {\bibfnamefont {M.}~\bibnamefont {Doucet}}, \bibinfo {author} {\bibfnamefont {N.}~\bibnamefont {Draper}}, \bibinfo {author} {\bibfnamefont {R.}~\bibnamefont {Ferraz~Leal}}, \bibinfo {author} {\bibfnamefont {M.}~\bibnamefont {Gigg}}, \bibinfo {author} {\bibfnamefont {V.}~\bibnamefont {Lynch}}, \bibinfo {author} {\bibfnamefont {A.}~\bibnamefont {Markvardsen}}, \bibinfo {author} {\bibfnamefont {D.}~\bibnamefont {Mikkelson}}, \bibinfo {author} {\bibfnamefont {R.}~\bibnamefont {Mikkelson}}, \bibinfo {author} {\bibfnamefont {R.}~\bibnamefont {Miller}}, \bibinfo {author} {\bibfnamefont {K.}~\bibnamefont
  {Palmen}}, \bibinfo {author} {\bibfnamefont {P.}~\bibnamefont {Parker}}, \bibinfo {author} {\bibfnamefont {G.}~\bibnamefont {Passos}}, \bibinfo {author} {\bibfnamefont {T.}~\bibnamefont {Perring}}, \bibinfo {author} {\bibfnamefont {P.}~\bibnamefont {Peterson}}, \bibinfo {author} {\bibfnamefont {S.}~\bibnamefont {Ren}}, \bibinfo {author} {\bibfnamefont {M.}~\bibnamefont {Reuter}}, \bibinfo {author} {\bibfnamefont {A.}~\bibnamefont {Savici}}, \bibinfo {author} {\bibfnamefont {J.}~\bibnamefont {Taylor}}, \bibinfo {author} {\bibfnamefont {R.}~\bibnamefont {Taylor}}, \bibinfo {author} {\bibfnamefont {R.}~\bibnamefont {Tolchenov}}, \bibinfo {author} {\bibfnamefont {W.}~\bibnamefont {Zhou}},\ and\ \bibinfo {author} {\bibfnamefont {J.}~\bibnamefont {Zikovsky}},\ }\bibfield  {title} {\bibinfo {title} {Mantid---{{Data}} analysis and visualization package for neutron scattering and {$\mu$} {{SR}} experiments},\ }\href {https://doi.org/10.1016/j.nima.2014.07.029} {\bibfield  {journal} {\bibinfo  {journal} {Nuclear
  Instruments and Methods in Physics Research Section A: Accelerators, Spectrometers, Detectors and Associated Equipment}\ }\textbf {\bibinfo {volume} {764}},\ \bibinfo {pages} {156} (\bibinfo {year} {2014})}\BibitemShut {NoStop}%
\end{thebibliography}%


\begin{thebibliography}{2}%
\makeatletter
\providecommand \@ifxundefined [1]{%
 \@ifx{#1\undefined}
}%
\providecommand \@ifnum [1]{%
 \ifnum #1\expandafter \@firstoftwo
 \else \expandafter \@secondoftwo
 \fi
}%
\providecommand \@ifx [1]{%
 \ifx #1\expandafter \@firstoftwo
 \else \expandafter \@secondoftwo
 \fi
}%
\providecommand \natexlab [1]{#1}%
\providecommand \enquote  [1]{``#1''}%
\providecommand \bibnamefont  [1]{#1}%
\providecommand \bibfnamefont [1]{#1}%
\providecommand \citenamefont [1]{#1}%
\providecommand \href@noop [0]{\@secondoftwo}%
\providecommand \href [0]{\begingroup \@sanitize@url \@href}%
\providecommand \@href[1]{\@@startlink{#1}\@@href}%
\providecommand \@@href[1]{\endgroup#1\@@endlink}%
\providecommand \@sanitize@url [0]{\catcode `\\12\catcode `\$12\catcode `\&12\catcode `\#12\catcode `\^12\catcode `\_12\catcode `\%12\relax}%
\providecommand \@@startlink[1]{}%
\providecommand \@@endlink[0]{}%
\providecommand \url  [0]{\begingroup\@sanitize@url \@url }%
\providecommand \@url [1]{\endgroup\@href {#1}{\urlprefix }}%
\providecommand \urlprefix  [0]{URL }%
\providecommand \Eprint [0]{\href }%
\providecommand \doibase [0]{https://doi.org/}%
\providecommand \selectlanguage [0]{\@gobble}%
\providecommand \bibinfo  [0]{\@secondoftwo}%
\providecommand \bibfield  [0]{\@secondoftwo}%
\providecommand \translation [1]{[#1]}%
\providecommand \BibitemOpen [0]{}%
\providecommand \bibitemStop [0]{}%
\providecommand \bibitemNoStop [0]{.\EOS\space}%
\providecommand \EOS [0]{\spacefactor3000\relax}%
\providecommand \BibitemShut  [1]{\csname bibitem#1\endcsname}%
\let\auto@bib@innerbib\@empty
%</preamble>
\bibitem [{\citenamefont {Pet{\v r}{\'i}{\v c}ek}\ \emph {et~al.}(2023)\citenamefont {Pet{\v r}{\'i}{\v c}ek}, \citenamefont {Palatinus}, \citenamefont {Pl{\'a}{\v s}il},\ and\ \citenamefont {Du{\v s}ek}}]{2023_Petricek_ZFurKrist-CrystMater}%
  \BibitemOpen
  \bibfield  {author} {\bibinfo {author} {\bibfnamefont {V.}~\bibnamefont {Pet{\v r}{\'i}{\v c}ek}}, \bibinfo {author} {\bibfnamefont {L.}~\bibnamefont {Palatinus}}, \bibinfo {author} {\bibfnamefont {J.}~\bibnamefont {Pl{\'a}{\v s}il}},\ and\ \bibinfo {author} {\bibfnamefont {M.}~\bibnamefont {Du{\v s}ek}},\ }\bibfield  {title} {\bibinfo {title} {Jana2020 -- a new version of the crystallographic computing system {{Jana}}},\ }\href {https://doi.org/10.1515/zkri-2023-0005} {\bibfield  {journal} {\bibinfo  {journal} {Zeitschrift f{\"u}r Kristallographie - Crystalline Materials}\ }\textbf {\bibinfo {volume} {238}},\ \bibinfo {pages} {271} (\bibinfo {year} {2023})}\BibitemShut {NoStop}%
\bibitem [{\citenamefont {Freeman}\ \emph {et~al.}(1976)\citenamefont {Freeman}, \citenamefont {Desclaux}, \citenamefont {Lander},\ and\ \citenamefont {Faber}}]{1976_Freeman_PhysRevB}%
  \BibitemOpen
  \bibfield  {author} {\bibinfo {author} {\bibfnamefont {A.~J.}\ \bibnamefont {Freeman}}, \bibinfo {author} {\bibfnamefont {J.~P.}\ \bibnamefont {Desclaux}}, \bibinfo {author} {\bibfnamefont {G.~H.}\ \bibnamefont {Lander}},\ and\ \bibinfo {author} {\bibfnamefont {J.}~\bibnamefont {Faber}},\ }\bibfield  {title} {\bibinfo {title} {Neutron magnetic form factors of uranium ions},\ }\href {https://doi.org/10.1103/PhysRevB.13.1168} {\bibfield  {journal} {\bibinfo  {journal} {Phys. Rev. B}\ }\textbf {\bibinfo {volume} {13}},\ \bibinfo {pages} {1168} (\bibinfo {year} {1976})}\BibitemShut {NoStop}%
\end{thebibliography}%

\end{document}

\title{Supplemental Information for ``Enhanced two-dimensional ferromagnetism in van-der Waals $\beta$-UTe$_3$ monolayers''}

\author{S. M. Thomas}
%\email{smthomas@lanl.gov}
\affiliation{Los Alamos National Laboratory, Los Alamos, NM 87545}
\author{A.E. Llacsahuanga}
\affiliation{Department of Physics and Astronomy, Purdue University, West
Lafayette, Indiana, 47907, USA}
\author{W. Simeth}
\affiliation{Los Alamos National Laboratory, Los Alamos, NM 87545}
\author{C. S. Kengle}
\affiliation{Los Alamos National Laboratory, Los Alamos, NM 87545}
\author{F. Orlandi}
\affiliation{ISIS Facility, Rutherford Appleton Laboratory, Chilton, Didcot, OX11 0QX, United Kingdom}
\author{D. Khalyavin}
\affiliation{ISIS Facility, Rutherford Appleton Laboratory, Chilton, Didcot, OX11 0QX, United Kingdom}
\author{P. Manuel}
\affiliation{ISIS Facility, Rutherford Appleton Laboratory, Chilton, Didcot, OX11 0QX, United Kingdom}
\author{F. Ronning}
\affiliation{Los Alamos National Laboratory, Los Alamos, NM 87545}
\author{E. D. Bauer}
\affiliation{Los Alamos National Laboratory, Los Alamos, NM 87545}
\author{J. D. Thompson}
\affiliation{Los Alamos National Laboratory, Los Alamos, NM 87545}
\author{Jian-Xin Zhu}
\affiliation{Los Alamos National Laboratory, Los Alamos, NM 87545}
\author{A. O. Scheie}
\affiliation{Los Alamos National Laboratory, Los Alamos, NM 87545}
\author{Yong P. Chen}
\affiliation{Department of Physics and Astronomy, Purdue University, West
Lafayette, Indiana, 47907, USA}
\affiliation{School of Electrical and Computer Engineering and Purdue Quantum
Science and Engineering Institute, Purdue University, West Lafayette,
Indiana, 47907, USA}
\author{P. F. S. Rosa}
%\email{pfsrosa@lanl.gov}
\affiliation{Los Alamos National Laboratory, Los Alamos, NM 87545}

\date{\today}

\maketitle

\newpage

\section{AFM analysis of exfoliated samples}

\begin{figure*}[!ht]
	\includegraphics[width=1.0\textwidth]{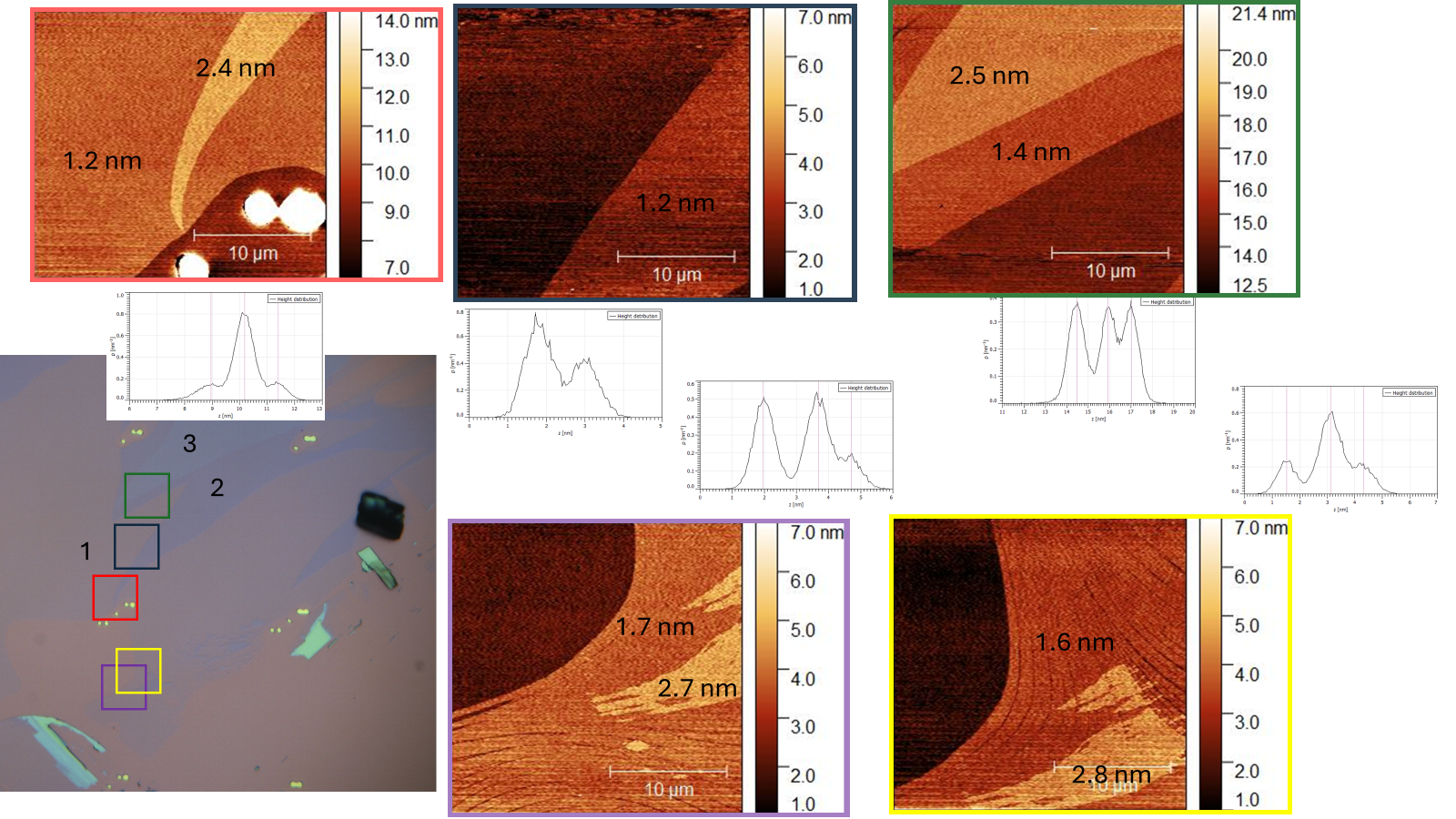}
	\caption{
        Bottom-right shows an optical image of exfoliated UTe$_3$ along with colored squares showing the field-of-view for matching AFM measurements.
        The colored squares correspond with the border color of the height maps shown in the remainder of the panel.
    }
	\label{fig:afm}
\end{figure*}

Layer thickness for the polar Kerr effect (PKE) measurements was determined by optical contrast.
In order to determine the absolute scale, we compare optical images to atomic force microscopy (AFM) measurements in the same region, as shown in Figure~\ref{fig:afm}.
The step height between regions is 1.1--1.2~nm, which corresponds to half of a unit cell along the $b$ axis (2.4743~nm).
This is consistent with the $Cmcm$ crystal structure shown in Fig.~1a in the main text, which displays two identical slabs per unit cell separated by a van der Waals gap along the $b$ axis.

\newpage

\section{Hysteresis in surface ferromagnetism above bulk transition temperature}
To confirm the ferromagnetic nature of the enhanced surface magnetism, Kerr rotation was measured as a function of applied field at a temperature above the bulk transition temperature.
Although the magnitude of the Kerr rotation shows thickness dependence, likely due to an optical interaction with the substrate, there is no thickness-dependence of the coercive field.
This is somewhat surprising as it suggests there is minimal interaction between the top and bottom surfaces above the bulk transition.
This is strikingly different from the behavior below the bulk transition temperature as presented in the main text at 2~K.
At low temperatures, $\beta$-UTe$_3$ becomes a softer ferromagnet as the thickness is increased above one unit cell.

\begin{figure*}[!ht]
	\includegraphics[width=0.5\textwidth]{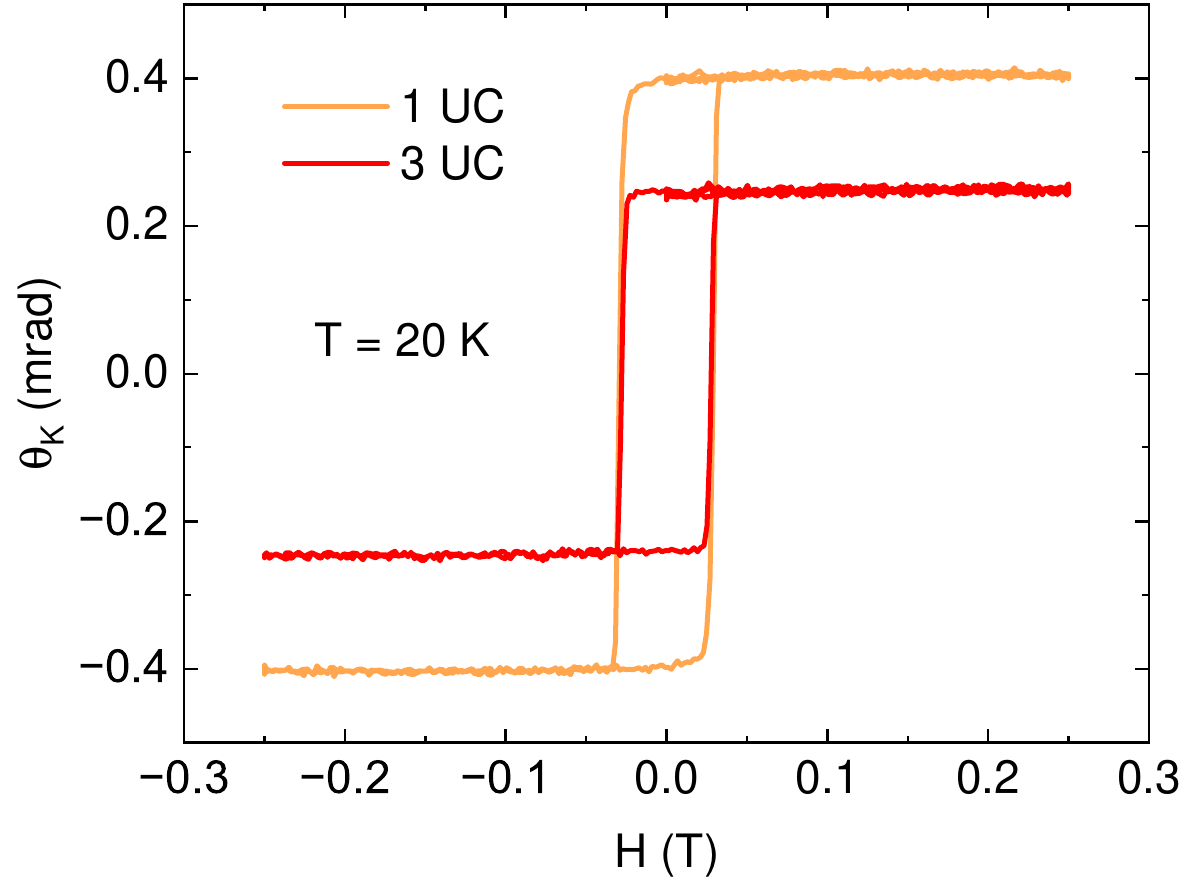}
	\caption{
        Kerr rotation versus field at 20~K for two different thicknesses.
        A linear background (due to the effect of the magnetic field on the optics) was subtracted by fitting the data in the fully-polarized region (H $>$ 0.1~T).
        The coercivity is the same for both thickness ($\pm{}$28~mT).
    }
	\label{fig:20K}
\end{figure*}

\newpage

\section{Evidence for enhanced surface magnetism in bulk samples}
\begin{figure*}[!ht]
	\includegraphics[width=1.0\textwidth]{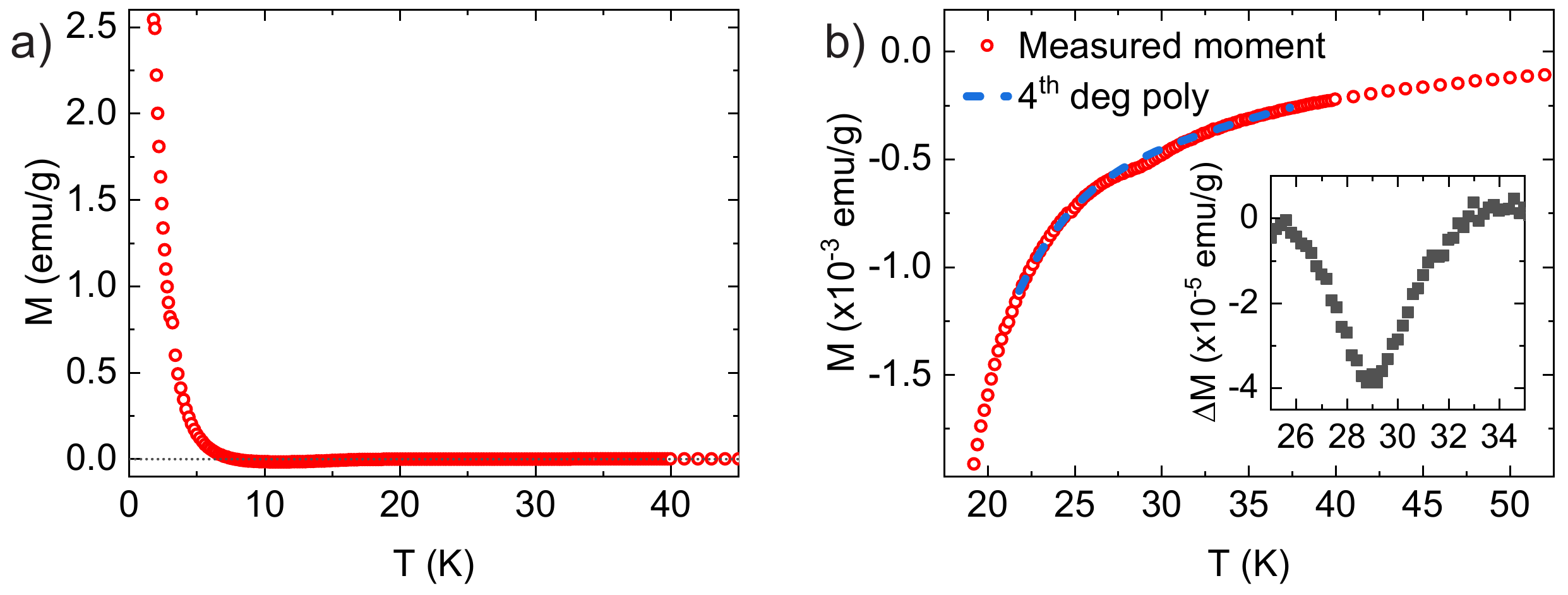}
	\caption{
        Magnetic moment versus temperature while warming in zero field after cooling in a 0.2~T field applied along the $b$~axis.
        a) A zoomed-out view showing a moment of about 2.5~emu/g at 1.8~K.
        b) A zoomed-in view showing the small anomaly measured near 29~K. The inset shows the difference in the moment between a background polynomial fit and the measured data.
    }
	\label{fig:mag}
\end{figure*}

To determine whether the enhanced surface magnetism can be detected in bulk samples, measurements of magnetic susceptibility versus temperature were performed while warming from 1.8~K in zero applied field after cooling in a field of 0.2~T applied along the $b$ axis.
As shown in Figure~\ref{fig:mag}, the measured moment at 1.8~K is 2.54~emu/g, or $0.28$~$\mu_{B}$/U.
In comparison, a small feature above the background is observed at 29~K, which is about 5~K lower compared with the T$^{ML}_C$ measured on the flakes with 0.5 and 1.0 unit cell thickness.
This difference indicates that there may be a slight enhancement of the surface magnetism in the exfoliated flakes due to strain induced by the substrate.
The size of the deviation from background is approximately $4\times10^{-5}$~emu/g, or $4.4\times10^{-6}$~$\mu_{B}$/U. As a result, the moment at 1.8~K is $6\times10^4$ times larger than the high-temperature anomaly.
This ratio is the same order of magnitude as the number of unit cells along the $b$ axis in a 100~$\mu$m-thick sample.
This suggests that the feature observed at 29~K is coming from only the top and bottom layers.

\newpage
\section{Additional specific heat data}

The magnetic contribution to specific heat was determined by subtracting off the value of non-magnetic LaTe$_3$ from the value on $\beta$-UTe$_3$.
Fitting C/T versus T$^2$ yields a Sommerfeld coefficient ($\gamma$) of 131~mJ~mol$^{-1}$~K$^{-2}$, which indicates the presence of electronic correlations.
In comparison, the value for $\gamma$ obtained on LaTe$_3$ is less than 1~mJ~mol$^{-1}$~K$^{-2}$.

\begin{figure*}[!ht]
	\includegraphics[width=1.0\textwidth]{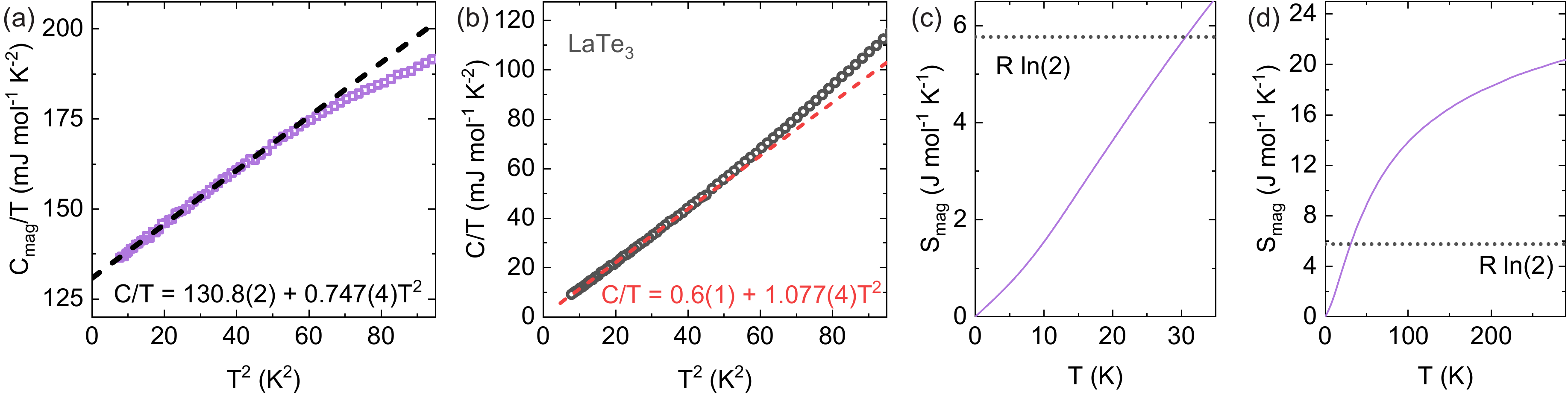}
	\caption{
        (a)~Magnetic specific heat (C$_{\mathrm{\beta{}UTe_3}}-$C$_{\mathrm{LaTe_3}}$ )  divided by temperature versus temperature squared.
        (b)~Specific heat of LaTe$_3$ divided by temperature versus temperature squared.
        (c)~Integrated magnetic entropy versus temperature. The dashed line indicates the value R$\ln{2}$.
        (d)~Same as (c), but over a wider temperature range.
    }
	\label{fig:additional-cp}
\end{figure*}

\newpage
\section{Additional electrical transport data}
\begin{figure*}[!ht]
	\includegraphics[width=1.0\textwidth]{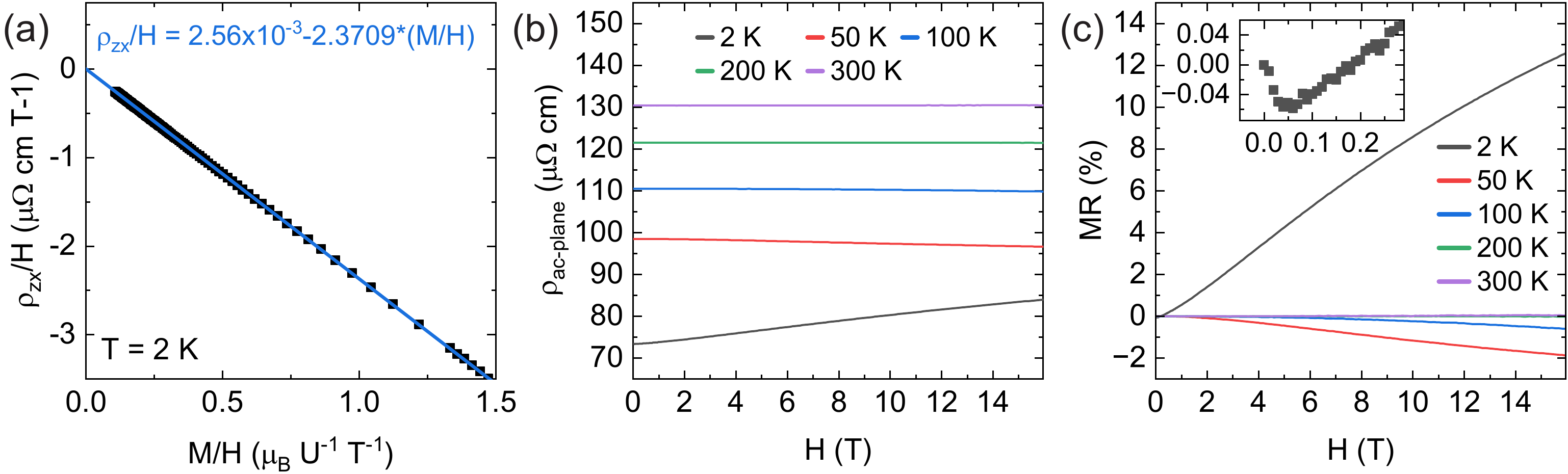}
	\caption{
        (a)~$\rho{}_{zx}/H$ versus $M/H$ at 2~K.
        %(b)~Extracted values for ordinary ($R_{0}$) and anomalous ($R_{s}$) Hall coefficients.
        (b)~In-plane magnetoresistance versus field.
        (c)~Magnetoresistance percentage $[\rho(H)-\rho(0)]/\rho(0)$ versus field. The inset shows a zoomed-in view for the 2~K data at low field.
        The magnetic field is applied parallel to the $b$ axis for all plots.
    }
	\label{fig:xport}
\end{figure*}

Generally, $\rho_{zx}$ can be written as $R_{0}H+R_{s}M$, where the first term is the ordinary Hall component due to the Lorentz force, and $R_{0}$ depends on the carrier density.
Fig.~\ref{fig:xport}(a) shows $\rho{}_{zx}/H$ versus $M/H$ at 2~K, which allows for the determination of the ordinary and anomalous contributions.
Assuming a single electron band at 2~K would give a carrier concentration of $2.4\times{}10^{23}$~$h^{+}$/cm$^3$.
At intermediate temperatures, $\rho{}_{zx}/H$ versus $M/H$ acquires curvature.
This hinders the determination of the ordinary Hall contribution, especially considering the large ratio between $R_s$ and $R_0$.
At even higher temperatures, $M/H$ collapses to a single value because there is a strictly linear relationship between $M$ and $H$.

%Figure~\ref{fig:xport}(b) shows the extracted ordinary ($R_{0}$) and anomalous ($R_{s}$) Hall coefficients.
%Although the ordinary and anomalous contributions were only extracted for three temperatures, there is a clear change in the sign of the ordinary Hall contribution.
%This sign change makes it impossible to infer a carrier density from the Hall data alone.
%The sign change indicated a clear competition between electron and hole contributions to the Hall resistivity, however, so this value is not reliable.
%It is therefore necessary to rely on the density of states as determined by DFT to get an estimate for the carrier concentration.

Figures~\ref{fig:xport}(b),(c) show the in-plane magnetoresistance in absolute units and as a percentage, respectively.
The magnetoresistance is small and negative above T$_{\mathrm{C}}$.
At 2~K, it remains negative up to the field where the magnetic domains are aligned, then becomes positive in the polarized region.

\newpage
\section{Thermal expansion}

\begin{figure*}[!ht]
	\includegraphics[width=0.45\textwidth]{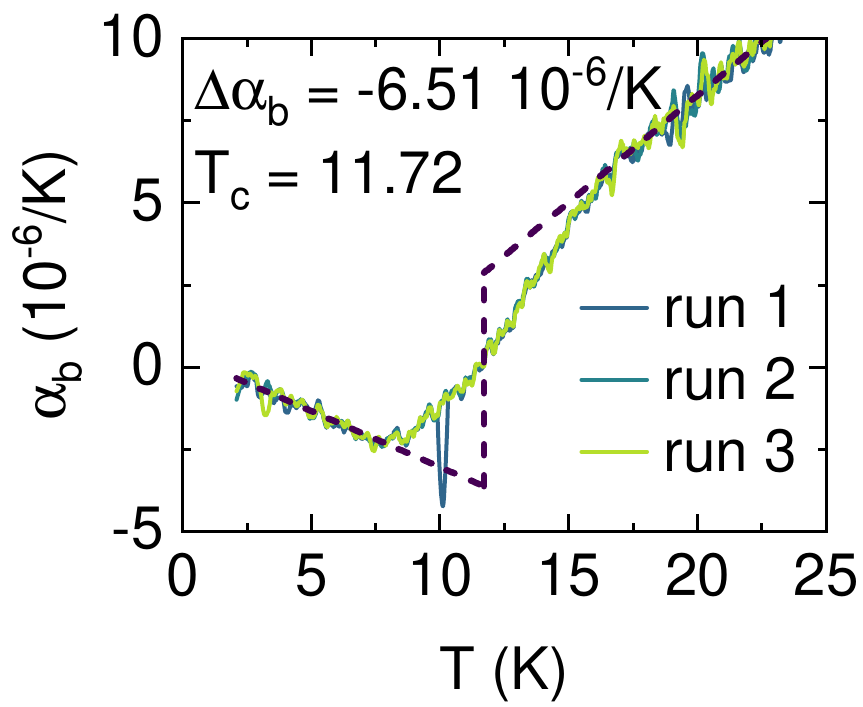}
	\caption{
        Thermal expansion coefficient for the $b$ axis versus temperature.
        The dashed line is the result of an equal area construction to determine T$_C$.
    }
	\label{fig:thermal_exp}
\end{figure*}

Thermal expansion was measured along the $b$ axis as a function of temperature, as shown in Fig.~\ref{fig:thermal_exp}.
Compared to other measurements, T$_{\mathrm{C}}$ is reduced to about 12~K.
This is not surprising because there is a reasonably large negative jump in thermal expansion along the $b$~axis, and any strain induced from mounting the crystal will result in a reduction in T$_{\mathrm{C}}$.
The Ehrenfest relation can be used to determine the pressure dependence of a second-order phase transition through the ratio of the volumetric change ($\Delta\beta{}V_m$) and the heat capacity jump ($\Delta{}C_p/T_C$):
\begin{equation}
    \frac{dT_C}{dp}=\frac{\Delta{}\beta{} V_m}{\Delta{}C_p/T_c}.
\end{equation}
This is only an approximation for uniaxial pressure, but it is still useful for determining the relative sensitivity of the transition temperature to uniaxial pressure.
Using the value of $\Delta{}C_{\mathrm{mag}}/T_C$ from above (5 mJ mol$^{-1}$ K$^{-2}$) and the value of $\Delta\alpha_b$ from Fig.~\ref{fig:thermal_exp} (10$^{-6}$ K$^{-1}$) yields a $\frac{dT_C}{dp}$ of -91~K/GPa.
Mounting the sample in the thermal expansion cell induces a small, non-uniform uniaxial pressure along the $b$ axis that will both suppress and broaden the transition as observed here.

\newpage
\section{Arrott plot}

\begin{figure*}[!ht]
	\includegraphics[width=0.4\textwidth]{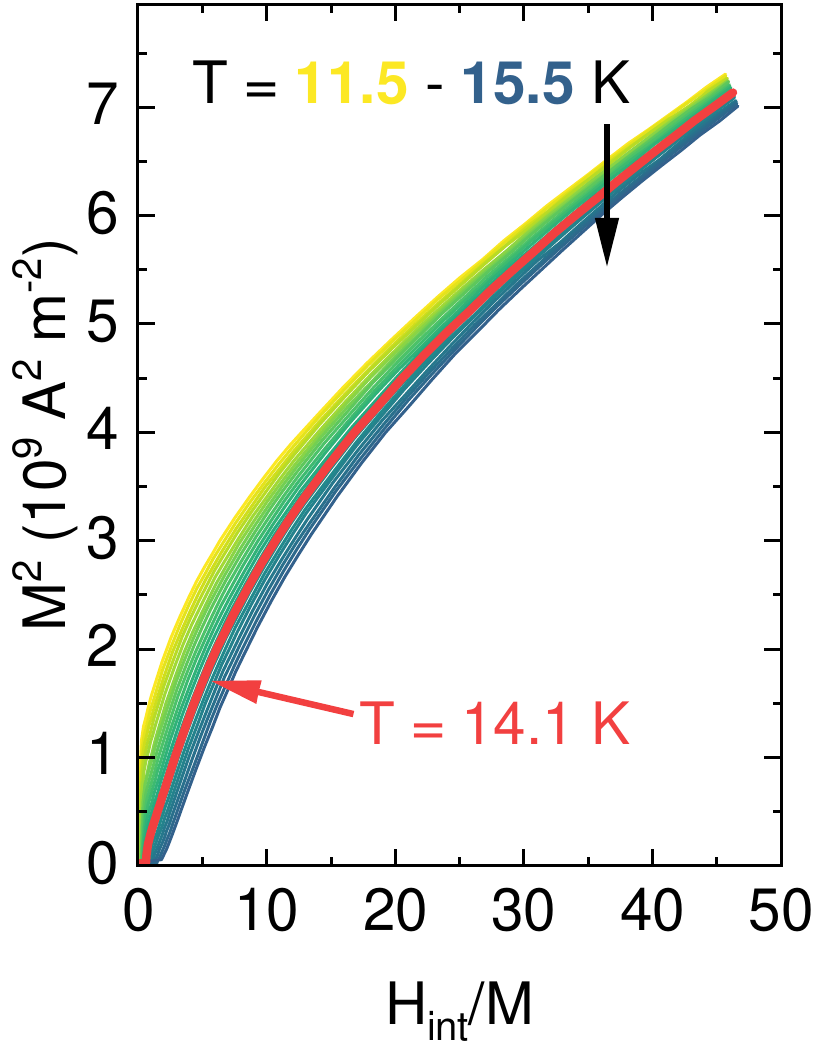}
	\caption{
        An Arrott plot for temperatures near T$_\textrm{C}$ indicates that the transition temperature is approximately 14.1 K (red curve).
    }
	\label{fig:arrott}
\end{figure*}

\newpage
\section{Additional density functional theory (DFT) calculations}

\begin{figure*}[!ht]
	\includegraphics[width=0.8\textwidth]{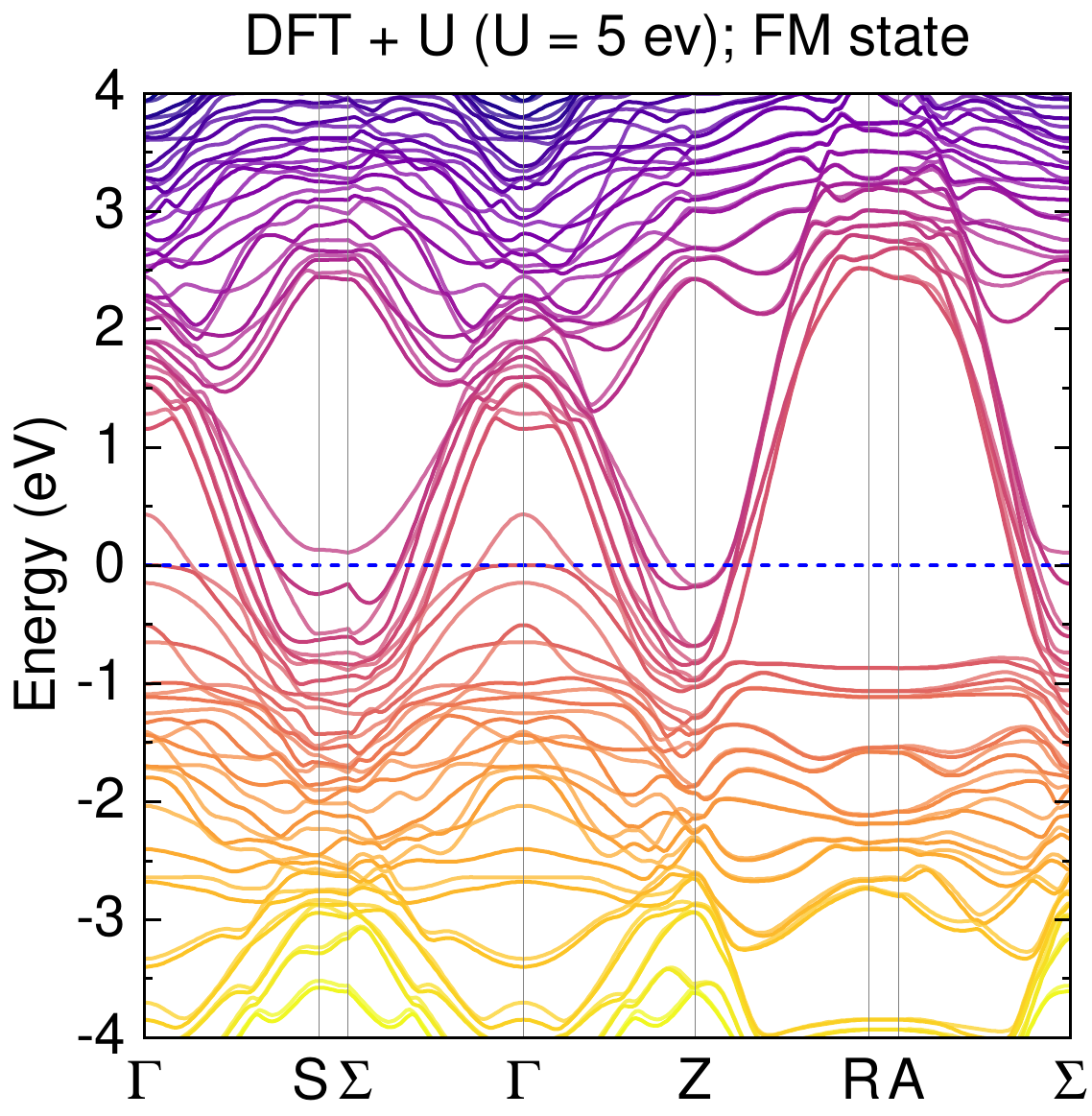}
	\caption{DFT+U band structure calculation in the FM state with a Coulomb term $U=5$~eV.}
	\label{fig:dft-fm}
\end{figure*}

Figure~\ref{fig:dft-fm} shows the DFT+U band structure calculation in the FM state with a Coulomb term $U=5$~eV.

\newpage
\section{Bulk single crystal image}

\begin{figure*}[!ht]
	\includegraphics[width=0.6\textwidth]{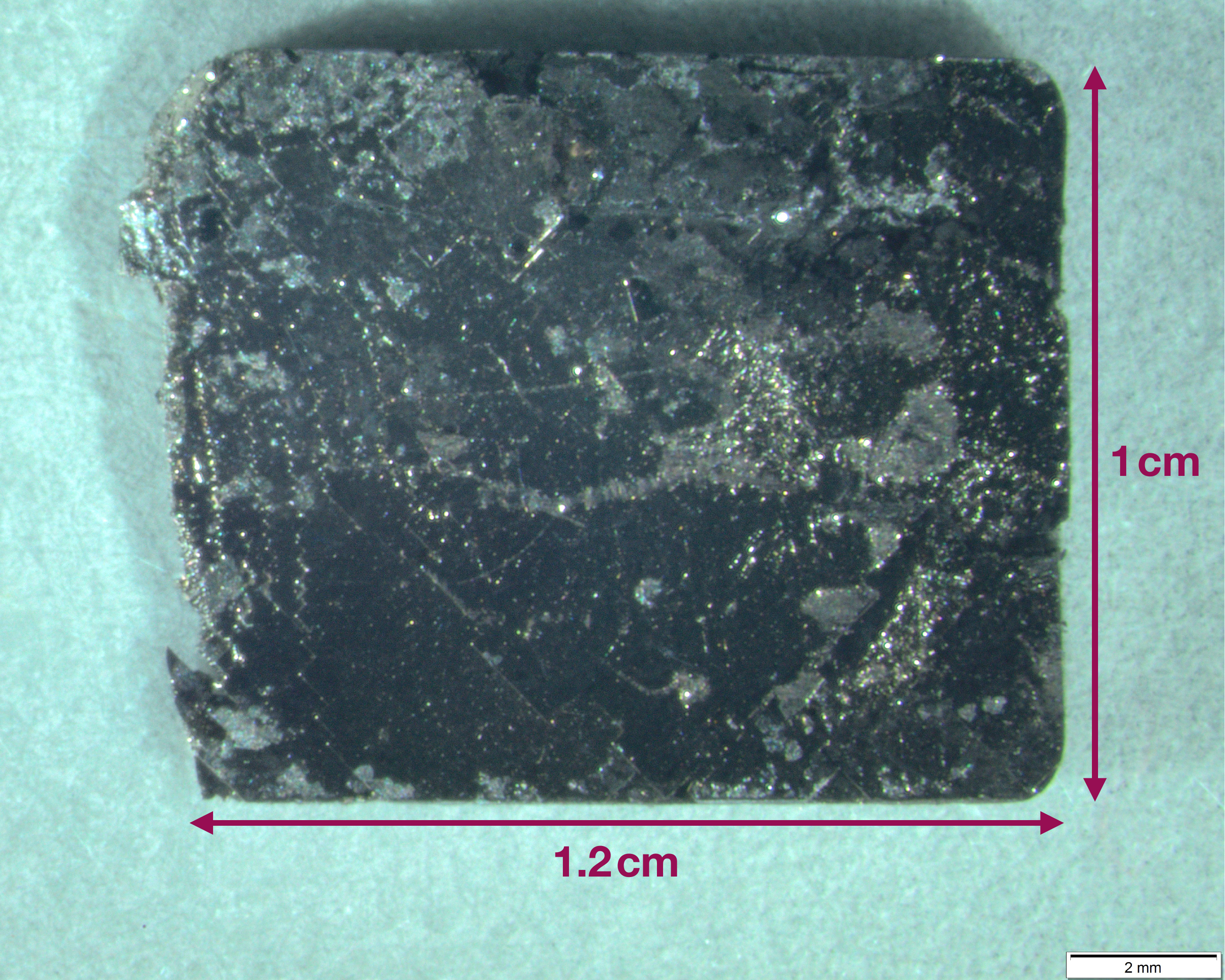}
	\caption{Image of a large bulk single crystal of $\beta$-UTe$_{3}$.}
	\label{fig:image}
\end{figure*}

\clearpage\newpage

\section{Crystal Structure Refinement}
To confirm the crystal structure of UTe$_3$, which belongs to orthorhombic space group Cmcm, we recorded neutron diffraction data at $T=25\,$K on WISH (ISIS). The data were first formally indexed in the reciprocal space of an orthorhombic lattice with parameters $a=4.338\,$\AA, $b=24.743\,$\AA, and $c=4.338\,$\AA. According to X-ray diffraction, lattice parameters $a$ and $c$ are equal up to the third digit after the comma. Therefore, the definition of axes $H$ and $L$ is not directly clear from the $d$-spacing of Bragg peaks in time-of flight data and Bragg peaks cannot be directly assigned to the twins in the sample via the $d$-spacing. For initial analysis steps, we therefore made an arbitrary choice of $H$ and $L$ and indexed Bragg peaks in this coordinate frame.

To integrate Bragg peaks, we utilized the automatic peak detection of the Mantid software. Peaks with ratio of uncertainty $\sigma$ to integrated intensity $I$ larger than 0.33 were discarded. We eventually proceeded with a list of 111 integrated Bragg reflections.

In principle, the Bragg peak reflection rules of the Cmcm space group permit to distinguish the two axes $H$ and $L$ from each other. But our data did not allow us to make this decision consistently due to crystal twins that are related by a 90 deg angle around the long van-der-Waals axis. To illustrate this, we performed crystal structure refinements on the integrated data using the software Jana \cite{2023_Petricek_ZFurKrist-CrystMater}. We refined an overall scale parameter, a Debye-Waller parameter, an isotropic Gaussian extinction parameter, and the atomic positions. As refinements with anisotropic harmonic Debye-Waller parameters did not improve the refinements, we used isotropic Debye-Waller parameters.

\begin{figure*}[h]
    \includegraphics[]{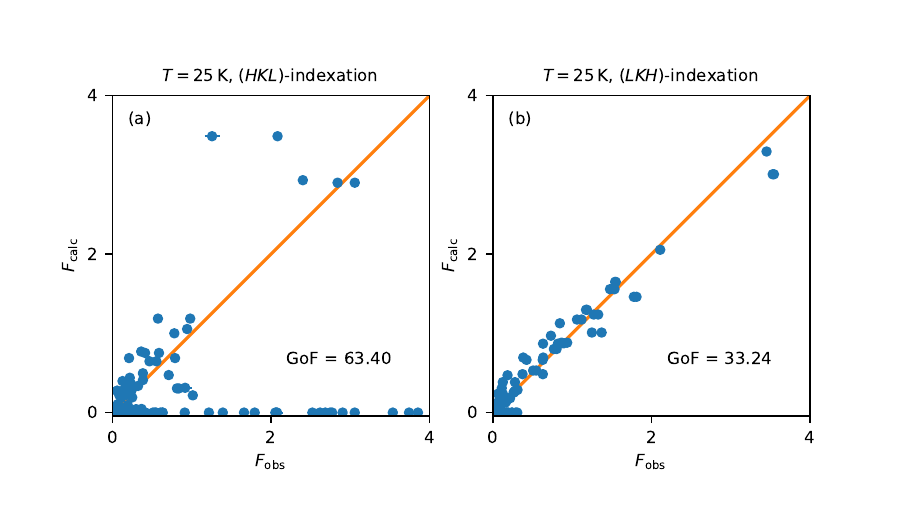} 
    \caption{Crystal structure refinement of a single-grain Cmcm structure. Experimentally observed structure factor inferred from Bragg peaks, $F_{\text{obs}}$, is compared with the calculated structure factor of a Cmcm structure with (a) momentum-space coordinates $(HKL)$ and (b) with coordinates $(LKH)$. The orange line denotes the region, where $F_{\text{calc}}=F_{\text{obs}}$.  }
    \label{fig:FigureNeutronS2}
\end{figure*}

Fig. \ref{fig:FigureNeutronS2}(a) presents the comparison of observed structure factor with the calculated structure factor for the crystal structure model as defined by our initial choice of the axes $H$ and $L$. The x-axis shows for each Bragg peak the structure factor seen experimentally and the $y$-axis the structure factor calculated for space group Cmcm in the coordinate frame $(HKL)$ that we initially defined. The large deviation of peaks from $F_{\text{calc}}=F_{\text{obs}}$ (orange line) illustrates the bad agreement of the model. Further, the peaks on the x-axis, which possess odd values of $H$, are forbidden by the reflection rules of the Cmcm crystal structure and therefore reflect the inadequacy of the refined structural model.

Fig. \ref{fig:FigureNeutronS2}(b) shows the results for alternative definition of momentum space, where the $H$ and $L$ axis are swapped. Although the agreement is much better than before, there are again peaks on the x-axis that possess odd $H$-coordinate and that are therefore forbidden in the refined structural model.

\begin{figure*}[h]
    \includegraphics[]{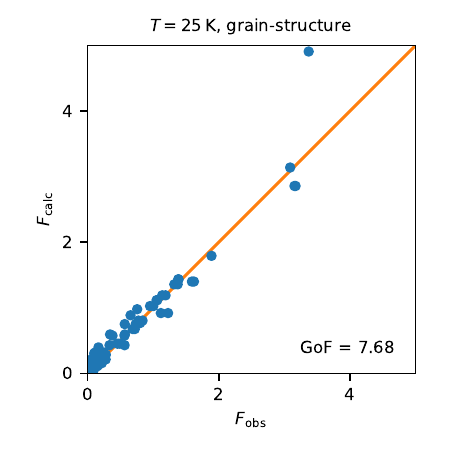} 
    \caption{Calculated vs. observed structure factor for a crystal structure refinement with twin structure. The twins are related by a 90 deg rotation around the vdW axis.}
    \label{fig:FigureNeutronS3}
\end{figure*}

A considerably better fit is obtained assuming that the sample possesses twins that are related by a 90 deg rotation around the van-der-Waals axis.  Fig.~\ref{fig:FigureNeutronS3} shows the results of the refinement, optimizing the scale parameter, the isotropic displacement, extinction, the twin ratio, and the uranium occupancy. The model with twin structure can now account for all peaks that were forbidden for a single-grain structural model and allover the agreement is relatively good. Quantitatively, the refinement is characterized by the goodness of fit parameters $GoF=7.68$, $R=8.88$, and $wR2=19.98$. The twin ratio determined in the refinement indicates that 13 percent are in the $(HKL)$-domain, see \ref{fig:FigureNeutronS2}(a), and  87 percent in the $(LKH)$ domain, see (b). The uranium occupancy is given by 92(1) percent. Isotropic displacement and extinction parameters refined to $G_{\text{iso}}=0.015(2)$ and $G_{\text{iso}}=0.0030(8)$.

\clearpage\newpage

\section{Magnetic Structure Determination}
To determine the ordered magnetic moment of UTe$_3$, we performed neutron diffraction at $T=1.5\,$K. Magnetic Bragg peaks were observed at the same momentum transfers as structural Bragg peaks, therefore suggesting that the scattering is located at the $\Gamma$-point associated with wave-vector $\bm{k}_0=(0,0,0)$. 

In total, 108 Bragg reflections were indexed and integrated (as before, peaks with low statistics were discarded). The refinement of magnetic structures with wave-vector $(0,0,0)$ is frequently challenging, as magnetic intensities coincide with nuclear Bragg intensities in momentum space. In addition, magnetic symmetry analysis may provide many different structures that display almost the same diffraction pattern. Group theoretical magnetic symmetry analysis reveals six representations for the wave-vector $\bm{k}_0=(0,0,0)$, which are denoted mGM2+, mGM3+, mGM4+, mGM1-, mGM2-, and mGM3-. The first three representations are ferro- and the last three antiferromagnetic.

As bulk measurements on UTe$_3$ suggested ferromagnetism with magnetic moments along the $b$-axis and, in addition, signatures of antiferromagnetic components are absent in our diffraction data, we probed three different magnetic structures, namely moments along the $c$-axis (mGM2+), along the $a$-axis (mGM3+), and along the $b$-axis (mGM4+). For the refinements, we fixed the allover scale parameter, the extinction parameter and the twin fraction to the value inferred from structural refinements and fitted the the isotropic displacement parameter as well as all components of magnetic moments that are permitted in the respective irreducible representation.

We further assumed magnetism carried by magnetic U$^{3+}$ ions with formfactor:
\begin{align}
    f= \left\langle j_0 \right\rangle + c_2\cdot\left\langle j_2 \right\rangle \, .
 \end{align}
For the radial integrals of spherical Bessel functions, $\left\langle j_n \right\rangle$, we took the values tabulated in Ref. \cite{1976_Freeman_PhysRevB}. For the constant, $c_2$, we considered the value obtained for the Hunds-rule Russell-Saunders ground state given by $c_2=1.75$ (see also Ref. \cite{1976_Freeman_PhysRevB}).

\begin{table}[ht!]
\caption{Magnetic structure refinements. The table summarizes the values $R$, $wR2$, and GoF for the refinement of the three models mGM2+, mGM3+, and mGM4+ with data recorded at $T=1.5\,$K. Values $R$ and $wR2$ are shown separately for magnetic and nuclear Bragg peaks. Although mGM3+ and mGM4+ GoF are nearly identical, consideration of the (130) and (020) peak intensities identify mGM4+ as the correct magnetic order (see SI text).\label{tab:TableMagneticRefinements}}
\def\arraystretch{1}
\setlength\extrarowheight{0pt}
\begin{tabular}{c | c c c c c}
IR& $R_{\text{struc}}$ &$wR2_{\text{struc}}$      & $R_{\text{mag}}$&$wR2_{\text{mag}}$  & GoF  \\
\hline  
mGM2+ & 12.08  & 24.25   & 21.23 & 39.42  & 11.02   \\
mGM3+ & 10.44  & 20.75   & 19.77 & 38.21  & 10.36   \\
mGM4+ & 10.64  & 22.74   & 20.05 & 38.50  & 10.51   
\end{tabular}
\end{table}

\begin{figure*}[h]
    \includegraphics[]{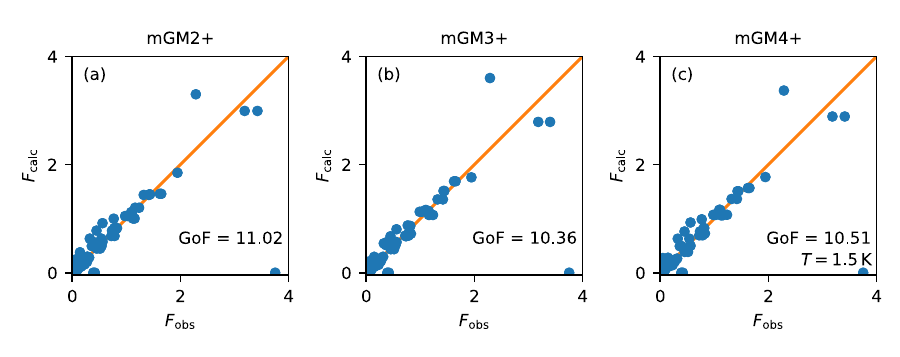} 
    \caption{Combined magnetic and crystal structure refinements at $T=1.5\,$K. Neutron diffraction data, comprising of a list of 108 integrated Bragg intensities, were compared to ferromagnetic spin textures with moments (a) along the $c$-axis, (b) along the $a$-axis, and (c) along the $b$-axis. The blue data points compare for each Bragg peak the recorded and calculated structure factor.}
    \label{fig:FigureNeutronS4}
\end{figure*}

Fig.~\ref{fig:FigureNeutronS4} compares the structure factors of the three refined models with the neutron diffraction data. A summary of all fit-parameters is further provided in Tab. \ref{tab:TableMagneticRefinements}. The plots and the goodness of fit parameters (GoF) look relatively similar and the refinements themselves do not permit to distinguish the magnetic structures.  However, as we explain below, the structure factor on well selected momentum space positions clearly shows that magnetic moments are essentially aligned along the $b$-axis with a small possible canting angle.

The refined magnetic moment obtained for the best fit with mGM4+ is given by $M=0.48(11)\,\mu_{\text{B}}$.

In order to further improve the refinement results, we superposed the irreducible representation mGM4+ with other irreducible representations associated with $\bm{k}=(0,0,0)$ and $\bm{k}=(1,0,0)$. But none of these superpositions of two IRs resulted in a converging refinement with better statistics. Taken together, refinements therefore suggest magnetic long-range order with wave-vector $\bm{k}=(0,0,0)$ and magnetic moments aligned along the $b$-axis.

Neutron diffraction data at carefully selected momentum-space positions show further that the magnetic ground state has almost negligible projection perpendicular to the $b$-axis. Therefore, the ground state corresponds essentially to ferromagnetic order with moments aligned along the $b$-axis.

\begin{figure*}[h]
    \includegraphics[]{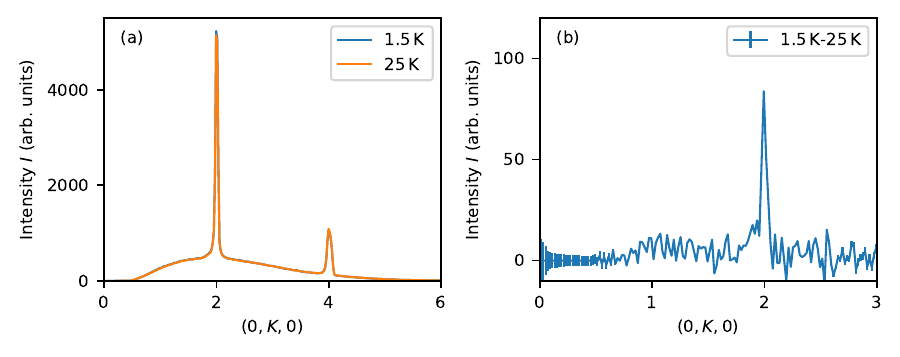} 
    \caption{Magnetic ordering vector at $T=1.5\,$K. Neutron diffraction data shown were recorded at $T=1.5\,$K and at $T=25\,$K. The thermal variation displays Bragg intensity at the same momentum transfers as nuclear Bragg diffraction, therefore suggesting zero momentum wave-vector.}
    \label{fig:FigureNeutronS9}
\end{figure*}

Fig. \ref{fig:FigureNeutronS9} shows temperature subtracted diffraction data on the line $(0,K,0)$. The peak at $K=2$ and the absence of intensity at $K=1$ indicate the presence of a weak $\bm{k}=0$ component perpendicular to the $b$-axis as well as the absence of an antiferromagnetic component.
 
To assess the magnitude of components perpendicular to $b$ (denoted $m_{\perp}$) and along $b$ (denoted $m_{\parallel}$), we consider integrated intensities of specific deliberately chosen magnetic Bragg peaks. The magnetic Bragg intensity at $(020)$ is purely due to components perpendicular to $b$, denoted $m_{\perp}$, and given by $I_{m}(020)=2454\pm1036$. The Bragg peak at (130), in turn, is due to both $m_{\perp}$ and $m_{\parallel}$ and has intensity $I_{m}(130)=194333\pm5448$.

Comparing these intensities with the magnetic structure factor calculated for collinear ferromagnetic spin texture with moment projections both along $m_{\perp}$ and $m_{\parallel}$ (with $\text{U}^{3+}$ form factor and neglected Debye-Waller factor) yields a ratio $m_{\perp}=m_{\parallel}\cdot 0.07(4)$, where $m_{\perp}$ points either along $[100]$ or $[001]$. 

Taken together, the comparison of integrated intensities at (130) and (020) show that the ground state is essentially a $b$-axis ferromagnet, possibly with a tiny tilt of moments either along $[100]$ or $[001]$.

\clearpage\newpage

\section{Mosaicity and integration of Bragg peaks}
\begin{figure*}[h]
    \includegraphics[]{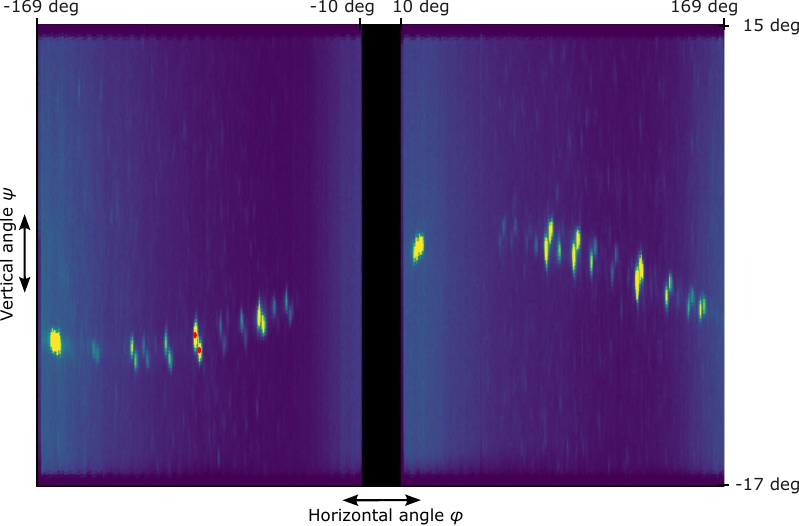} 
    \caption{Typical detector image recorded on Wish. The exemplary image was taken at $T=1.5\,$K. The bright spots are associated with Bragg peaks. The splitting of Bragg peaks is due to finite mosaicity and grains in our sample. The two red dots correspond to the same Bragg peaks in different grains.}
    \label{fig:FigureNeutronS5}
\end{figure*}

Our sample used in neutron diffractions showed two separate crystals misaligned by a couple of degrees each showing 90 degrees twinning. Fig.~\ref{fig:FigureNeutronS5} illustrates this showing an exemplary detector image taken on WISH. Peaks clearly display splitting (bright spots). The grains are spread over an angular range of the order 3 deg. For example the two red dots are separated by a horizontal angle $\Delta\phi=2.1\,$deg and a vertical angle $\Delta \Psi=0.7\,$deg.

For integration, we chose radii on the detector large enough to cover Bragg peak of all grains around a chosen position.

\clearpage\newpage

\section{Magnetic wave-vector inferred from neutron scattering}
To determine magnetic ordering vectors at low temperature, we performed neutron diffraction below the transition temperature (at $T=1.5\,$K) and above the transition temperature (at $T=25\,$K).

\begin{figure*}[h]
    \includegraphics[]{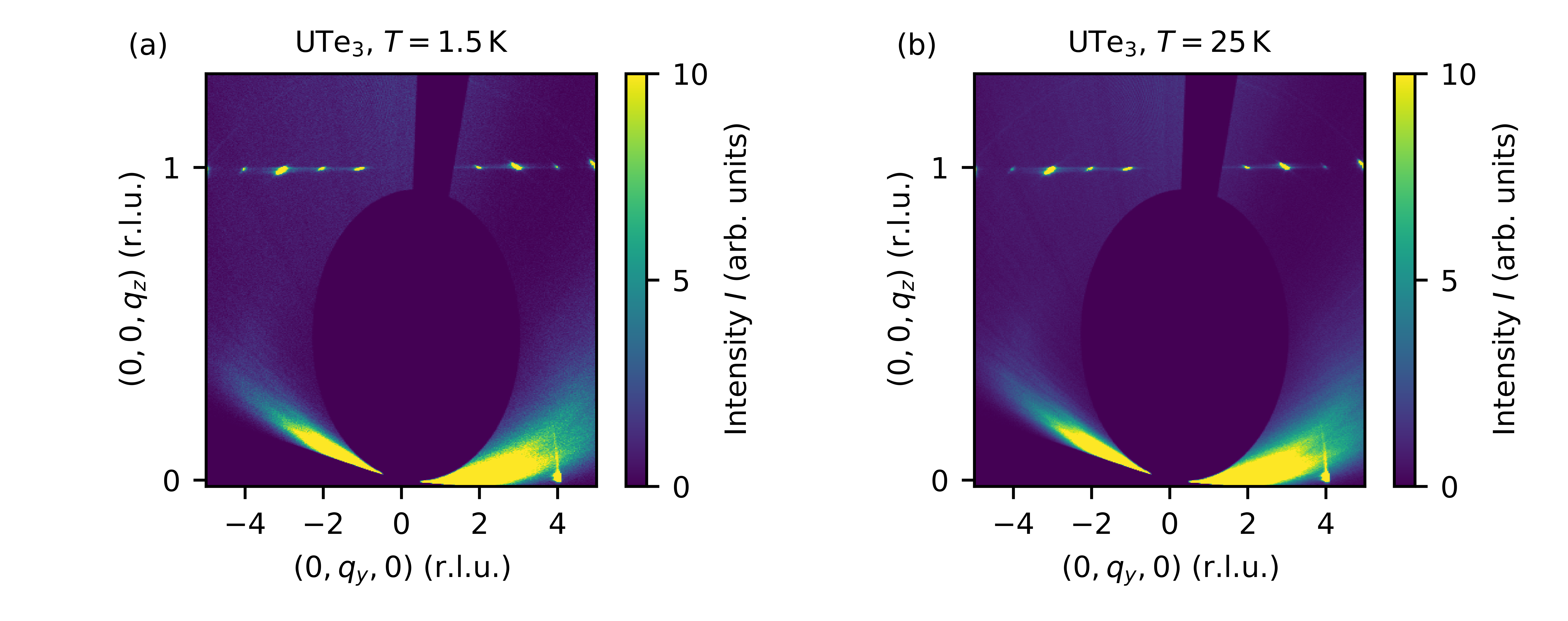} 
    \caption{Magnetic ordering vector at $T=1.5\,$K. Neutron diffraction data shown on the lines (a) $(0,K,1)$ and (a) $(0,K,2)$ were recorded at $T=1.5\,$K and at $T=25\,$K. The thermal variation displays Bragg intensity at the same momentum transfers as nuclear Bragg diffraction, therefore suggesting wave-vector at zero momentum.}
    \label{fig:FigureNeutronS13}
\end{figure*}

Data are presented in Fig. \ref{fig:FigureNeutronS13}. 

\begin{figure*}[h]
    \includegraphics[]{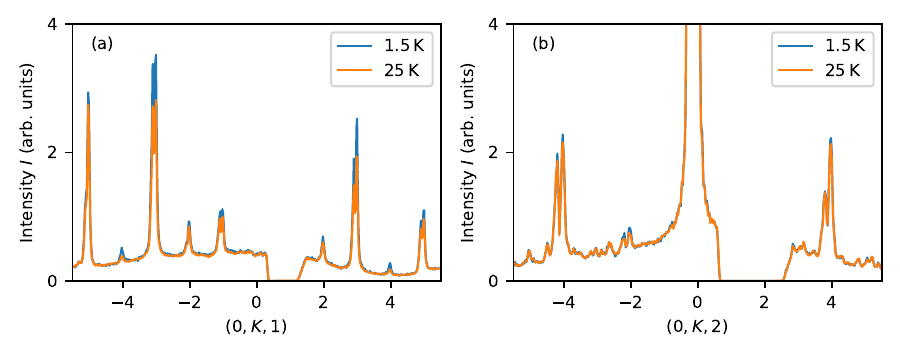} 
    \caption{Magnetic ordering vector at $T=1.5\,$K. Neutron diffraction data shown on the lines (a) $(0,K,1)$ and (a) $(0,K,2)$ were recorded at $T=1.5\,$K and at $T=25\,$K. The thermal variation displays Bragg intensity at the same momentum transfers as nuclear Bragg diffraction, therefore suggesting wave-vector at zero momentum.}
    \label{fig:FigureNeutronS6}
\end{figure*}

Fig. \ref{fig:FigureNeutronS6} compares the two diffraction data-sets. Magnetic scattering intensity is observed at integer-valued momentum transfers $(h,k,l)$, therefore suggesting magnetic ordering wave-vector $\bm{k}=(0,0,0)$ ($\Gamma$-point). Magnetic ordering at the $Y$-point would be antiferromagnetic and would result in diffraction intensity at momentum-transfers, where structural Bragg peaks are absent, such as $(030)$.  In our data-set, we only observed magnetic scattering at the same momentum-transfers, where structural Bragg peaks appear, whereas $(0,3,0)$ do not display any magnetic scattering, therefore suggesting that the magnetic ordering wave-vector is $\bm{k}=(0,0,0)$ ($\Gamma$-point). As bulk characterization suggested ferromagnetism, we conclude, that the wave-vector is indeed at the $\Gamma$-point.

\begin{figure*}[h]
    \includegraphics[]{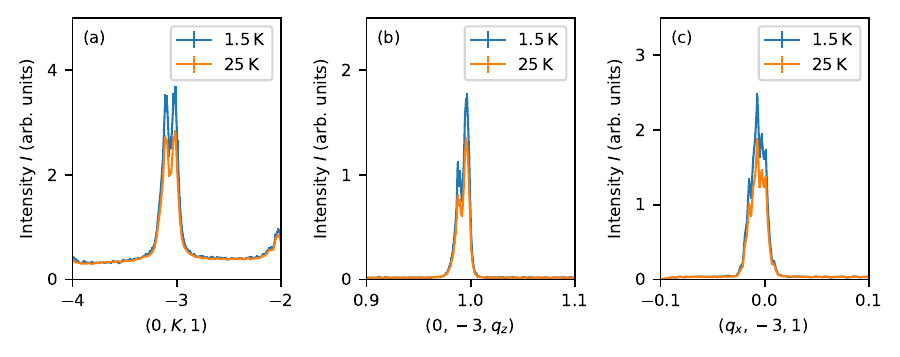} 
    \caption{Cut through neutron diffraction data around the Bragg peak at $(0,-3,1)$. Along all three momentum-space directions (a) $H$, (b) $q_z$, and (c) $q_x$, the magnetic Bragg peak displays the same width as the structural Bragg peak.}
    \label{fig:FigureNeutronS7}
\end{figure*}

Magnetic Bragg peaks display the same width in all momentum space-directions as structural Bragg peaks, indicating the three-dimensional nature of the magnetic long-range order. Fig. \ref{fig:FigureNeutronS7} illustrates this, showing cuts through $\bm{Q}=(0,-3,1)$ along the three different momentum-space directions $K$, $L$, and $H$.

\clearpage\newpage

\section{Transition temperature inferred from neutron scattering}

The temperature dependence of magnetic order parameter was inferred from the magnetic Bragg peak at $(310)$, where magnetic scattering intensity was strongest.

The thermal variation of the magnetic intensity is shown in Fig. 2\textbf{b} of the main text and was fitted with order parameter curves:
\begin{align}
    I(T) := I_0\cdot \left( 1-\frac{T}{Tc}  \right)^{2\beta}
\end{align}
by means of least-squares fits.

\begin{figure*}[h]
    \includegraphics[]{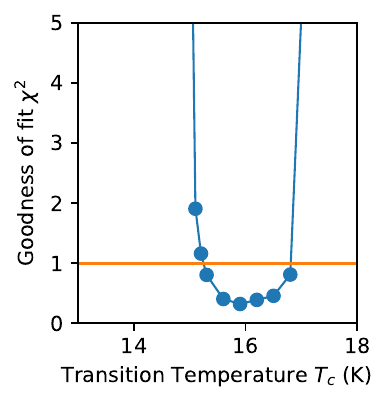} 
    \caption{Goodness of order-parameter fit to neutron diffraction data. The circle symbols denote the $\chi^2$ values obtained for different transition temperatures. The orange line corresponds to the border of $\chi^2\leq1$, which we considered as acceptable range for the order-parameter fit.}
    \label{fig:FigureNeutronS10}
\end{figure*}

In order to determine the transition temperature $T_c$, we repeated the fit for different values of $T_c$ and made an assessment of the fit based on the $\chi^2$ values calculated on all data-points below the respective $T_c$. Fig. \ref{fig:FigureNeutronS10} shows the goodness of the fit as a function of $T_c$. The best fit (lowest $\chi^2$) is obtained for $T_c=15.9$\,K. We further find that $\chi^2$ is smaller than 1 for temperatures $15.3\leq T_c\leq16.8$ and take the respective temperature range as uncertainty for $T_c$. Taken together, the transition temperature is given by $T_c=15.9(9)$\,K

\clearpage\newpage

\section{Correlation length of diffuse magnetic scattering}
\begin{figure*}[h]
    \includegraphics[]{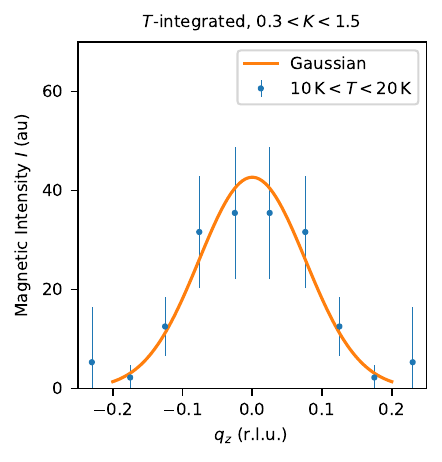} 
    \caption{Diffuse magnetic scattering intensity integrated over the temperature range from 10 K to 20 K for different values of $q_z$. The orange line corresponds to a Gaussian peak shape with FWHM $f_1=3.6\cdot10^{-2}\,$r.l.u..}
    \label{fig:FigureNeutronS11}
\end{figure*}

In order to obtain an estimate of the correlation length of the diffuse magnetic scattering observed along the line $\bm{Q}=(0,K,0)$, we integrated the diffuse scattering presented in Fig. 2\textbf{c} of the main text along the temperature-axis. The resulting intensity as a function of $\Delta L$ is presented in Fig. \ref{fig:FigureNeutronS11}.

The profile displays a Gaussian shape with full width at half maximum $f_1=1.79\cdot10^{-1}\,$r.l.u.. The peak profile is a convolution of intrinsic peak-width, $f_0$, and experimental resolution, $f_R$. Assuming a Gaussian profile for resolution and intrinsic peak width, we obtain $f_1=\sqrt{f_0^2+f_R^2}$.

The experimental resolution can be inferred from the peak at $\bm{Q}=(0,2,0)$, which along the $L$-axis exhibts a full width at half maximum given by $f_R=6.9\cdot10^{-3}\,$r.l.u.. Intrinsic peak-width is therefore given by $f_0=1.70\cdot10^{-1}\,$r.l.u. and the correlation length along the $L$-axis is given by $\kappa_0 = 2/f_0=7.7\,$\AA .

\bibliography{lib}